\documentclass[superscriptaddress,twocolumn,secnumarabic,
nobibnotes,aps,prd,showpacs,nofootinbib]{revtex4}
\usepackage{eurosym}
\usepackage{graphicx}
\usepackage{epsf}
\usepackage{bm}
\usepackage{amsmath}
\usepackage{amsfonts}
\usepackage{amssymb}
\usepackage{epstopdf}
\usepackage{color}
\providecommand{\U}[1]{\protect\rule{.1in}{.1in}}

\newcommand{\be}{\begin{equation}}
\newcommand{\ee}{\end{equation}}

\newcommand{\mincir}{\raise
-3.truept\hbox{\rlap{\hbox{$\sim$}}\raise4.truept\hbox{$<$}\ }}
\newcommand{\magcir}{\raise
-3.truept\hbox{\rlap{\hbox{$\sim$}}\raise4.truept\hbox{$>$}\ }}

\begin{document}

\title{Imprints of an extended Chevallier-Polarski-Linder parametrization on
the large scale of our universe}
\author{Supriya Pan}
\email{supriya.maths@presiuniv.ac.in}
\affiliation{Department of Mathematics, Presidency University, 86/1 College Street,
Kolkata 700073, India}
\author{Weiqiang Yang}
\email{d11102004@163.com}
\affiliation{Department of Physics, Liaoning Normal University, Dalian, 116029, P. R.
China}
\author{Andronikos Paliathanasis}
\email{anpaliat@phys.uoa.gr}
\affiliation{Institute of Systems Science, Durban University of Technology, PO Box 1334,
Durban 4000, Republic of South Africa}

\begin{abstract}
In the present work we perform a systematic analysis of a new dark energy
parametrization and its various corrections at first and higher orders
around the presence epoch $z=0$, where the first order correction of this
dark energy parametrization recovers the known Chevallier-Polarski-Linder
model. We have considered up to the third order corrections of this
parametrization and investigate the models at the level of background and
perturbations. The models have been constrained using the latest
astronomical datasets from  a series of potential astronomical data, such as the cosmic microwave background observations, baryon
acoustic oscillations measurements, recent Pantheon sample of the supernova
type Ia and the Hubble parameter measurements. 
From the analyses we found that all parametrization favor
the quintessential character of the dark energy equation of state where the
phantom crossing is marginally allowed (within 68\% CL). Finally, we perform the
Bayesian analysis using \texttt{MCEvidence} to quantify the
statistical deviations of the parametrizations compared to  the standard $\Lambda$CDM cosmology. The Bayesian analysis reports that  $\Lambda $CDM is favored over all the DE parametrizations.
\end{abstract}

\maketitle
\date{\today}

%---------------------------------------

%\title{Imprints of a new dark energy parametrization and its extensions}

%%%%%%%%%%%%%%%%%%%%%%%%%%%%%%%%%%%%%%%%%%%%%%%%%%%%%%%%%%%%%%%%%

\section{Introduction}

According to the theory of general relativity, one possible way to describe
the recent observational evidences is to introduce the dark energy, a
hypothetical fluid with large negative pressure \cite{Copeland:2006wr}.
However, apart from this negativity condition on the pressure of dark energy, no one knows what
exactly this particular fluid is. The simplest explanation to the dark
energy theory comes through the introduction of positive cosmological constant, $\Lambda$, which does not evolve with the time. But, the cosmological constant already suffers
from two major problems, one which is recognized as the cosmological
constant problem and the other is the cosmic coincidence problem. Thus,
although as stated by a series of observational data, the $\Lambda $%
-cosmology is an elegant version to model the recent observational features
of the universe, the problems associated with the above motivate us to think
of the scenarios beyond the standard $\Lambda $-cosmology paradigm. 

The simplest extension to $\Lambda$-cosmology is the $w_x$-cosmology in
which $w_x$ is the dark energy equation-of-state quantified as the ratio of
pressure to its density, mathematically which is $w_x = p_x /\rho_x$. One
can identify that $p_x$ and $\rho_x$ are respectively the pressure and
energy density of the dark energy fluid. The equation-of-state $w_x$ being $%
-1$ recovers the $\Lambda$-cosmology. In general one can assume $w_x $ ($%
\neq -1$) to be either time independent or dependent while the latter
scenario is the most general one. Thus, in the present work we shall focus on
the alternative cosmologies to the $\Lambda$-cosmology in which the dark
energy equation-of-state is evolving with the expansion of the universe.

The parametrization of $w_{x}$ could be any function of the redshift $z$ or
the scale factor $a(t)$ of the Friedmann-Lema\^{\i}tre-Robertson-Walker
universe; note that, $1+z=a_{0}/a(t)$, where $a_{0}$ is the present value of
the scale factor in this universe. Thus, since $w_{x}\equiv w_{x}(z)\equiv
w_{x}(a)$ could be any arbitrary function of the redshift or the scale
factor, therefore, in principle this gives us a complete freedom to pick up
any particular model of interest and test it with the observational data in
order to see whether that model is able to correctly describe the evolution
of the universe. In fact one can realize that the introduction of the dark
energy equation-of-state is a reverse mechanism to probe the expansion
history of the universe. Going back to literature, one can find that this
particular area of cosmology has been investigated well both at the level of background
and perturbations where various parametrizations for $w_{x}$ were
introduced earlier \cite{Chevallier:2000qy, Linder:2002et, Cooray:1999da,
Efstathiou:1999tm, Astier:2000as, Weller:2001gf, Wetterich:2004pv, Hannestad:2004cb, Jassal:2005qc,Gong:2005de,Feng:2004ff} and later 
\cite{Nojiri:2006ww, Zhao:2006qg, Kurek:2007bu, Barboza:2008rh, Saridakis:2008fy,Lazkoz:2010gz, Ma:2011nc,Li:2011dr,Feng:2011zzo,Sendra:2011pt,
Feng:2012gf,DiValentino:2016hlg, Zhao:2017cud, DiValentino:2017zyq,
DiValentino:2017gzb, Yang:2017amu, Rezaei:2017yyj, Marcondes:2017vjw, Yang:2017alx,Jaber:2017bpx,
Pan:2017zoh, Vagnozzi:2018jhn, Li:2018nlh, Panotopoulos:2018sso, Jaime:2018ftn, Yang:2018qmz, Yang:2018prh, Pace:2018xqy, Du:2018tia,Tamayo:2019gqj}. Precisely, the dark energy parametrization with only a single
free parameter, with two free parameters, with three free parameters and
finally with more than three parameters have been rigorously studied by
various investigators.

The aim of the present work is slightly different. Here, we are considering
an exponential dark energy parametrization that in its first order
approximation around $z=0$ recovers the CPL parametrization, and further we allow its
higher order corrections in order to investigate how such extended 
corrections affect the evolution of the universe both at the level of
background and perturbations. More specifically, we consider upto the third
order expansion of the exponential dark energy model. We remark that in
general every analytic function for the equation-of-state parameter around
the $z=0$ describes  the CPL
parametrization in the first correction;
however, while we want to assume a general Taylor
expansion of an analytic function $f\left( a\right) $ around %
$a=1$, i.e. $f\left( a\right) =\sum_{i=0}^{\infty
}w_{i}\left( a-1\right) ^{i}$, every new term which is introduced in
the correction provides a new degree of freedom, a free parameter, in the
model. Consequently, the models will have different degrees of freedom and
they will not be in comparison. Hence, special relations amount the
constants $w_{i}$ should be considered, and for our analysis we
assume that $w_{0}$ is free while $w_{j}=\frac{w_{1}}{j!}$%
, which $j\neq 0$, in which $f\left( a\right) $
is now the exponential function. However, by this approach we will get a
remarkable information on how the nonlinear terms in the parametrizations of the
equation-of-state affect the viability of the model in higher-redshifts. 

The work has been organized in the following way. In section \ref{sec-efe}
we introduce the models for $w_{x}(z)$ and describe the general equations at
the level of background and perturbations. After that in section \ref{sec-sf}
we provide an equivalence of the present dark energy parametrizations with the 
scalar field theory. Then 
in section \ref{sec-data} we
describe the observational data and the statistical analysis that are used
to constrain the models. After that in section \ref{sec-results} we describe
the observational constraints extracted from the models using the
astronomical data described in section \ref{sec-data}. Then in section \ref{sec-bayesian} we compute the evidences of the dark energy parametrizations through the 
\texttt{MCEvidence}. Finally, we close the
work in section \ref{sec-conclu} with a brief summary of everything.

\section{Basic Equations and the Models}

\label{sec-efe}

Considering a spatially flat Friedmann-Lema\^{\i}tre-Robertson-Walker
line element $ds^2 = -dt^2 + a^2 (t) \sum_{i=1}^{3} dx_i^2$ (where $a(t)$
is the expansion scale factor of the universe),
in the context of the Einstein gravity, we assume that (i) matter
is minimally coupled to gravity, (ii) there is no interaction between any
two fluids under consideration and (iii) all the fluids satisfy barotropic
equation of state, i.e., $p_{i}=w_{i}\rho _{i}$ in which $w_{i}$ being the
barotropic state parameter for the $i$-th fluid having $(\rho _{i},p_{i})$
as its the energy density and pressure, respectively.
Precisely, we consider that the total energy density of the universe is, $%
\rho _{tot}=\rho _{r}+\rho _{b}+\rho _{c}+\rho _{x}$ and the total pressure
thus becomes $p_{tot}=p_{r}+p_{b}+p_{c}+p_{x}$. Here, the subscripts $r$, $b$%
, $c$ and $x$ respectively stands for radiation, baryons, cold dark matter
and dark energy. Thus, the barotropic indices are, $w_{r}=1/3$, $%
w_{b}=w_{c}=0$ and we assume $w_{x}$ to be dynamical. The Einstein's field equations 
for the above FLRW universe can be written down as 

\begin{eqnarray}
&& H^2 = \frac{8 \pi G}{3} \rho_{tot}, \label{efe1}\\
&& \dot{H} = -4 \pi G (p_{tot} + \rho_{tot}),\label{efe2}
\end{eqnarray}
in which an overhead dot represents the cosmic time differentiation and 
$H \equiv \dot{a}/a$ is the Hubble rate of this universe. Now, using (\ref{efe1})
and (\ref{efe1}) (or alternatively the Bianchi's identity), one can 
find the balance equation 

\begin{eqnarray}\label{balance}
\dot{\rho}_{tot} + 3 H (p_{tot} + \rho_{tot}) = 0.
\end{eqnarray} 
Now, since as we assumed that we don't have any interaction between any two fluids 
of the universe, thus, they should satisfy their own conservation equation leading to 
\begin{eqnarray}
\dot{\rho}_{i} + 3 H (p_{i} + \rho_{i}) = 0,
\end{eqnarray}
from which using the relation between pressure and energy density for the radiation, baryons, and cold (pressureless-) dark matter,  one can find that $\rho_r = \rho_{r0}a^{-4}$, $\rho_m = \rho_b +\rho_c = (\rho_{b0} +\rho_{c0}) a^{-3}$. Here, $\rho_{i0}$ is the present value of $\rho_i$. And finally, the evolution of the dark energy fluid can be given by,
\begin{eqnarray}
\rho_{x}=\rho_{x0}\,\left(  \frac{a}{a_{0}}\right)  ^{-3}\,\exp\left(
-3\int_{a_{0}}^{a}\frac{w_{x}\left(  a'\right)  }{a'}\,da'
\right),
\end{eqnarray}
where $\rho_{x0}$ being the current value of $\rho_x$ and $a_0$ is the 
present value of the scale factor that we set to be unity ($a_0 =1$) 
without any loss of generality. We further note that the scale factor 
is related to the redhisft that we shall frequently use hereafter via
$1+z = a_0/a = 1/a$. Thus, once the dark energy equation of state is prescribed,
the evolution of the dark energy density can be found.   

As we discussed above, we consider that the dark energy fluid follows a
general parametrization in the following way:
\begin{equation}\label{model-general}
w_{x}(z)=(w_{0}-w_{a})+w_{a}\exp \left( \frac{z}{1+z}\right) 
\end{equation}%
where $w_{0}$ is the present value of the dark energy equation of state,
that means, $w_{x}(z=0)=w_{0}$ and $w_{a}$ is another free parameter. The
model (\ref{model-general}) is very interesting by its construction since
one can easily recognize that it could return a number of interesting
parametrization that includes the classic Chevallier-Polarski-Linder
parametrization $w_{x}(z)=w_{0}+w_{a}z/(1+z)$ if we take the first
approximation of the exponential function in (\ref{model-general}).

We expand the exponential function of (\ref{model-general}) upto its first,
second and third order corrections leading to the following class of dark
energy parametrization:

\begin{widetext}
\begin{eqnarray}
&& w_x (a) = w_0 + w_a \, \frac{z}{1+z}, \label{extension1}\\
&& w_x (a) = w_0 + w_a \left[\frac{z}{1+z} + \frac{1}{2!}\left(\frac{z}{1+z}\right)^2 \right],\label{extension2}\\
&& w_x (a) = w_0 + w_a \left[\frac{z}{1+z} + \frac{1}{2!}\left(\frac{z}{1+z}\right)^2 + \frac{1}{3!}\left(\frac{z}{1+z}\right)^3 \right],\label{extension3}
\end{eqnarray}
\end{widetext}and for convenience we call the dark energy parametrization of
equations (\ref{extension1}), (\ref{extension2}) and (\ref{extension3}) as
\textquotedblleft Extension 1\textquotedblright\ (Ext1 in short),
\textquotedblleft Extension 2\textquotedblright\ (Ext2 in short) and
\textquotedblleft Extension 3\textquotedblright\ (Ext3 in short),
respectively. Let us note that in the above  cases  we have not considered the pivot redshift \cite{Yang:2018prh}. However, the consideration of pivoting redshift might be an interesting issue for investigations because as already commented in \cite{Yang:2018prh}, one can find a specific value of the pivot redshift where the parameters $w_0$ and $w_a$ are uncorrelated.  

\begin{figure*}
\includegraphics[width=0.34\textwidth]{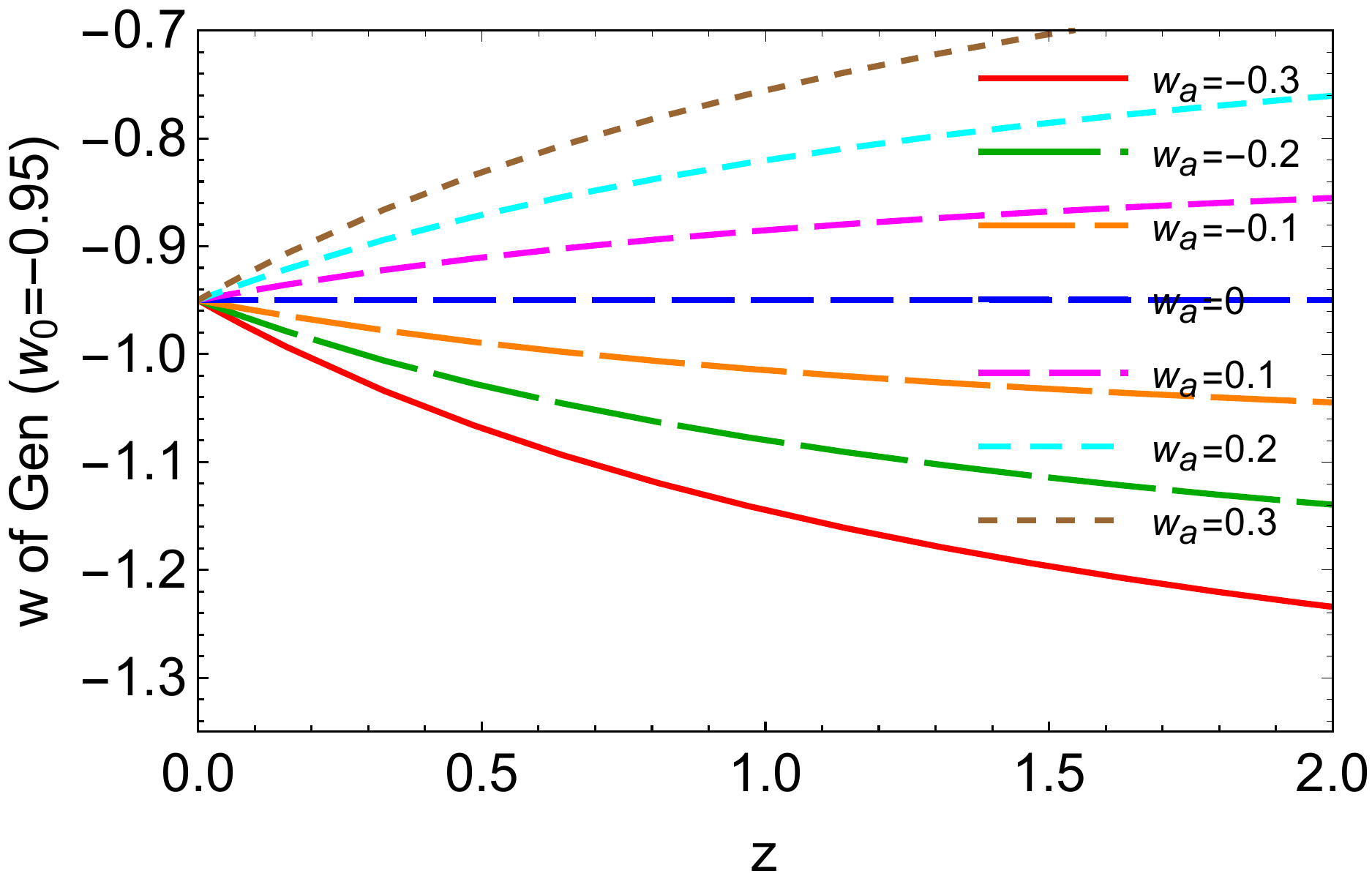}
\includegraphics[width=0.34\textwidth]{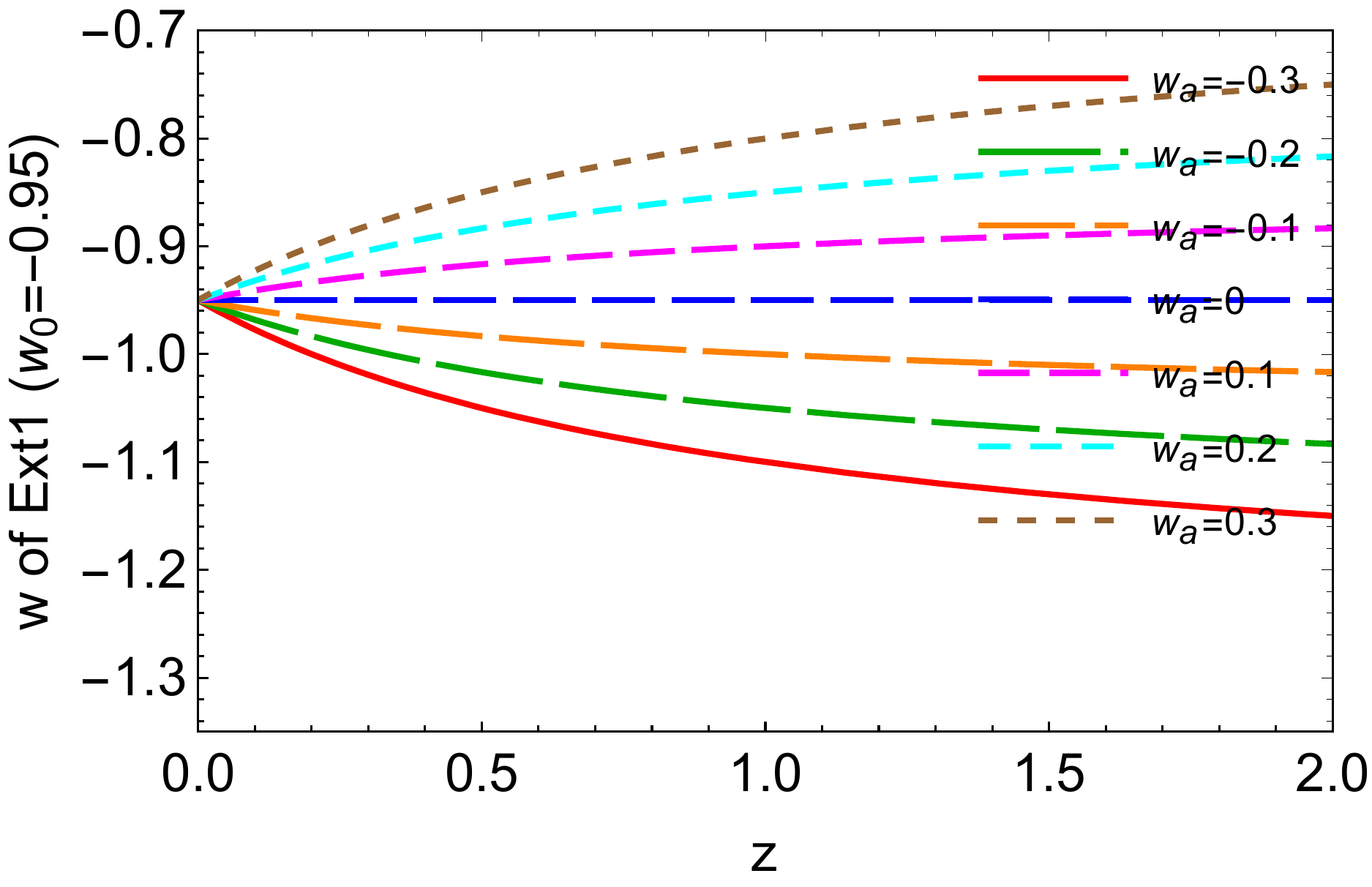}
\includegraphics[width=0.34\textwidth]{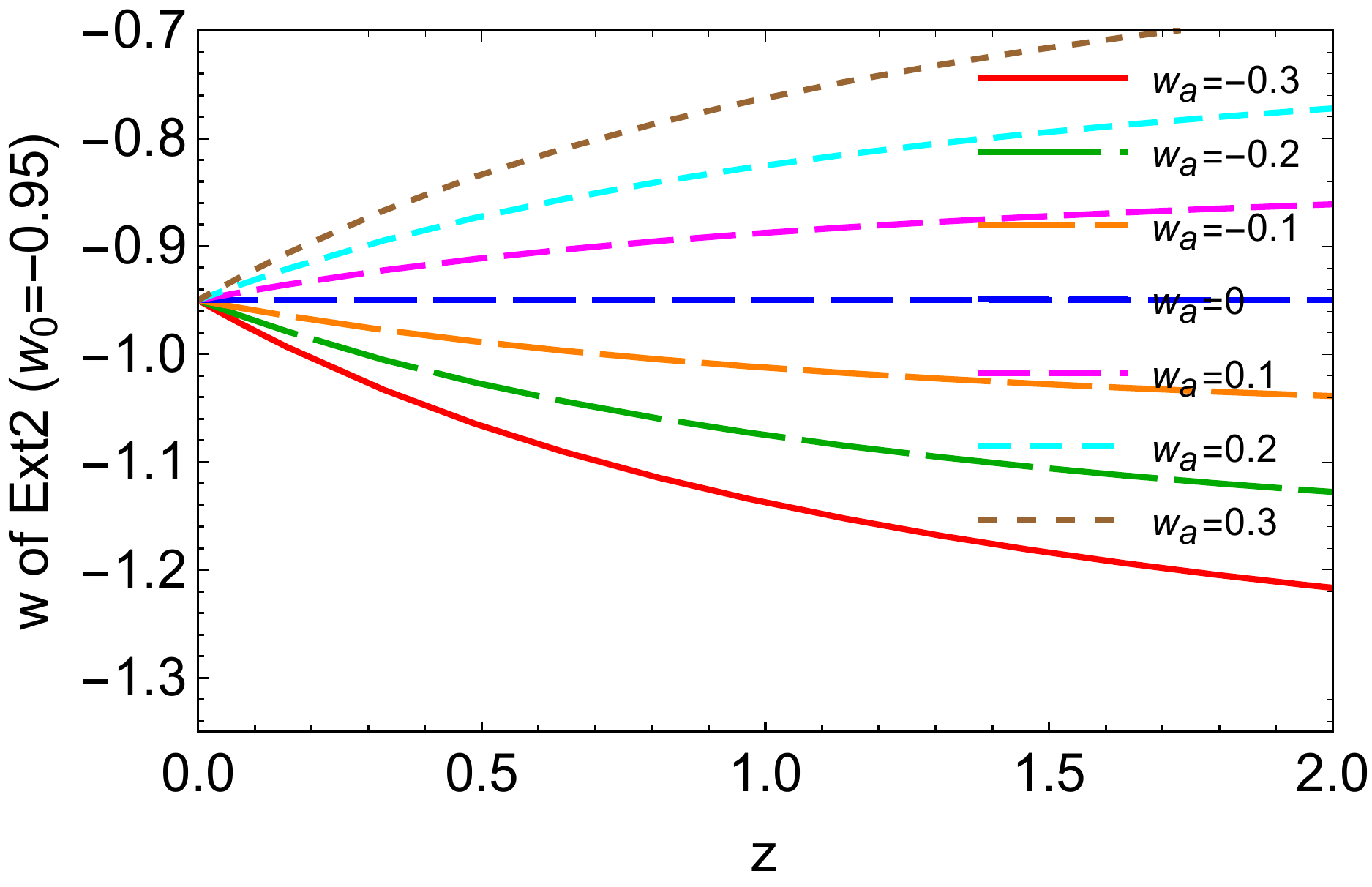}
\includegraphics[width=0.34\textwidth]{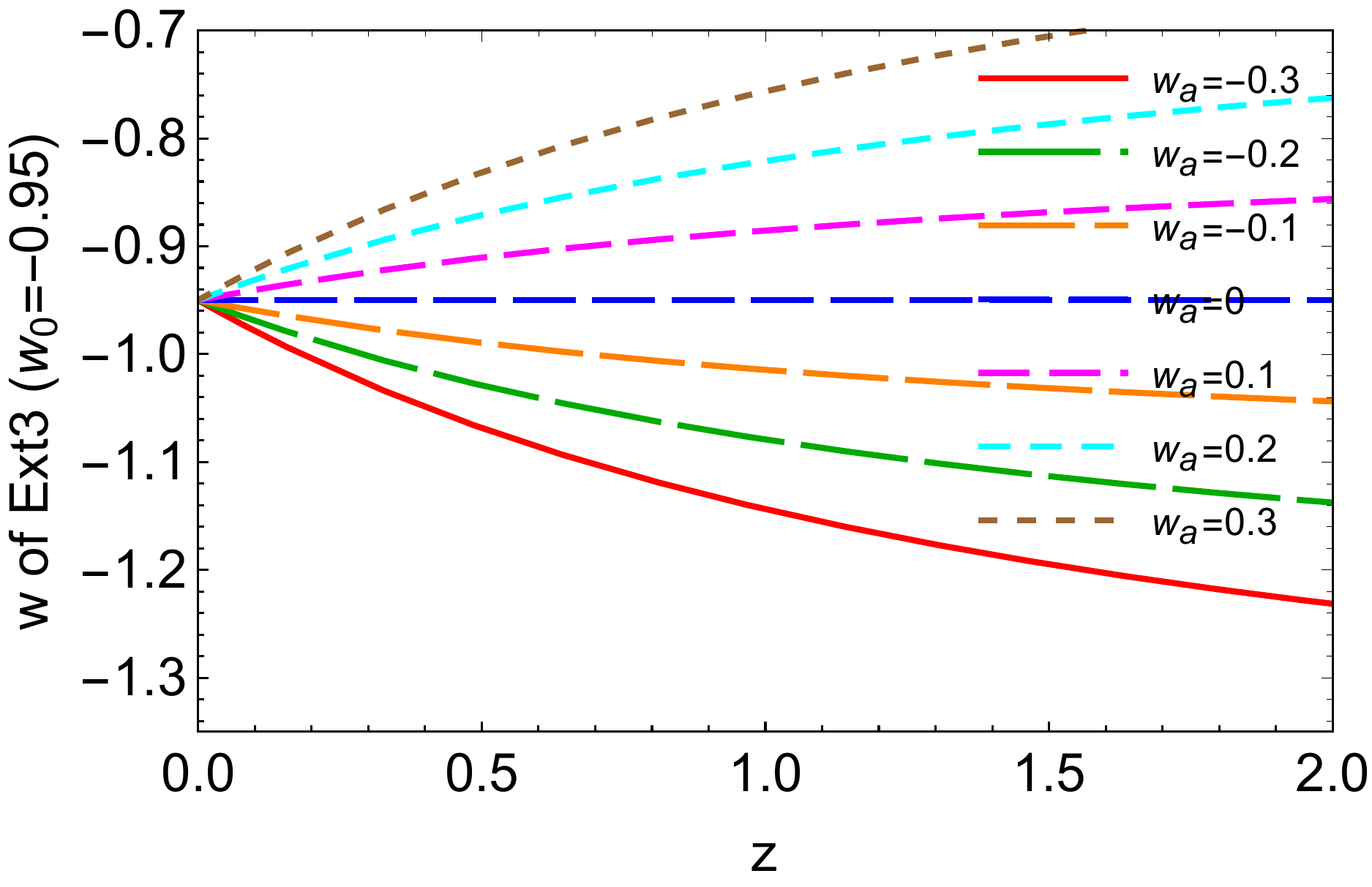}
\caption{We show the evolution of the dark energy parametrizations for different values of $w_a$ with a fixed value of $w_0 = -0.95$ }
\label{w-1}
\end{figure*}

At the end of this section, we would like to present the qualitative features of the present dark energy parametrizations in terms of the evolution of their equations of state and the deceleration parameters. In order to do so, we assumed three different values of $w_0$, namely, $w_0 = -0.95$, $w_0 =-1$ and $w_0 = -1.1$ and in each case we consider various values of $w_a$ to understand how the curves behave with the increasing of the $w_a$ parameter. In Fig. \ref{w-1} we show the evolution of the dark energy parameterizations (\ref{model-general}), (\ref{extension1}), (\ref{extension2}) and (\ref{extension3}) setting the present value of the dark energy equation of state at $w_0 =0.95$ where we allow different values of $w_a$ such as $w_a = -0.3, -0.2, -0.1, 0, 0.1, 0.2, 0.3$. The curve with $w_a =0$ simply returns $w = w_0$ and this has been kept to compare with other cruves having $w_a \neq 0$. From Fig. \ref{w-1}, we notice that for $w_a <0$, the dark energy equation of state allows its phantom character which is much pronounced at high redshifts, while for $w_a> 0$, the reverse scenario is found. In a similar fashion, we investiagted the other cases with $w_0 = -1$ and $w_0 = -1.1$, however, we did not observe any significant changes in the qualitative evolution of $w_x(z)$, so we did not include the other figures.

\begin{figure*}
\includegraphics[width=0.34\textwidth]{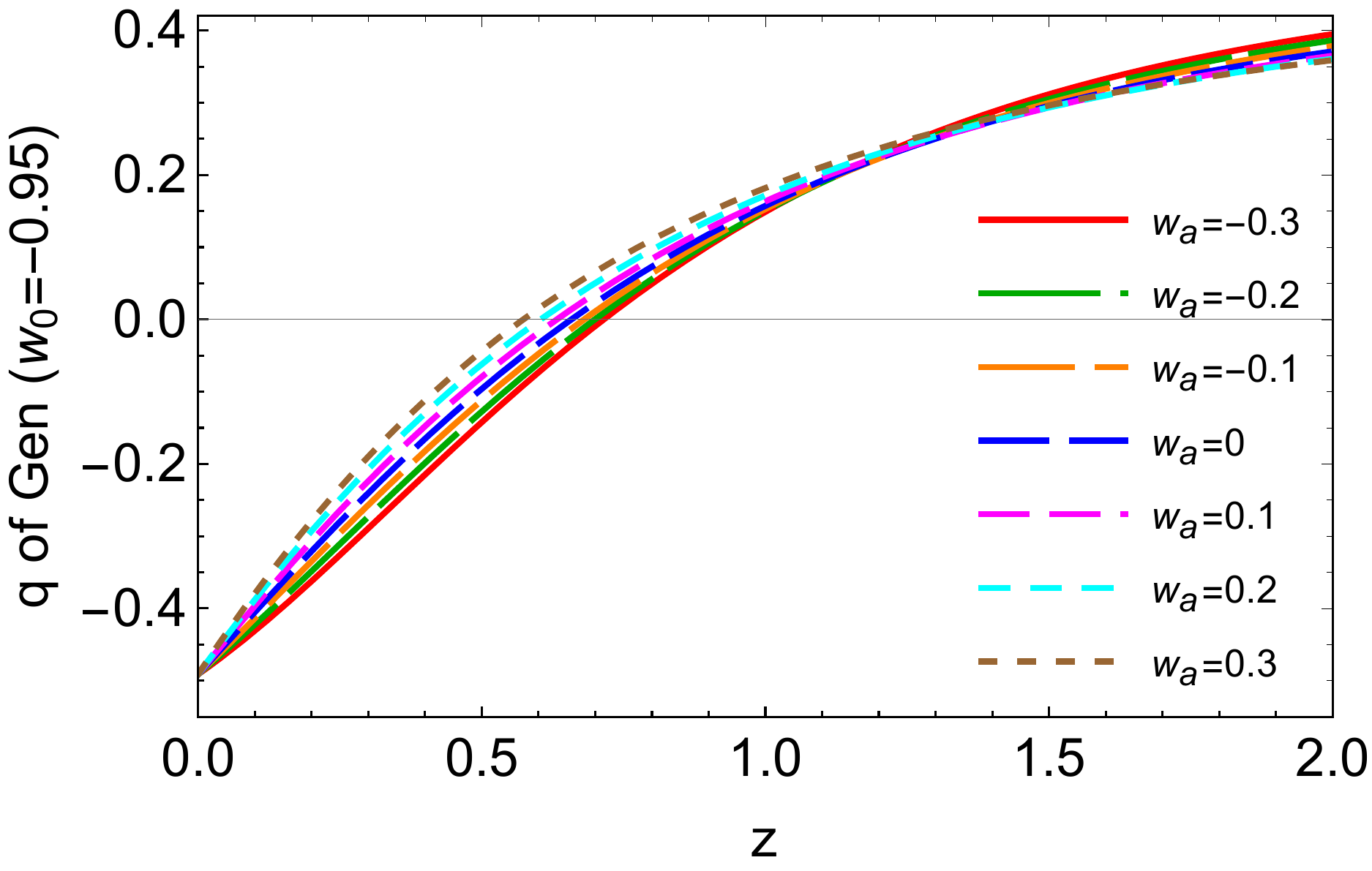}
\includegraphics[width=0.34\textwidth]{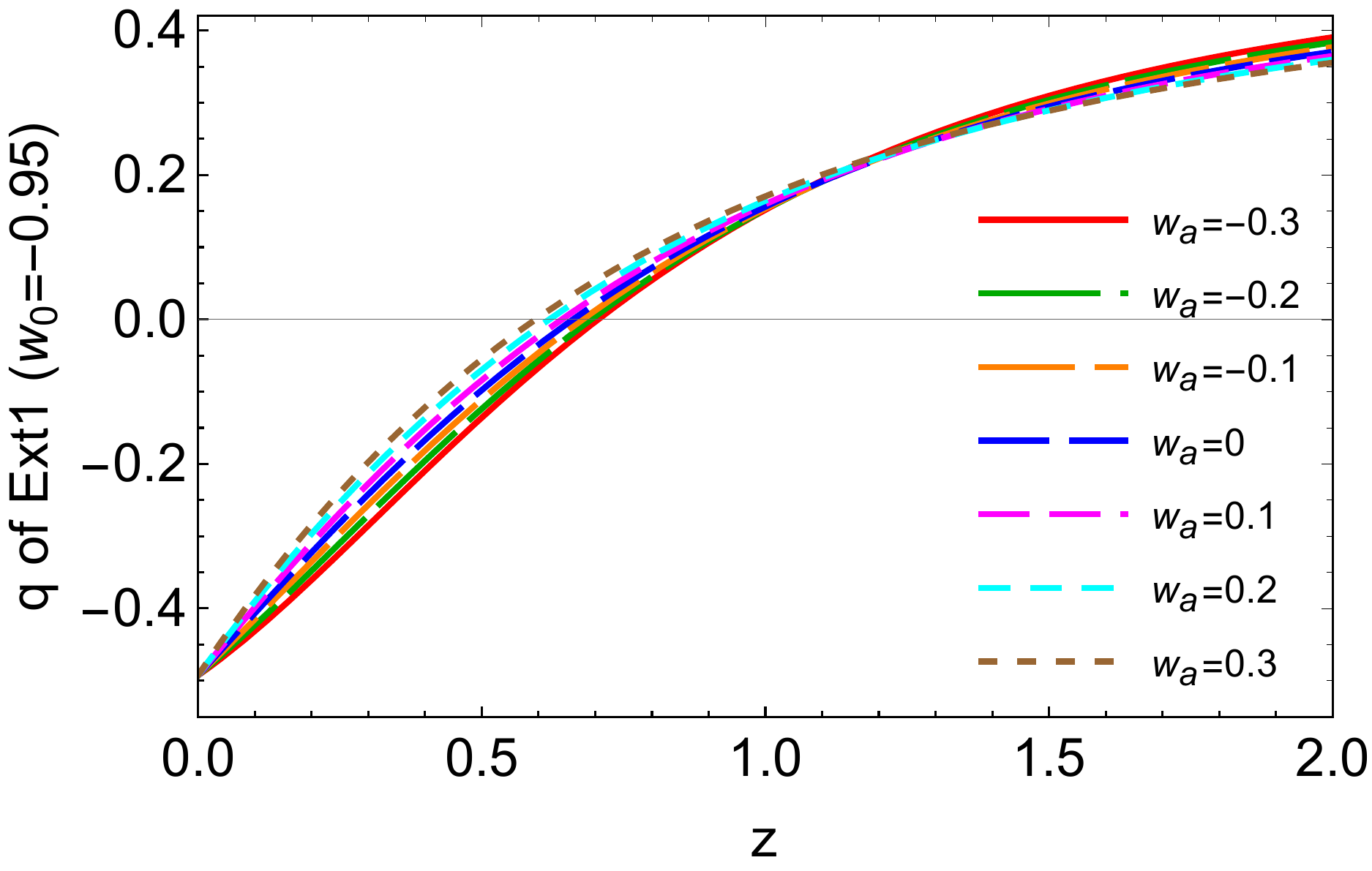}
\includegraphics[width=0.34\textwidth]{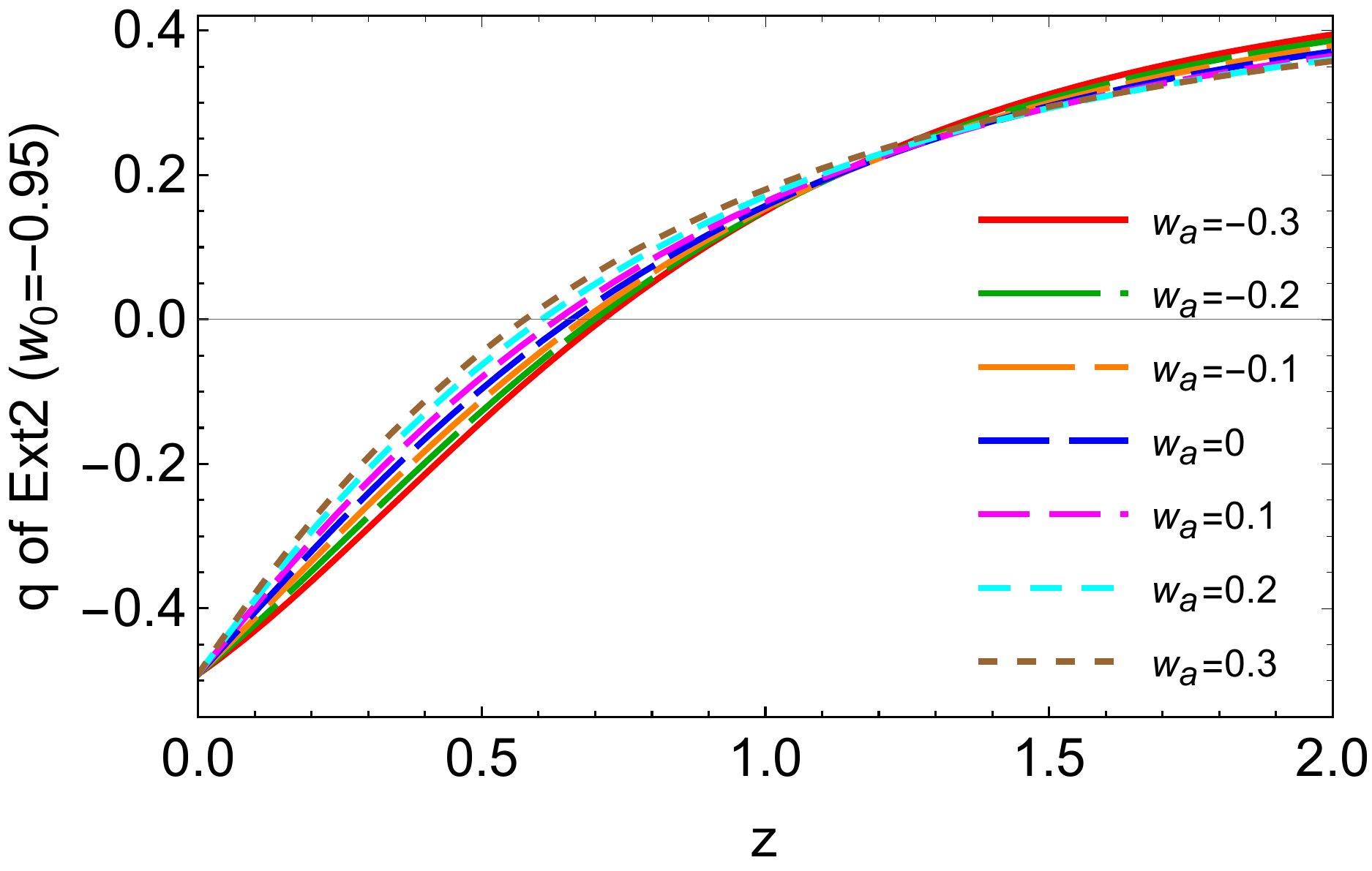}
\includegraphics[width=0.34\textwidth]{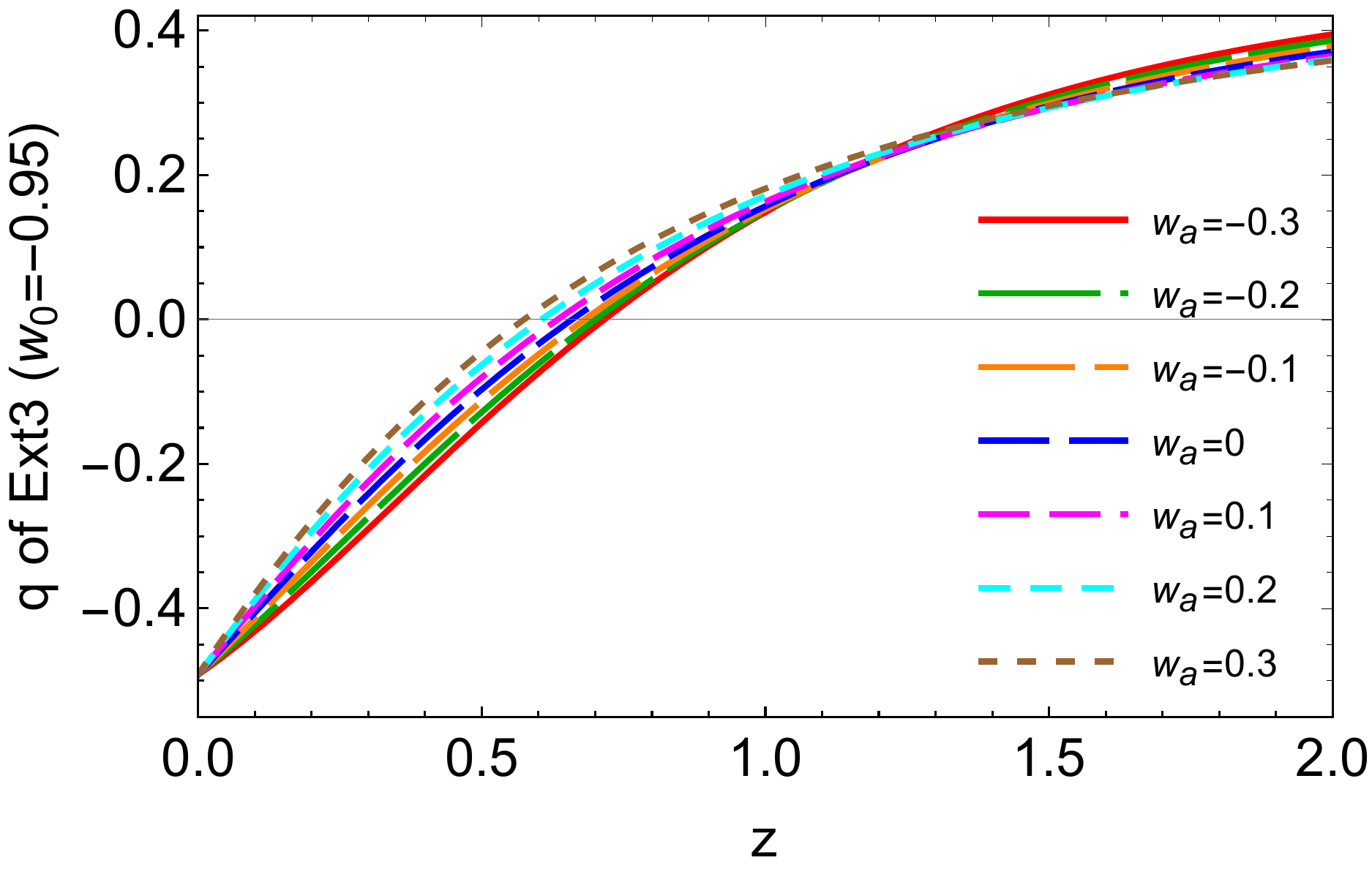}
\caption{The evolution of the deceleration parameter depicting a clear transition from the past decelerating phase to the current accelerating phase for all the dark energy models has been presented for different values of $w_a$ and with a fixed value of $w_0 =-0.95$. One can easily notice that as long as $w_a$ increase from its negative to positive values, the transition redshift shifts more closer to the present epoch. }
\label{q-1}
\end{figure*}

We then plot the evolution of the deceleration parameter for all the DE parametrizations, namely, (\ref{model-general}), (\ref{extension1}), (\ref{extension2}) and (\ref{extension3}). Here we again assumed three fixed values of $w_0$, namely, $w_0 = -0.95$, $-1$, $-1.1$ and in each case we  assume different values of $w_a$ similar to what we have shown in Fig. \ref{w-1}. Finally, we analyzed the evolution of the deceleration parameter for all the cases and found that all three cases return almost similar dynamics. That is why choose the case with for $w_0 = -0.95$ and exclude the others. The Fig. \ref{q-1} corresponds to the case  $w_0 = -0.95$. From this figure  one can see that irrespective of the values of $w_0$, a fine transition from the past decelerating phase to the current accelerating one is observed. The interesting and important point in Fig. \ref{q-1} is that, for negative values of $w_a$ the transition redshifts are shifting towards higher redshifts (although mild) while for positive values of $w_a$, we see the reverse, that means the transition redshifts are shifting towards lower values of the redhift.

Overall, we find that the models at the level of background do not exhibit any deviations from one another. This is not surprising because the deviations between the cosmological models are usually reflected from their analysis at the level of perturbations. In what follows we shall consider the perturbation equations for all the DE parametrizations in this work.

We start with the following metric which is the perturbed form of the FLRW line element: 
\begin{equation*}
ds^{2}=a^{2}(\eta )\left[ -d\eta ^{2}+(\delta _{ij}+h_{ij})dx^{i}dx^{j}%
\right] ,
\end{equation*}%
Here, $\eta $ denotes the conformal time; $\delta _{ij}$, $h_{ij}$ are the
unperturbed and the perturbative metric tensors, respectively. Now,
considering the perturbed Einstein's field equations, for a mode with
wave-number ${k}$ one can write down \cite{Mukhanov,Ma:1995ey, Malik:2008im}:

\begin{eqnarray}
\delta^{\prime }_{i} = - (1+ w_{i})\, \left(\theta_{i}+ \frac{h^{\prime }}{2}%
\right) - 3\mathcal{H}\left(\frac{\delta p_i}{\delta \rho_i} - w_{i}
\right)\delta_i  \notag \\
- 9 \mathcal{H}^2\left(\frac{\delta p_i}{\delta \rho_i} - c^2_{a,i} \right)
(1+w_i) \frac{\theta_i} {{k}^2},  \label{per1} \\
\theta^{\prime }_{i} = - \mathcal{H} \left(1- 3 \frac{\delta p_i}{\delta
\rho_i}\right)\theta_{i} + \frac{\delta p_i/\delta \rho_i}{1+w_{i}}\, {k}%
^2\, \delta_{i} -{k}^2\sigma_i,  \label{per2}
\end{eqnarray}
where $\delta_i = \delta \rho_i/\rho_i$ is the density perturbation for the $%
i$-th fluid; the prime associated to any quantity denotes the derivatives
with respect to conformal time; $\mathcal{H}= a^{\prime}/a$ is the conformal
Hubble parameter; $\theta_{i}\equiv i k^{j} v_{j}$ is the divergence of the $%
i$-th fluid velocity; $h = h^{j}_{j}$, is the trace of the metric
perturbations $h_{ij}$; $\sigma_i$ denotes the anisotropic stress related to
the $i$-th fluid. Let us also note that $c_{a,i}^2 = \dot{p}_i/\dot{\rho}_i$%
, is the adiabatic speed of sound of the $i$-th fluid which can also be
written in terms of other physical quantities as $c^2_{a,i} = w_i - \frac{%
w_i^{\prime}}{3\mathcal{H}(1+w_i)}$, where we fix the sound speed $c^2_{s} =
\delta p_i / \delta \rho_i$ to be unity. Finally, we also note that we have
neglected the anisotropic stress from the system for simplicity.


\section{Scalar-field description}
\label{sec-sf}

This section is devoted to provide with an equivalent field theoretic
description for the dark energy parametrizations.  
A method to construct the scalar field potential which describe a given
equation of state parameter was presented in \cite{dimakis}. Specifically, for
a spatially flat FLRW as in the case of consideration, with a line element

\begin{equation}
ds^{2}=-e^{F\left(  \omega\right)  }d\omega^{2}+e^{\omega/3} \left[\frac{dr^2}{1-Kr^2}+ r^2 d\theta^2 + r^2 \sin^2 \theta d \phi^2 \right],
\label{SF.12}%
\end{equation}
where $e^{F\left(  \omega\right)  }$ plays the role of a lapse function,
the scale factor $a\left(  \omega\right)  =e^{\omega/3}$ and $K$ is the curvature scalar of the universe. The exact solution
of the scalar field and the scalar field potential in  case of vacuum are%
\begin{equation}
\phi(\omega)=\pm\frac{\sqrt{6}}{6}\int\!\!\sqrt{F^{\prime}(\omega)}d\omega,
\label{so.01}%
\end{equation}%
\begin{equation}
V(\omega)=\frac{1}{12}e^{-F(\omega)}\left(  1-F^{\prime}(\omega)\right)
\label{so.02}%
\end{equation}
or, equivalently, for the physical parameters, such as energy density and
pressure%
\begin{equation}
\rho_{\phi}(\omega)=\frac{1}{12}e^{-F(\omega)}~,~P_{\phi}(\omega)=\frac{1}%
{12}e^{-F(\omega)}\left(  2F^{\prime}(\omega)-1\right). \label{so.03}%
\end{equation}

Conseqently, for the latter definitions it follows%
\begin{equation}
w_{\phi}\left(  \omega\right) =  \frac{P_{\phi} (\omega)}{\rho_{\phi}(\omega)}  =\left(  2F^{\prime}(\omega)-1\right)  .
\end{equation}
Hence, for a specific equation of state parameter$~w_{\phi}\left(
\omega\right)  $ the latter first-order equation can be solved and we can
determine the function $F\left(  \omega\right)  $. Subsequently, by replacing $F\left(  \omega\right)  $ in eqns. (\ref{so.01}), (\ref{so.02}) one can 
find the functional form of $V\left( \phi\right)  $. 
For the general functional form of the equation of 
state, namely, eqn. (\ref{model-general}), $w_x(\omega) = w_0 -w_a + w_a \exp\left(1- e^{\omega/3}\right)$, we find that 
\begin{align}
F(\omega) =  \frac{1}{2} (1+w_0-w_a) \omega + \frac{1}{2} w_a \int \exp\left(1- e^{\omega/3}\right) \; d \omega + F_0 ,
\end{align}
where $F_0$ is the constant of integration. Now, using the value of $F (\omega)$, one can find that
\begin{align}
\phi (\omega) = \frac{1}{\sqrt{2}} \int \sqrt{1+w_0 - w_a +  w_a \exp\left(1- e^{\omega/3}\right)}\; d \omega ,
\end{align}
and $V (\omega)$ can be solved as
\begin{widetext}
\begin{eqnarray}
V (\omega) = \frac{1}{24} \left(-e^{e^{\omega /3}} (w_0+ w_a-1)-e\;
   w_a\right) \exp \left(-\frac{3 e}{2}\; w_a\;
   \text{Ei}\left(-e^{\omega /3}\right)-\frac{\omega}{2} 
   (w_0+ w_a+1)-e^{\omega /3}\right),
\end{eqnarray} 
\end{widetext}
where the symbol `$\mbox{Ei}$' represents the exponential integral. 
For the CPL potential $w_x\left(  \omega\right)  =w_{0}+w_{a}\left(
1-e^{\omega/3}\right)  $ we find%
\[
F\left(  \omega\right)  =\frac{1}{2}\left(  1+w_{0}+w_{a}\right)  \omega
-\frac{3}{2}w_{a}e^{\omega/3}+F_{0},
\]
where $F_0$ is the integration constant and consequently
we find that%
\begin{eqnarray}
\frac{\phi\left(  \omega\right)  }{\sqrt{3}}=\sqrt{w_{a}\left(  e^{w/3}%
-1\right)  -\left(  1+w_{0}\right)  }-\sqrt{1+w_{0}+w_{a}} \nonumber \\ \times \arctan\sqrt
{\frac{w_{a}\left(  e^{w/3}-1\right)  -\left(  1+w_{0}\right)  }{1+w_{0}%
+w_{a}}},
\end{eqnarray}%
\begin{eqnarray}
V\left(  \omega\right)  =\frac{1}{24}e^{F_{0}}\exp\left(  \frac{3}{2}%
w_{a}e^{\omega/3}-\frac{1}{2}\left(  1+w_{0}+w_{a}\right)  \right) \nonumber\\ \times   \left(
1-w_{0}+w_{a}\left(  e^{\omega/3}-1\right)  \right) .
\end{eqnarray}

For the Ext2 and Ext3 models the corresponding functions $F\left(
\omega\right)  $ are derived to be%
\begin{eqnarray}
F_{Ext}^{2}\left(  \omega\right)  =F_{0}+\frac{1}{4}\Bigl(  \left(
2+2w_{0}+3w_{a}\right)  \omega \nonumber\\ - 12w_{a}e^{\omega/3}+\frac{3}{2}w_{a}%
e^{2\omega/3}\Bigr) ,
\end{eqnarray}%
\begin{eqnarray}
F_{Ext}^{3}\left(  \omega\right)  =  F_{0}+\frac{1}{12}\Bigl(  \left(
6\omega+6w_{0}+10w_{a}\right)  \omega \nonumber\\ -w_{a}\left(  e^{\omega}+45e^{\omega
/3}-9e^{2\omega/3}\right)  \Bigr) .
\end{eqnarray}

\begin{figure*}[ptb]
\includegraphics[height=5.5cm]{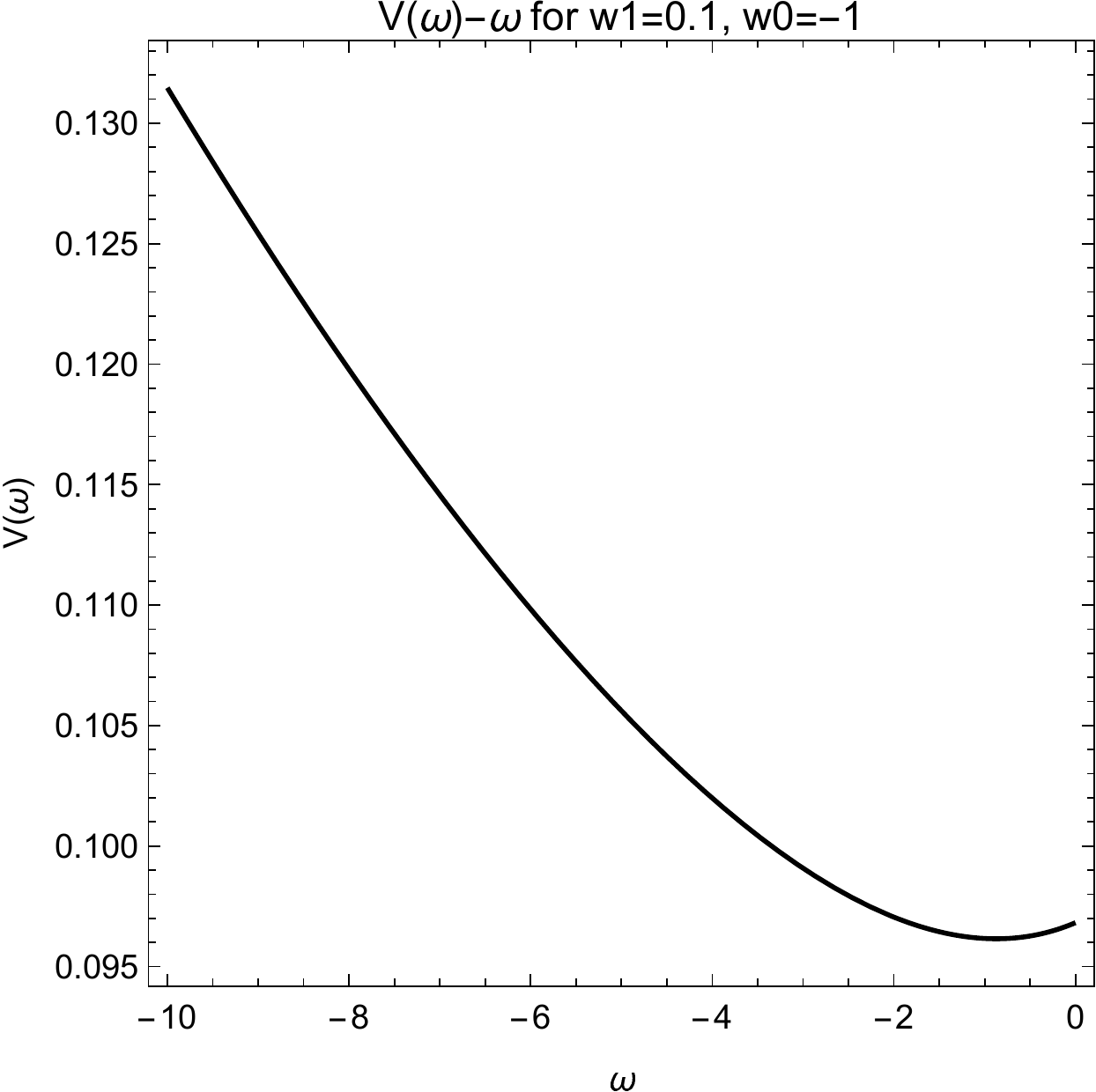}\centering
\includegraphics[height=5.5cm]{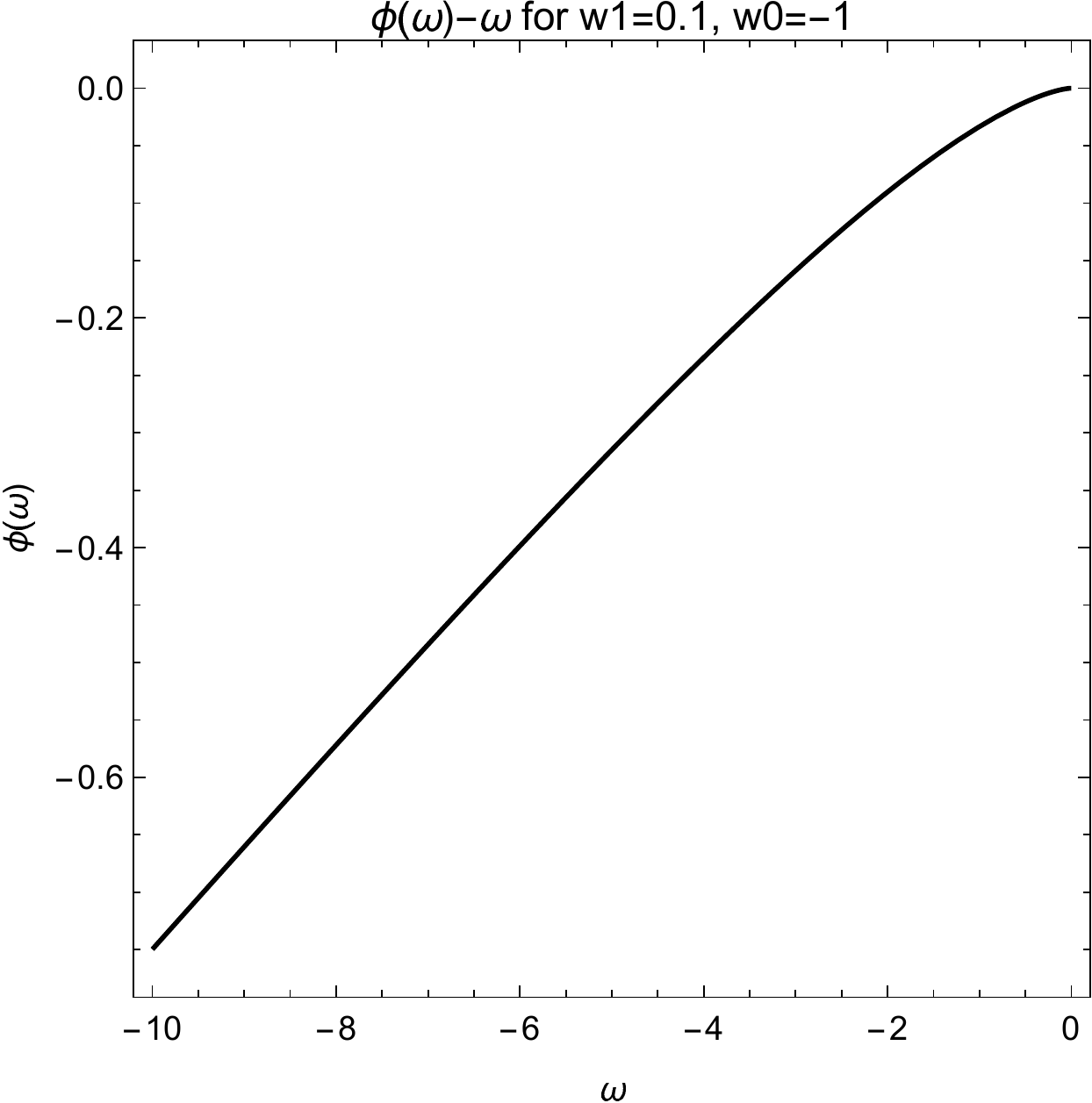}\centering
\includegraphics[height=5.5cm]{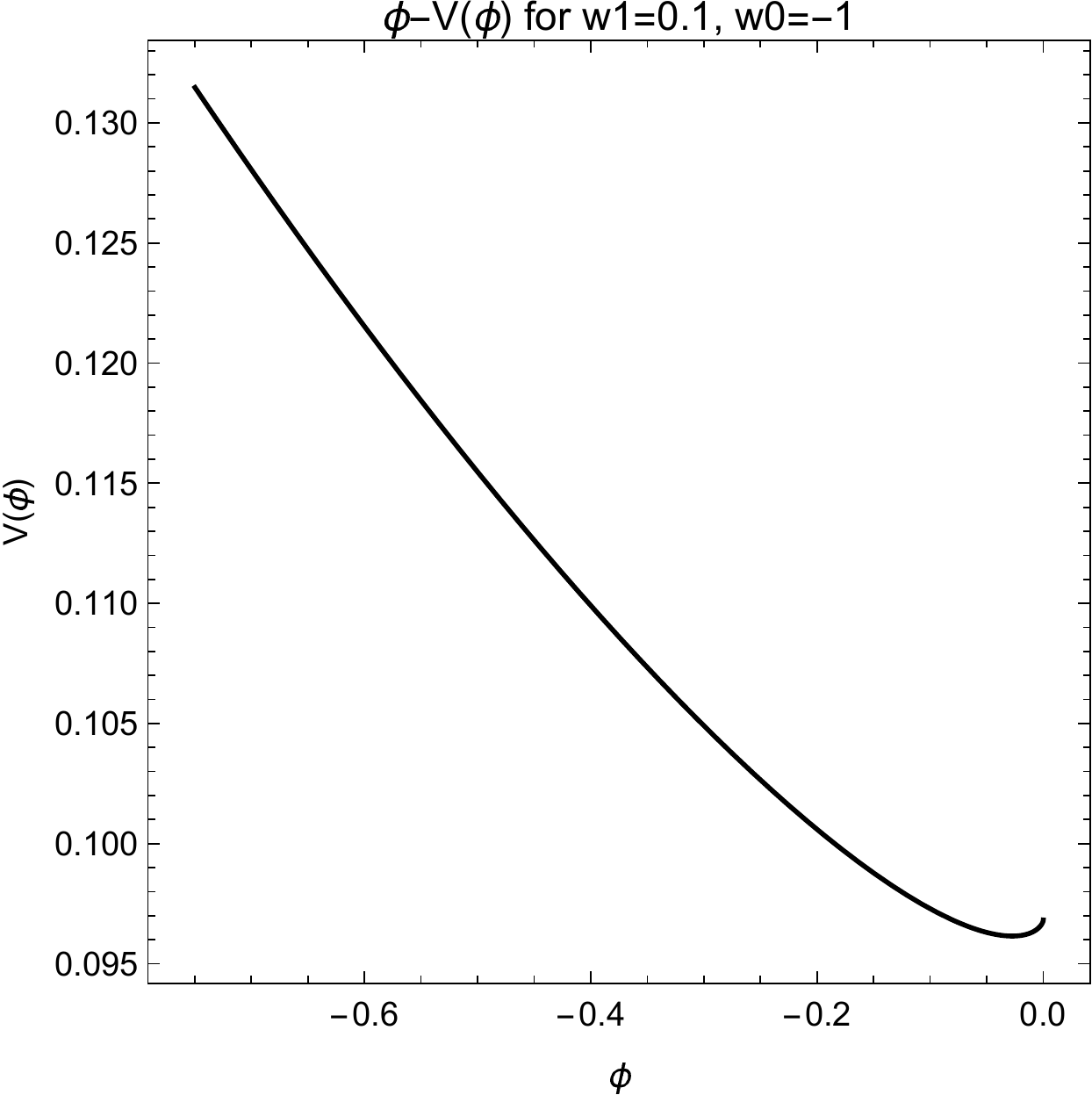}\centering
\caption{Qualitative evolution of the scalar field $\phi\left(  \omega\right)
,~$\ the scalar field potential $V\left(  \omega\right)  $ as also the
parametric plot $\phi-V\left(  \phi\right)  $ is given for the CPL (Ext1) model.}%
\label{figclp}%
\end{figure*}

\begin{figure*}[ptb]
\includegraphics[height=5.5cm]{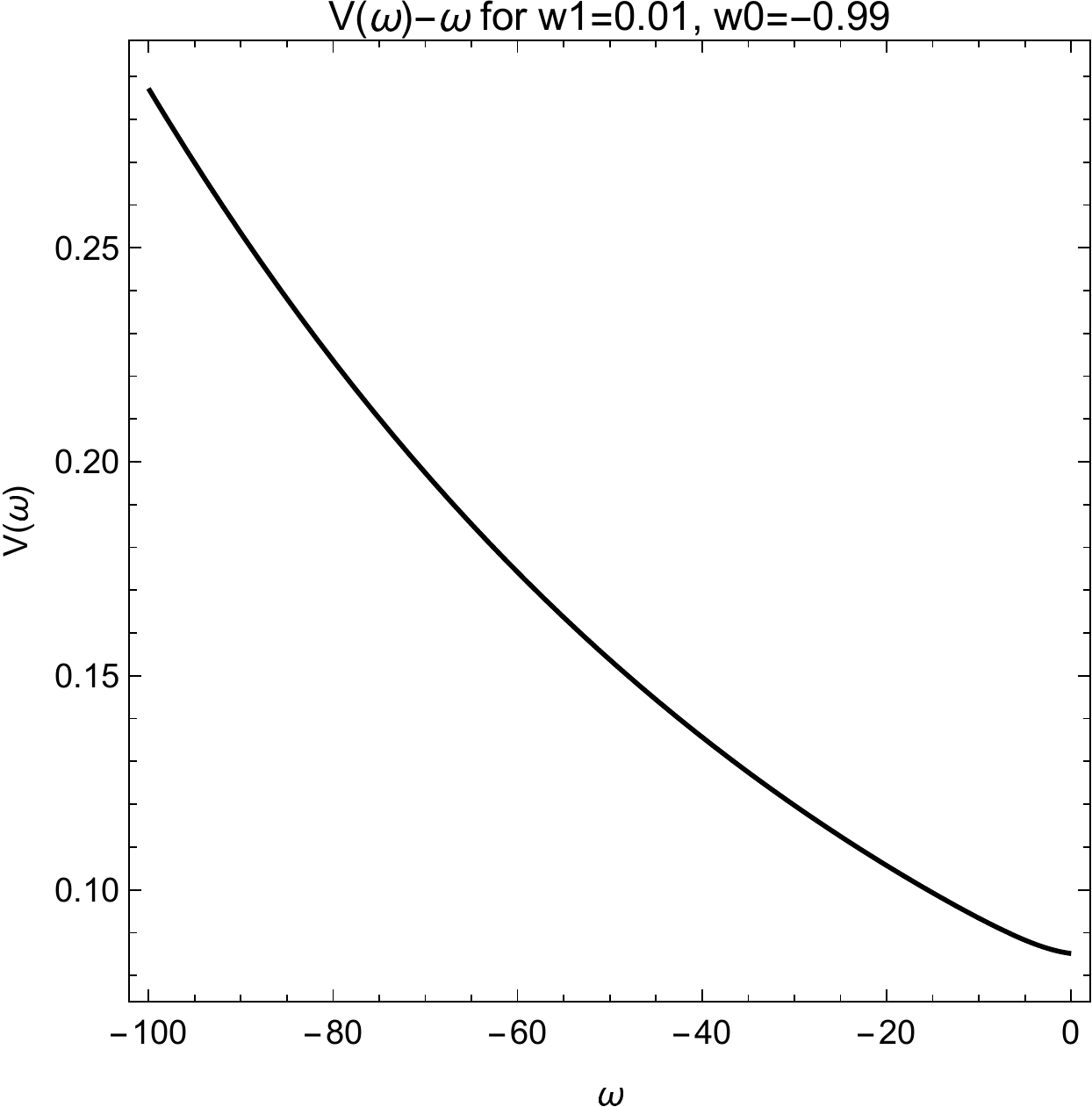}\centering
\includegraphics[height=5.5cm]{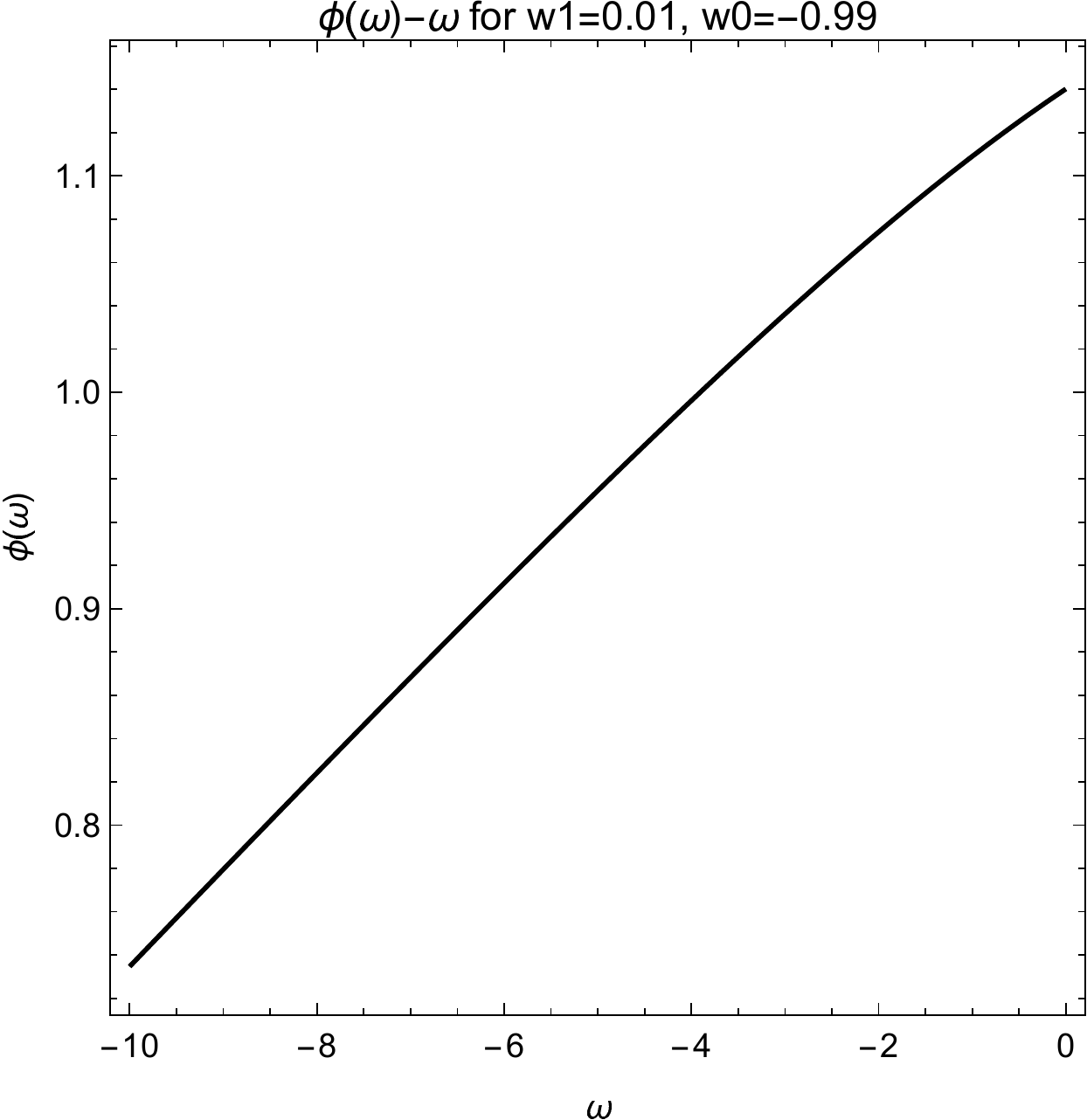}\centering
\includegraphics[height=5.5cm]{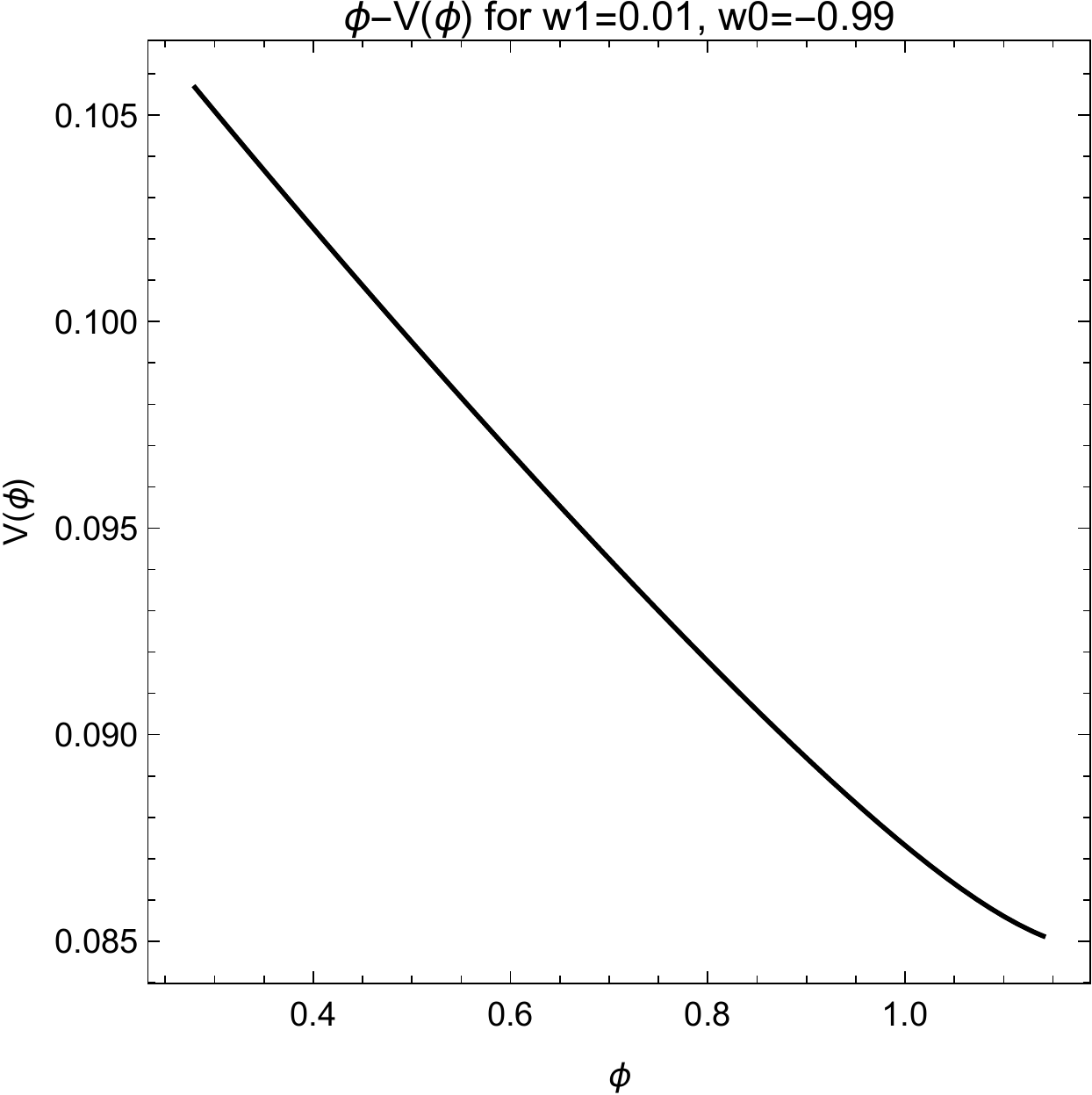}\centering
\caption{Qualitative evolution of the scalar field $\phi\left(  \omega\right)
,$\ the scalar field potential $V\left(  \omega\right)  $ as also the
parametric plot $\phi-V\left(  \phi\right)  $ is given for the Ext2
model.}%
\label{figclp1}%
\end{figure*}

\begin{figure*}[ptb]
\includegraphics[height=5.5cm]{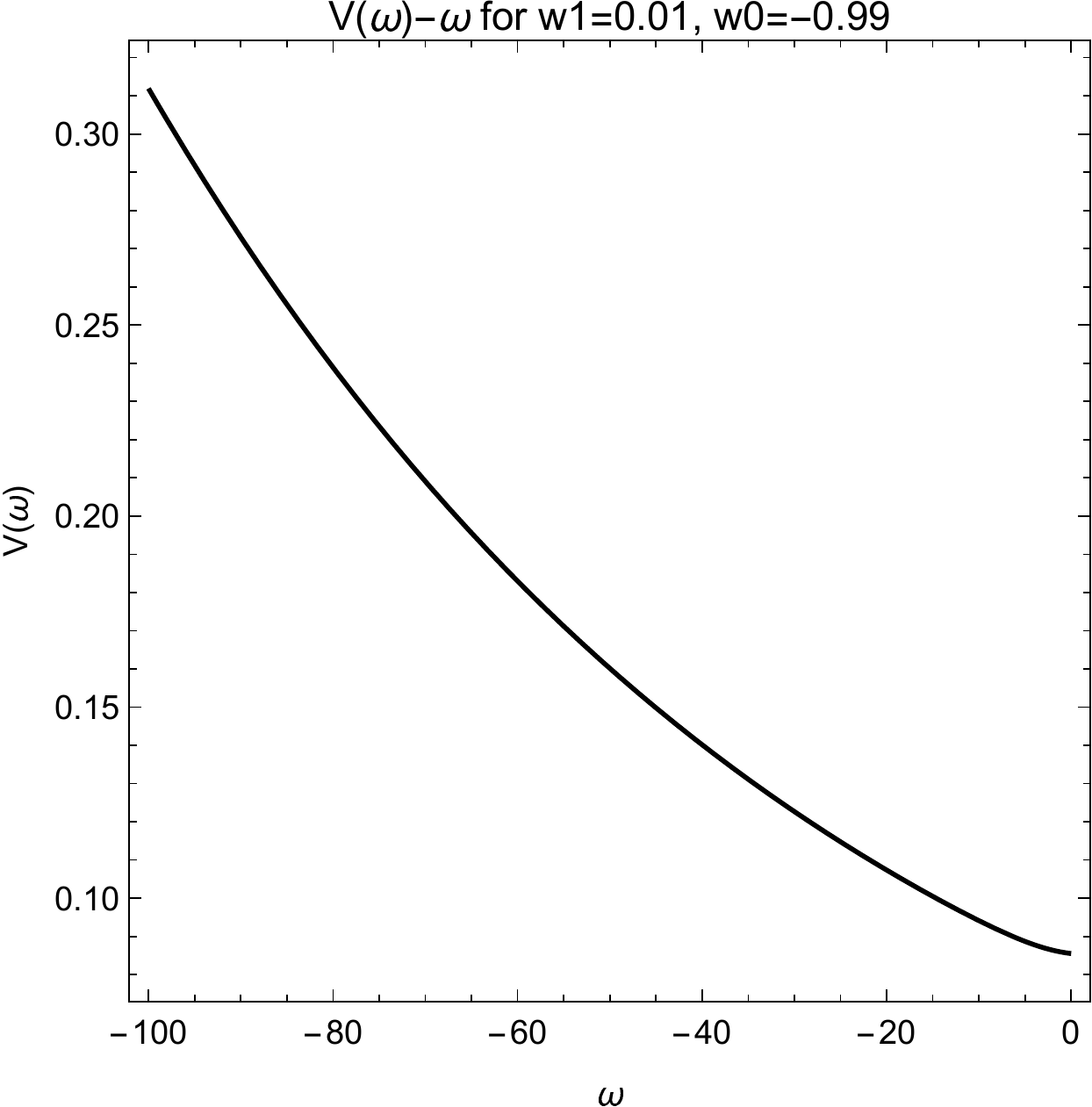}\centering
\includegraphics[height=5.5cm]{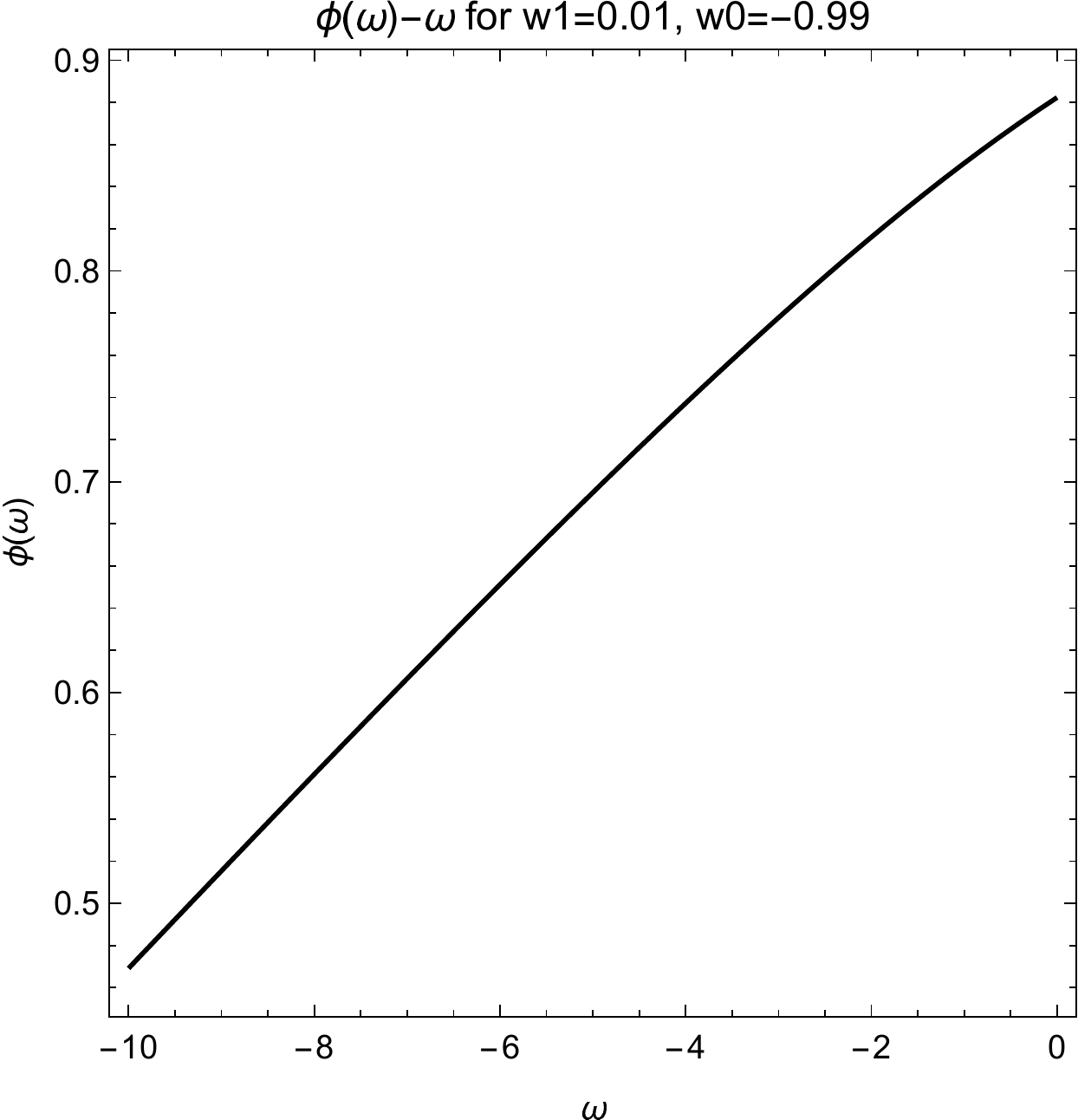}\centering
\includegraphics[height=5.5cm]{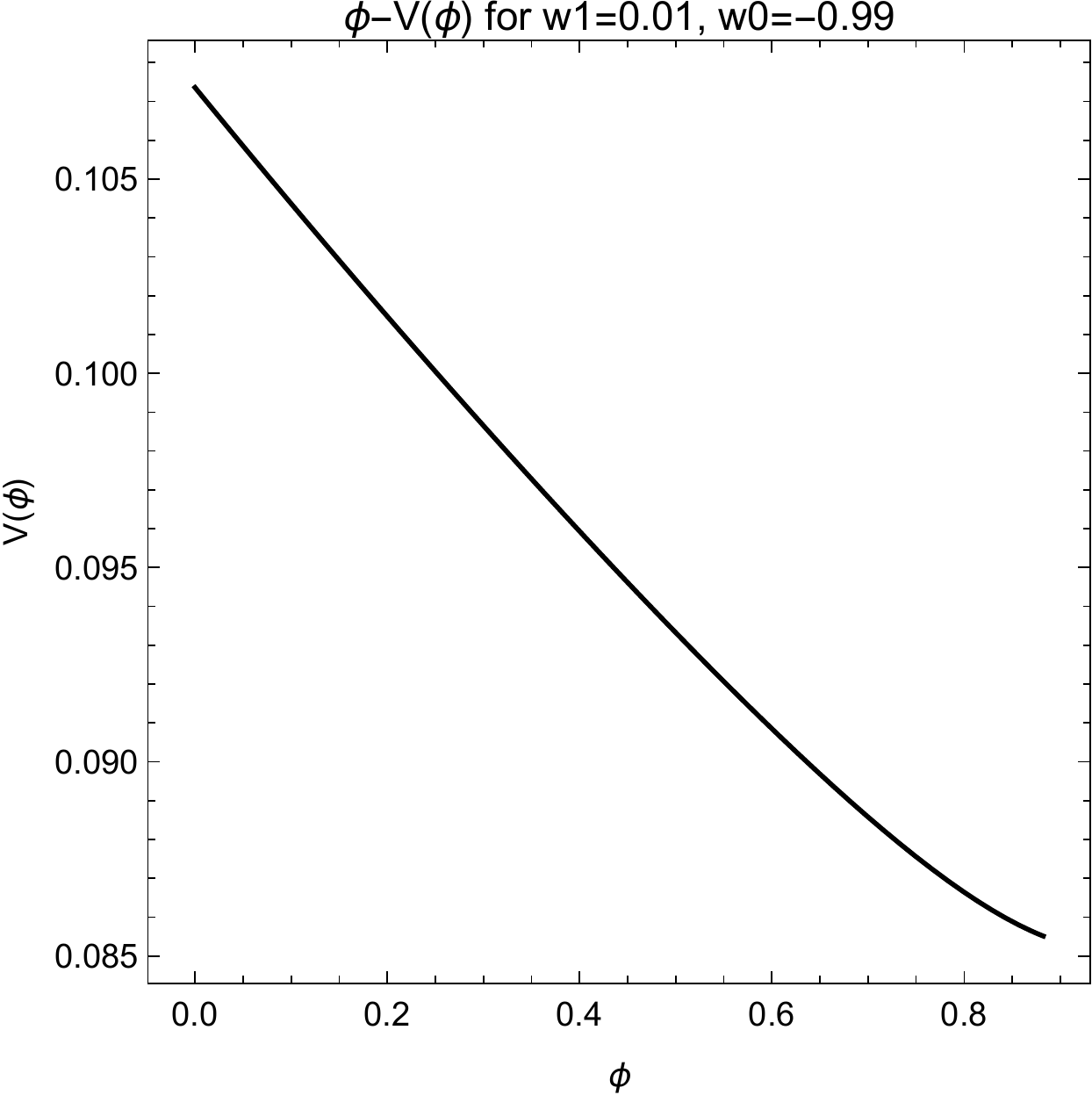}\centering
\caption{Qualitative evolution of the scalar field $\phi\left(  \omega\right)
,$\ the scalar field potential $V\left(  \omega\right)$ as also the
parametric plot $\phi-V\left(  \phi\right)  $ is given for the Ext3
model.}%
\label{figcpl2}%
\end{figure*}

In Figs. \ref{figclp}, \ref{figclp1}~and \ref{figcpl2} the qualitative evolution
of the scalar field equivalent $\phi\left(  \omega\right)$, the scalar field
potential $V\left(  \omega\right)  $ and the parametric plot $V\left(
\omega\right)  $ are presented respectively for the CPL parametrization (Ext1) and the other two extensions, namely, Ext2, and Ext3. 


\section{Observational data}

\label{sec-data}

For the convenience of the reader and for our presentation we provide the
details of the observational data used to constrain the dynamical dark
energy parametrization and also the methodology.

\begin{itemize}
\item Cosmic microwave background observations: the cosmic microwave
background (CMB) observations are one of the powerful data to probe the
nature of dark energy. Here we use the CMB from Planck 2015 \cite%
{Adam:2015rua, Aghanim:2015xee}. The high-$\ell$ temperature and
polarization data as well as the low-$\ell$ temperature and polarization
data from Planck 2015 (precisely the dataset: Planck TT, TE, EE + lowTEB) 
\cite{Adam:2015rua, Aghanim:2015xee} have been considered.

\item Baryon acoustic oscillations: The baryon acoustic oscillations (BAO)
data from different superovulation missions are used \cite{Beutler:2011hx,
Ross:2014qpa,Gil-Marin:2015nqa}.

\item Supernovae Type Ia: We also use latest released Pantheon sample \cite%
{Scolnic:2017caz} from the Supernovae Type Ia.

\item Hubble parameter measurements: Finally, we use the Hubble parameter
measurements from the Cosmic Chronometers (CC) \cite{Moresco:2016mzx}.
\end{itemize}

Now we come to the technical part of the statistical analysis. Thus, we have
performed the fitting analysis using the modified version of \texttt{cosmomc}
\cite{Lewis:2002ah, Lewis:1999bs}, an efficient markov chain monte carlo
package equipped with a convergence diagnostic given by the Gelman and Rubin
statistics \cite{Gelman-Rubin}. This package includes the support for the
Planck 2015 likelihood code \cite{Aghanim:2015xee} (see %
\url{http://cosmologist.info/cosmomc/}). In Table \ref{tab:priors} we have
shown the flat priors on the model parameters that have been used during the
observational analysis. Perhaps it might be important to mention here that 
in the present analysis we have used Planck 2015 likelihood \cite{Aghanim:2015xee}
instead of Planck 2018 likelihood (although the 
cosmological parameters from Planck 2018 are already available \cite{Aghanim:2018eyx}) because Planck 2018 likelihood code is not public yet. However, it will be worth to run the same codes that we use for the present models but with the new Planck 2018 likeliood which will enable us to understand any effective changes in the cosmological parameters and consequently more stringent constraints on them as well.  

\begin{table}%[tbp]
\begin{center}
\begin{tabular}{|c|c|c|}
\hline
Parameter & Prior  \\ \hline
$\Omega_{b} h^2$ & $[0.005,0.1]$   \\ 
$\Omega_{c} h^2$ & $[0.01,0.99]$   \\ 
$\tau$ & $[0.01,0.8]$   \\ 
$n_s$ & $[0.5, 1.5]$  \\ 
$\log[10^{10}A_{s}]$ & $[2.4,4]$   \\ 
$100\theta_{MC}$ & $[0.5,10]$   \\ 
$w_0$ & $[-2, 0]$   \\ 
$w_a$ & $[-3, 3]$   \\ 
\hline
\end{tabular}%
\end{center}
\caption{The table shows the flat priors on the model parameters used during
the statistical analysis. }
\label{tab:priors}
\end{table}

\begingroup                                                                                                                     
\squeezetable                                                                                                                   
\begin{center}
\begin{table*}%[tbp]
\begin{tabular}{ccccccccccccccccc}
\hline\hline
Parameters & CMB+BAO & CMB+BAO+Pantheon & CMB+BAO+Pantheon+CC \\ \hline
$\Omega_c h^2$ & $0.1194_{- 0.0014- 0.0026}^{+ 0.0013+ 0.0027}$ & $0.1193_{-
0.0013- 0.0025}^{+ 0.0013+ 0.0024}$ & $0.1195_{- 0.0013- 0.0026}^{+ 0.0013+
0.0026}$  \\ 
$\Omega_b h^2$ & $0.02222_{- 0.00016- 0.00030}^{+ 0.00015+ 0.00029}$ & $%
0.02223_{- 0.00015- 0.00029}^{+ 0.00015+ 0.00029}$ & $0.02222_{- 0.00015-
0.00030}^{+ 0.00014+ 0.00030}$  \\ 
$100\theta_{MC}$ & $1.04049_{- 0.00032- 0.00066}^{+ 0.00032+ 0.00063}$ & $%
1.04051_{- 0.00031- 0.00062}^{+ 0.00030+ 0.00063}$ & $1.04049_{- 0.00032-
0.00063}^{+ 0.00032+ 0.00066}$  \\ 
$\tau$ & $0.079_{- 0.018- 0.035}^{+ 0.018+ 0.033}$ & $0.080_{- 0.017-
0.034}^{+ 0.018+ 0.034}$ & $0.079_{- 0.018- 0.034}^{+ 0.018+ 0.034}$  \\ 
$n_s$ & $0.9738_{- 0.0044- 0.0085}^{+ 0.0044+ 0.0086}$ & $0.9740_{- 0.0041-
0.0082}^{+ 0.0041+ 0.0083}$ & $0.9735_{- 0.0042- 0.0078}^{+ 0.0042+ 0.0080}$
  \\ 
$\mathrm{ln}(10^{10} A_s)$ & $3.100_{- 0.034- 0.068}^{+ 0.035+ 0.064}$ & $%
3.103_{- 0.034- 0.067}^{+ 0.034+ 0.066}$ & $3.101_{- 0.035- 0.066}^{+ 0.034+
0.066}$   \\ 
$w_0$ & $-0.537_{- 0.441- 0.526}^{+ 0.442+ 0.532}$ & $-0.963_{- 0.082-
0.127}^{+ 0.060+ 0.158}$ & $-0.933_{- 0.070- 0.135}^{+ 0.071+ 0.140}$ \\
 
$w_a$ & $-1.154_{- 0.807- 1.298}^{+ 0.894+ 1.340}$ & $-0.231_{- 0.143-
0.499}^{+ 0.291+ 0.416}$ & $-0.337_{- 0.204- 0.493}^{+ 0.288+ 0.431}$ \\
 
$\Omega_{m0}$ & $0.348_{- 0.035- 0.056}^{+ 0.034+ 0.055}$ & $0.306_{- 0.007-
0.014}^{+ 0.007+ 0.015}$ & $0.306_{- 0.008- 0.015}^{+ 0.008+ 0.015}$  \\
 
$\sigma_8$ & $0.804_{- 0.033- 0.050}^{+ 0.026+ 0.056}$ & $0.839_{- 0.016-
0.032}^{+ 0.018+ 0.031}$ & $0.840_{- 0.017- 0.032}^{+ 0.016+ 0.033}$  \\
 
$H_0$ & $64.14_{- 3.80- 5.12}^{+ 2.51+ 5.60}$ & $68.24_{- 0.80- 1.51}^{+
0.78+ 1.56}$ & $68.23_{- 0.83- 1.62}^{+ 0.84+ 1.67}$ \\ \hline\hline
\end{tabular}%
\caption{Observational constraints on various free parameters at 68\% and
95\% CL for the dynamical dark energy state parameter $w_x (z) = (w_0 -w_a )
+ w_a \exp\left(\frac{z}{1+z}\right)$ (Gen) using different astronomical
datasets have been presented. Let us note that $\Omega_{m0}$ is the present
value of $\Omega_m = \Omega_c+\Omega_b$ and $H_0$ is in the units of
km/sec/Mpc. }
\label{tab:model-gen}
\end{table*}
\end{center}

\endgroup                                                                                                                       
\begin{figure*}%[tbp]
\includegraphics[width=0.65\textwidth]{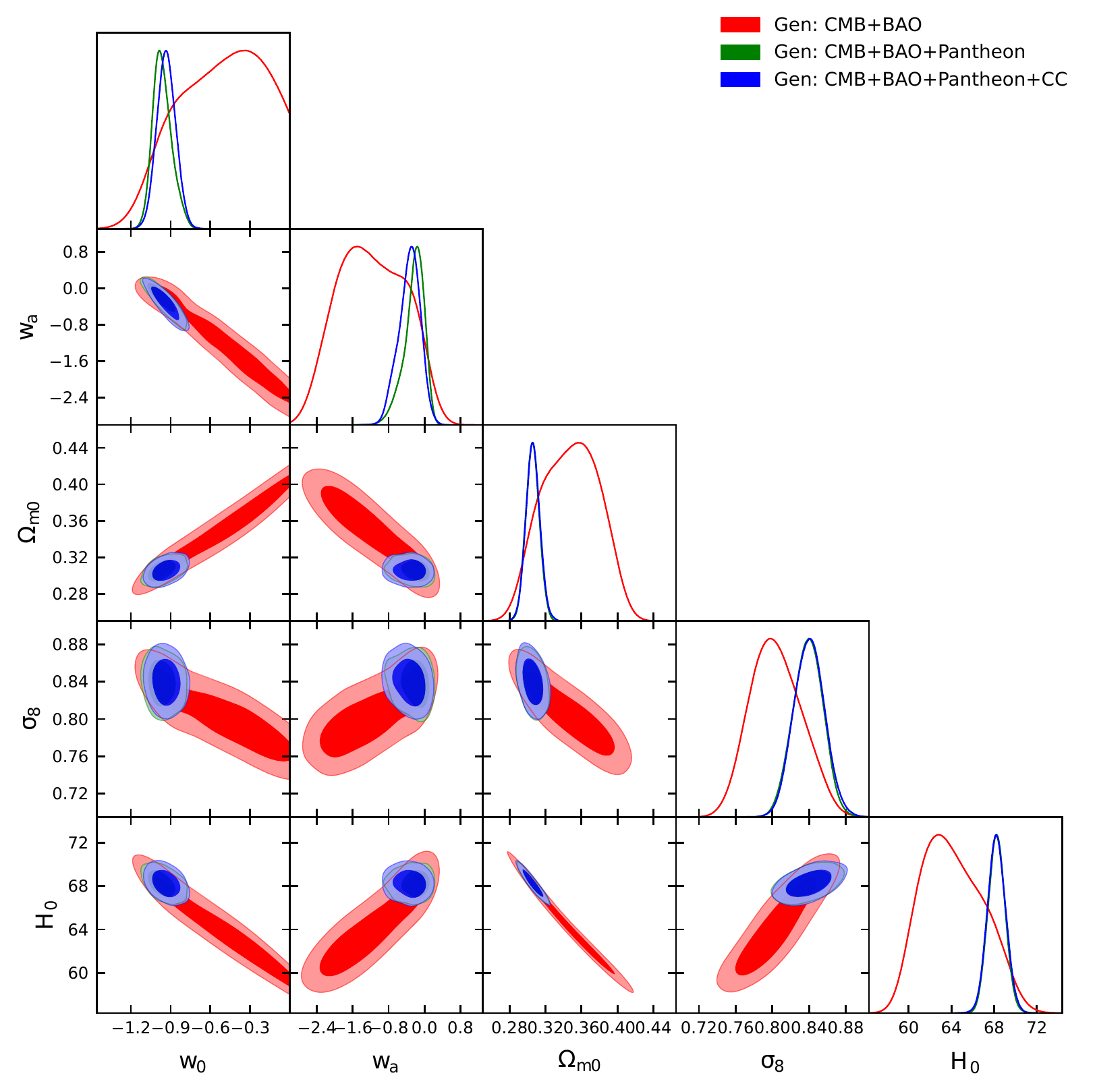}
\caption{68\% and 95\% CL contour plots for various combinations of the
model parameters of the general parametrization of (\protect\ref%
{model-general}) (Gen) have been shown for different observational
combinations. The figure also contains the one dimensional marginalized
posterior distributions for the parameters shown in the two dimensional
contour plots. }
\label{fig:contour-gen}
\end{figure*}
\begingroup                                                                                                                     
\squeezetable 

\begin{center}
\begin{table*}%[tbp]
\begin{tabular}{cccccccccccc}
\hline\hline
Parameters & CMB+BAO & CMB+BAO+Pantheon & CMB+BAO+Pantheon+CC \\ \hline
$\Omega_c h^2$ & $0.1191_{- 0.0013- 0.0027}^{+ 0.0014+ 0.0026}$ & $0.1191_{-
0.0013- 0.0026}^{+ 0.0013+ 0.0025}$ & $0.1191_{- 0.0013- 0.0026}^{+ 0.0013+
0.0026}$   \\ 
$\Omega_b h^2$ & $0.02226_{- 0.00015- 0.00029}^{+ 0.00015+ 0.00029}$ & $%
0.02227_{- 0.00015- 0.00030}^{+ 0.00015+ 0.00030}$ & $0.02227_{- 0.00015-
0.00030}^{+ 0.00015+ 0.00030}$  \\ 
$100\theta_{MC}$ & $1.04078_{- 0.00032- 0.00064}^{+ 0.00033+ 0.00063}$ & $%
1.04080_{- 0.00031- 0.00062}^{+ 0.00032+ 0.00064}$ & $1.04080_{- 0.00032-
0.00062}^{+ 0.00031+ 0.00062}$   \\ 
$\tau$ & $0.078_{- 0.017- 0.034}^{+ 0.017+ 0.034}$ & $0.080_{- 0.017-
0.035}^{+ 0.017+ 0.035}$ & $0.080_{- 0.018- 0.034}^{+ 0.018+ 0.034}$  \\ 
$n_s$ & $0.9665_{- 0.0044- 0.0084}^{+ 0.0044+ 0.0091}$ & $0.9667_{- 0.0044-
0.0086}^{+ 0.0044+ 0.0089}$ & $0.9666_{- 0.0044- 0.0087}^{+ 0.0045+ 0.0087}$
  \\ 
$\mathrm{ln}(10^{10} A_s)$ & $3.090_{- 0.033- 0.066}^{+ 0.034+ 0.066}$ & $%
3.092_{- 0.033- 0.068}^{+ 0.034+ 0.066}$ & $3.092_{- 0.034- 0.066}^{+ 0.034+
0.066}$  \\ 
$w_0$ & $-0.524_{- 0.236- 0.514}^{+ 0.374+ 0.524}$ & $-0.947_{- 0.088-
0.156}^{+ 0.076+ 0.165}$ & $-0.950_{- 0.084- 0.152}^{+ 0.075+ 0.152}$   \\
 
$w_a$ & $-1.403_{- 1.021- 1.466}^{+ 0.731+ 1.570}$ & $-0.308_{- 0.273-
0.677}^{+ 0.367+ 0.619}$ & $-0.291_{- 0.256- 0.588}^{+ 0.338+ 0.555}$   \\
 
$\Omega_{m0}$ & $0.344_{- 0.026- 0.054}^{+ 0.032+ 0.051}$ & $0.304_{- 0.007-
0.015}^{+ 0.008+ 0.016}$ & $0.304_{- 0.008- 0.015}^{+ 0.008+ 0.015}$ \\
 
$\sigma_8$ & $0.803_{- 0.030- 0.051}^{+ 0.024+ 0.053}$ & $0.838_{- 0.016-
0.033}^{+ 0.018+ 0.032}$ & $0.837_{- 0.016- 0.032}^{+ 0.016+ 0.032}$ \\
 
$H_0$ & $64.36_{- 3.23- 4.67}^{+ 2.05+ 5.26}$ & $68.34_{- 0.82- 1.63}^{+
0.81+ 1.70}$ & $68.31_{- 0.83- 1.66}^{+ 0.84+ 1.64}$ 
\\ \hline\hline
\end{tabular}%
\caption{Observational constraints on the dark energy parametrization,
namely, the Ext1 of (\protect\ref{extension1}) using various observational
datasets. We note that $H_0$ is in the units of km/Mpc/sec and $\Omega_{m0}$
is the present value of $\Omega_m = \Omega_b +\Omega_c$. }
\label{tab:ext1CPL}
\end{table*}
\end{center}

\endgroup  
\begin{figure*}%[tbp]
\includegraphics[width=0.65\textwidth]{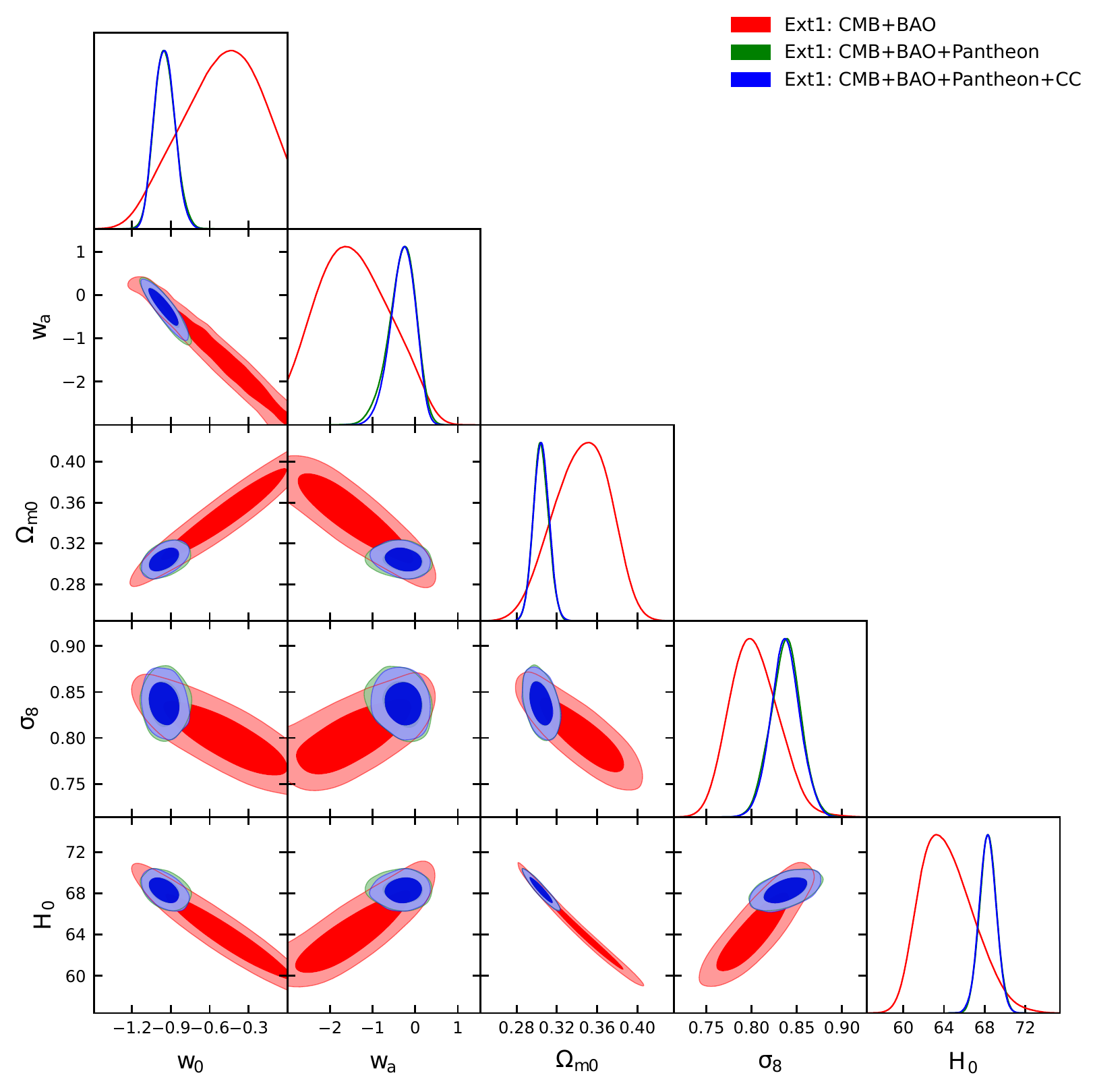}
\caption{68\% and 95\% CL contour plots for various combinations of the
model parameters of Ext1 of (\protect\ref{extension1}) [the CPL
parametrization] have been shown for different observational combinations.
The figure also contains the one dimensional marginalized posterior
distributions for the parameters shown in the two dimensional contour plots.}
\label{fig:contour-ext1CPL}
\end{figure*}
\begingroup                                                                                                                     
\squeezetable                                                                                                                   

\begin{center}
\begin{table*}%[tbp]
\begin{tabular}{ccccccccccccccc}
\hline\hline
Parameters & CMB+BAO & CMB+BAO+Pantheon & CMB+BAO+Pantheon+CC   
\\ \hline
$\Omega_c h^2$ & $0.1192_{- 0.0013- 0.0027}^{+ 0.0014+ 0.0026}$ & $0.1194_{-
0.0013- 0.0025}^{+ 0.0013+ 0.0026}$ & $0.1195_{- 0.0014- 0.0027}^{+ 0.0013+
0.0027}$  \\ 
$\Omega_b h^2$ & $0.02223_{- 0.00015- 0.00030}^{+ 0.00015+ 0.00030}$ & $%
0.02223_{- 0.00015- 0.00028}^{+ 0.00015+ 0.00028}$ & $0.02222_{- 0.00015-
0.00029}^{+ 0.00015+ 0.00028}$   \\ 
$100\theta_{MC}$ & $1.04051_{- 0.00034- 0.00063}^{+ 0.00032+ 0.00064}$ & $%
1.04051_{- 0.00033- 0.00059}^{+ 0.00031+ 0.00061}$ & $1.04050_{- 0.00033-
0.00065}^{+ 0.00033+ 0.00062}$   \\ 
$\tau$ & $0.079_{- 0.018- 0.035}^{+ 0.018+ 0.034}$ & $0.079_{- 0.017-
0.035}^{+ 0.017+ 0.033}$ & $0.081_{- 0.018- 0.035}^{+ 0.017+ 0.036}$  \\ 

$n_s$ & $0.9744_{- 0.0043- 0.0088}^{+ 0.0043+ 0.0083}$ & $0.9740_{- 0.0045-
0.0082}^{+ 0.0042+ 0.0086}$ & $0.9740_{- 0.0043- 0.0086}^{+ 0.0045+ 0.0085}$
  \\ 
$\mathrm{ln}(10^{10} A_s)$ & $3.100_{- 0.035- 0.068}^{+ 0.035+ 0.067}$ & $%
3.100_{- 0.033- 0.067}^{+ 0.036+ 0.064}$ & $3.104_{- 0.034- 0.070}^{+ 0.034+
0.069}$   \\ 
$w_0$ & $-0.616_{- 0.428- 0.501}^{+ 0.326+ 0.560}$ & $-0.949_{- 0.076-
0.139}^{+ 0.074+ 0.149}$ & $-0.946_{- 0.088- 0.150}^{+ 0.072+ 0.155}$   \\
 
$w_a$ & $-1.000_{- 0.609- 1.427}^{+ 1.039+ 1.262}$ & $-0.290_{- 0.217-
0.534}^{+ 0.277+ 0.460}$ & $-0.302_{- 0.219- 0.533}^{+ 0.310+ 0.500}$  \\
 
$\Omega_{m0}$ & $0.339_{- 0.037- 0.054}^{+ 0.031+ 0.058}$ & $0.305_{- 0.008-
0.015}^{+ 0.008+ 0.015}$ & $0.306_{- 0.007- 0.015}^{+ 0.008+ 0.015}$ \\
 
$\sigma_8$ & $0.810_{- 0.032- 0.053}^{+ 0.030+ 0.055}$ & $0.839_{- 0.016-
0.031}^{+ 0.016+ 0.031}$ & $0.841_{- 0.016- 0.033}^{+ 0.016+ 0.033}$  \\
 
$H_0$ & $64.98_{- 3.35- 5.47}^{+ 3.11+ 5.55}$ & $68.29_{- 0.82- 1.59}^{+
0.81+ 1.65}$ & $68.24_{- 0.78- 1.50}^{+ 0.79+ 1.56}$   \\ \hline\hline
\end{tabular}%
\caption{Observational constraints on the dark energy parametrization,
namely, the Ext2 of (\protect\ref{extension2}) using various observational
datasets. We note that $H_0$ is in the units of km/Mpc/sec and $\Omega_{m0}$
is the present value of $\Omega_m = \Omega_b +\Omega_c$.}
\label{tab:ext2}
\end{table*}
\end{center}

\endgroup        
\begin{figure*}%[tbp]
\includegraphics[width=0.65\textwidth]{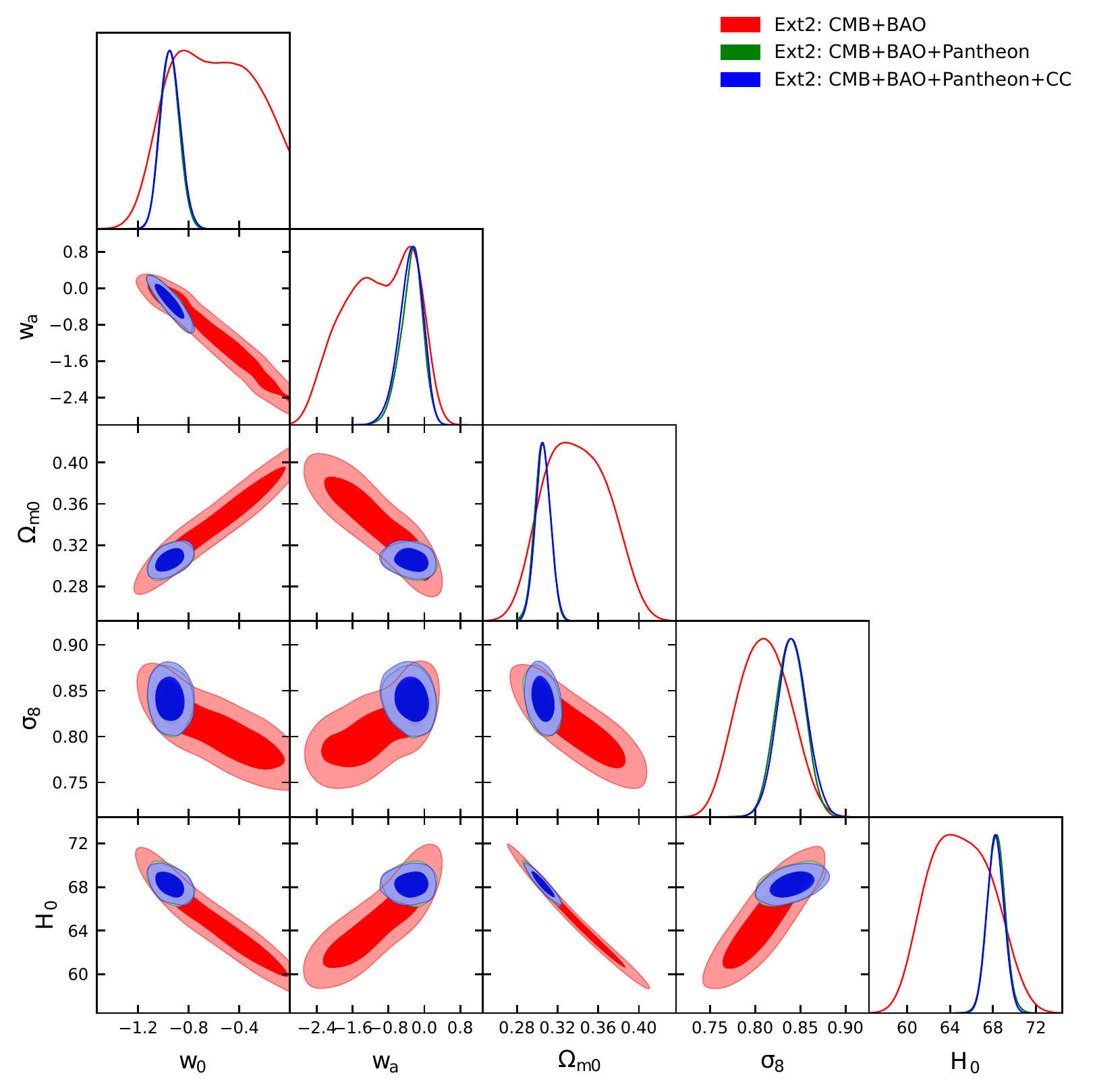}
\caption{68\% and 95\% CL contour plots for various combinations of the
model parameters of the Ext2 of (\protect\ref{extension2}) have been shown
for different observational combinations. The figure also contains the one
dimensional marginalized posterior distributions for the parameters shown in
the two dimensional contour plots.}
\label{fig:contour-ext2}
\end{figure*}
\begingroup                                                                                                                     
\squeezetable                                                                                                                   
\begin{center}
\begin{table*}%[tbp]
\begin{tabular}{ccccccccccccccc}
\hline\hline
Parameters & CMB+BAO & CMB+BAO+Pantheon & CMB+BAO+Pantheon+CC   
\\ \hline

$\Omega_c h^2$ & $    0.1195_{-    0.0013-    0.0027}^{+    0.0013+    0.0027}$ & $    0.1194_{-    0.0013-    0.0027}^{+    0.0013+    0.0026}$  & $    0.1195_{-    0.0013-    0.0025}^{+    0.0013+    0.0027}$ \\

$\Omega_b h^2$ & $    0.02222_{-    0.00014-    0.00029}^{+    0.00014+    0.00029}$ &  $    0.02222_{-    0.00014-    0.00029}^{+    0.00015+    0.00030}$ & $    0.02221_{-    0.00015-    0.00029}^{+    0.00015+    0.00029}$ \\

$100\theta_{MC}$ & $    1.04049_{-    0.00032-    0.00065}^{+    0.00031+    0.00065}$ &  $    1.04050_{-    0.00032-    0.00065}^{+    0.00033+    0.00062}$ & $    1.04047_{-    0.00032-    0.00063}^{+    0.00032+    0.00061}$ \\

$\tau$ & $    0.078_{-    0.018-    0.036}^{+    0.018+    0.035}$ &  $    0.080_{-    0.017-    0.035}^{+    0.018+    0.034}$ & $    0.079_{-    0.017-    0.034}^{+    0.018+    0.033}$ \\

$n_s$ & $    0.9737_{-    0.0048-    0.0083}^{+    0.0043+    0.0089}$ & $    0.9738_{-    0.0043-    0.0086}^{+    0.0043+    0.0086}$  & $    0.9737_{-    0.0043-    0.0084}^{+    0.0043+    0.0084}$ \\

${\rm{ln}}(10^{10} A_s)$ & $    3.099_{-    0.036-    0.069}^{+    0.035+    0.067}$ &  $    3.104_{-    0.033-    0.067}^{+    0.034+    0.066}$  & $    3.101_{-    0.034-    0.067}^{+    0.034+    0.065}$  \\

$w_0$ & $   -0.582_{-    0.501-    0.564}^{+    0.490+    0.581}$ & $   -0.946_{-    0.080-    0.137}^{+    0.070+    0.147}$  & $    -0.936_{-    0.071-    0.141}^{+    0.072+    0.138}$ \\

$w_a$ & $   -1.077_{-    1.010-    1.450}^{+    1.176+    1.422}$ &  $   -0.295_{-    0.215-    0.508}^{+    0.284+    0.483}$ & $   -0.335_{-    0.222-    0.472}^{+    0.264+    0.471}$  \\

$\Omega_{m0}$ & $    0.343_{-    0.045-    0.059}^{+    0.042+    0.058}$ & $    0.306_{-    0.0075-    0.015}^{+    0.0075+    0.015}$ & $    0.306_{-    0.0083-    0.015}^{+    0.0076+    0.016}$  \\

$\sigma_8$ & $    0.808_{-    0.036-    0.052}^{+    0.028+    0.059}$ & $    0.840_{-    0.016-    0.031}^{+    0.016+    0.031}$ & $    0.841_{-    0.017-    0.034}^{+    0.016+    0.032}$ \\

$H_0$ & $   64.67_{-    4.37-    5.42}^{+    3.73+    6.18}$ &  $   68.26_{-    0.82-    1.58}^{+    0.80+    1.57}$ & $   68.27_{-    0.85-    1.62}^{+    0.83+    1.73}$  \\

\hline\hline
\end{tabular}%
\caption{Observational constraints on the dark energy parametrization,
namely, the Ext3 of (\protect\ref{extension3}) using various observational
datasets. We note that $H_0$ is in the units of km/Mpc/sec and $\Omega_{m0}$
is the present value of $\Omega_m = \Omega_b +\Omega_c$. }
\label{tab:ext3}
\end{table*}
\end{center}

\endgroup         
\begin{figure*}%[tbp]
\includegraphics[width=0.65\textwidth]{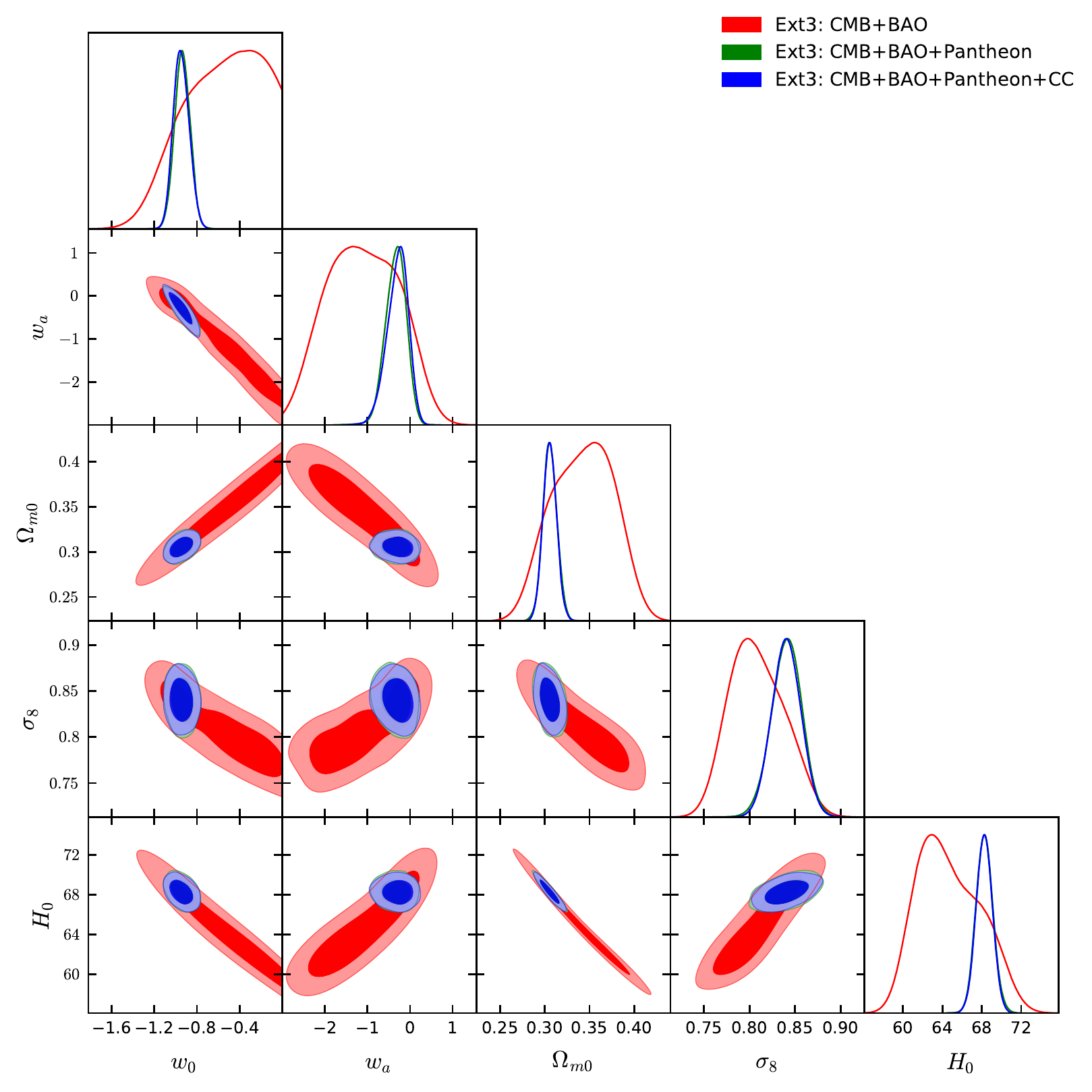}
\caption{68\% and 95\% CL contour plots for various combinations of the
model parameters of the Ext3 of (\protect\ref{extension3}) have been shown
for different observational combinations. The figure also contains the one
dimensional marginalized posterior distributions for the parameters shown in
the two dimensional contour plots. }
\label{fig:contour-ext3}
\end{figure*}
\begin{figure*}%[tbp]
\includegraphics[width=0.32\textwidth]{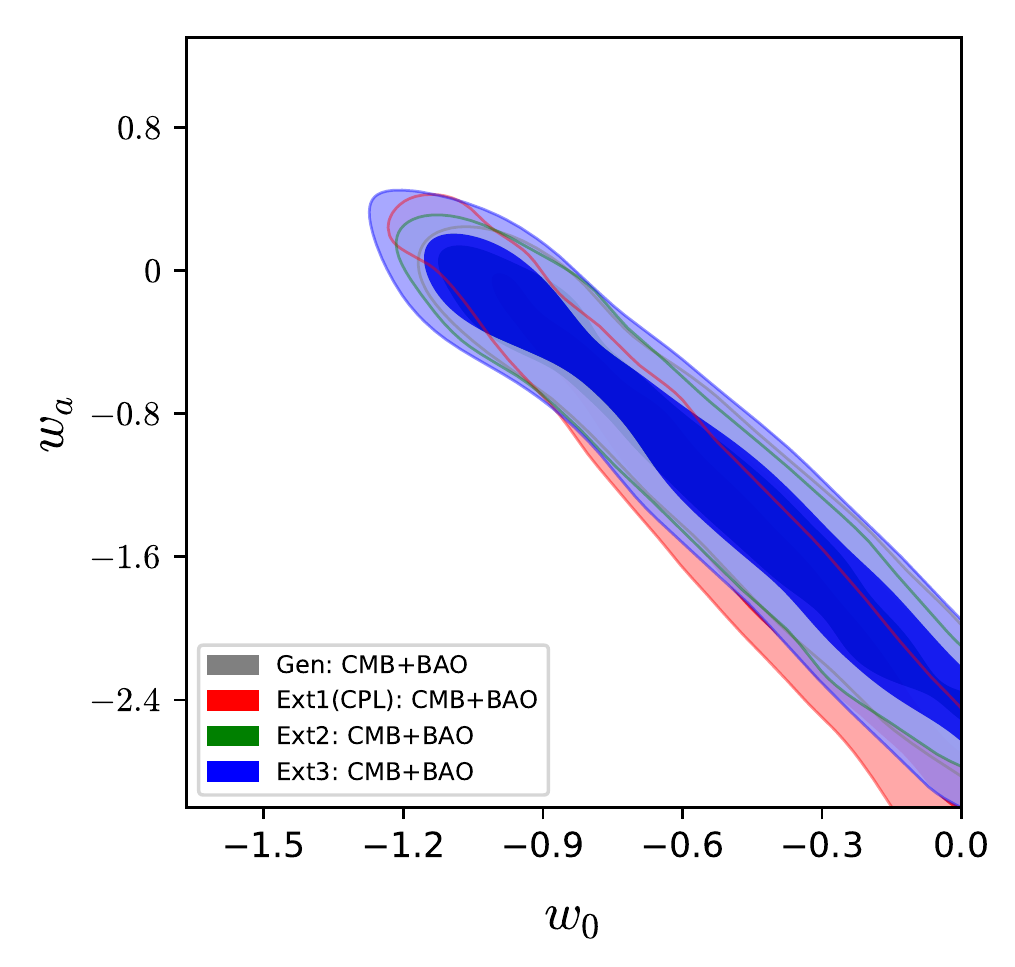} %
\includegraphics[width=0.32\textwidth]{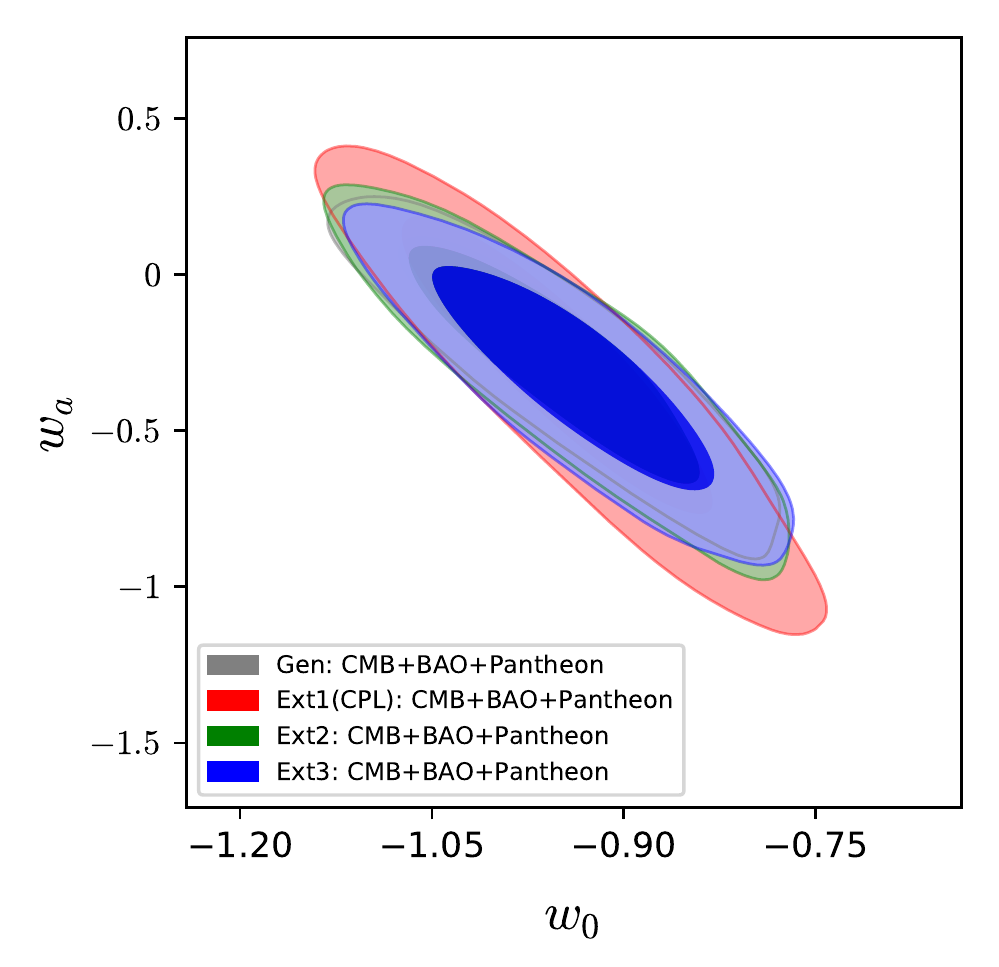} %
\includegraphics[width=0.32\textwidth]{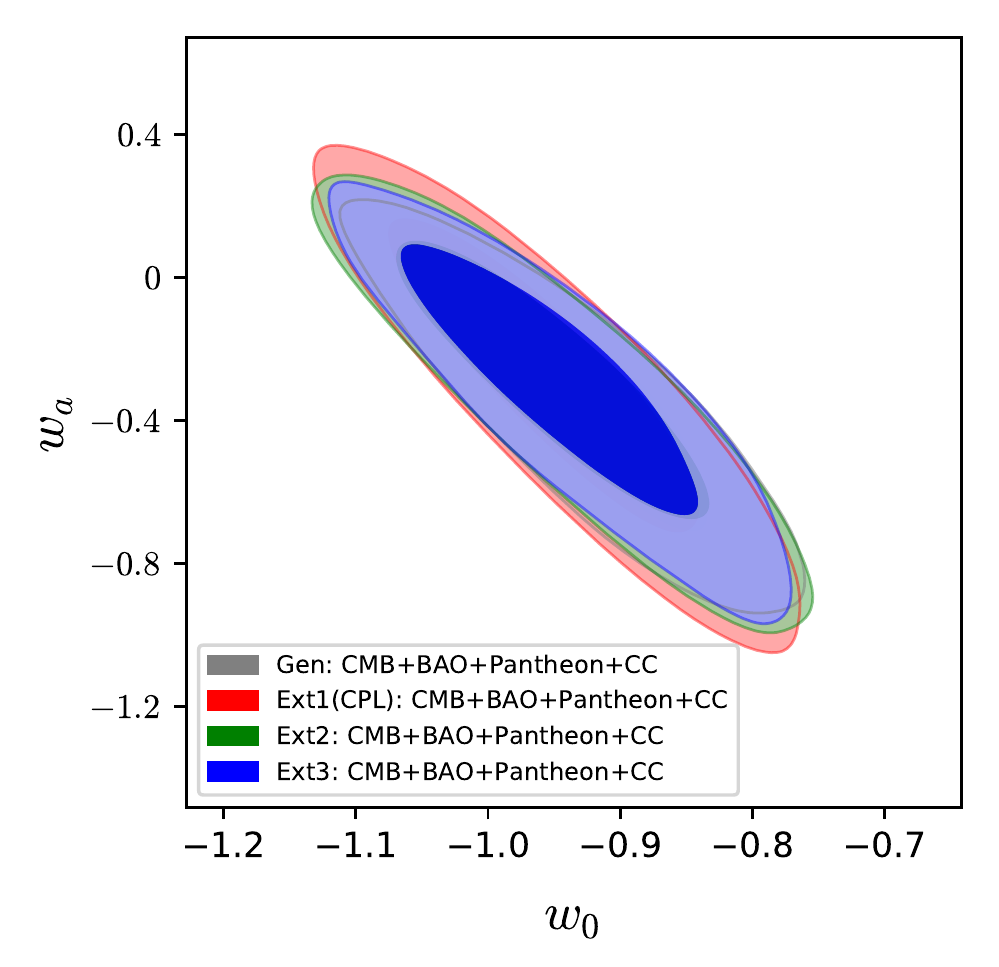}
\caption{We show the $(w_0, w_a)$ plane for the present dynamical dark
energy parametrizations using different observational datasets. The
left graph for the dataset CMB+BAO, the middle graph for the dataset
CMB+BAO+Pantheon and the right graph stands for the observational dataset
CMB+BAO+Pantheon+CC. }
\label{fig:w0-wa}
\end{figure*}
\begingroup                                          
\squeezetable                                                                                                 
\tiny
\begin{center}
\begin{table*}%[tbp]
\begin{tabular}{ccccccccccccc}
\hline\hline
Datasets & Parameters & Gen & Ext1 (CPL) & Ext2 & Ext3  \\ \hline
CB & ($w_0$, $w_a$) & ($-0.537_{-0.441}^{+ 0.442}$, $-1.154_{- 0.807}^{+
0.894}$) & ($-0.524_{- 0.236}^{+ 0.374}$, $-1.403_{- 1.021}^{+ 0.731}$) & ($%
-0.616_{- 0.428}^{+ 0.326}$, $-1.000_{- 0.609}^{+ 1.039}$) & $(-0.582_{-    0.501}^{+    0.490}, -1.077_{- 1.010}^{+ 1.176})$  \\ 
CBP & ($w_0$, $w_a$) & ($-0.963_{-0.082}^{+ 0.060}$, $-0.231_{- 0.143}^{+
0.291}$) & ($-0.947_{- 0.088}^{+ 0.076}$, $-0.308_{- 0.273}^{+ 0.367}$) & ($%
-0.949_{- 0.076}^{+ 0.074}$, $-0.290_{- 0.217}^{+ 0.277}$) &  $( -0.946_{-    0.080}^{+    0.070},  -0.295_{- 0.215}^{+    0.284})$ \\ 
CMPC & ($w_0$, $w_a$) & ($-0.933_{-0.070}^{+ 0.071}$, $-0.337_{- 0.204}^{+
0.288}$) & ($-0.950_{- 0.084}^{+ 0.075}$, $-0.291_{- 0.256}^{+ 0.338}$) & ( $%
-0.946_{- 0.088}^{+ 0.072}$, $-0.302_{- 0.219}^{+ 0.310}$) &  $(-0.936_{- 0.071}^{+    0.072}, -0.335_{-    0.222}^{+    0.264})$ \\ 
\hline
\end{tabular}%
\caption{For a clear understandig and comparison of the dark energy key
parameters ($w_0$, $w_a$) extracted from all the parametrizations, we show
their estimations at 68\% CL. Here, CB = CMB+BAO, CBP = CMB+BAO+Pantheon,
CBPC = CMB+BAO+Pantheon+CC. }
\label{tab:comparison}
\end{table*}
\end{center}

\endgroup  

\section{Observational constraints and the analysis}

\label{sec-results}

In this section we describe the observational constraints on all the dark
energy parametrization, namely the general parametrization of eqn. (\ref%
{model-general}), Extension 1 or the CPL parametrization of eqn. (\ref%
{extension1}), Extension 2 of eqn. (\ref{extension2}) and extension 3 of
eqn. (\ref{extension3}) using various astronomical datasets summarized in
section \ref{sec-data}. In particular, we focus on the two key parameters of
the dark energy parametrization, namely, $w_{0}$ and $w_{a}$ in order to
investigate the qualitative changes in the parametrization as long as
nonlinear terms are considered. In what follows we describe the
observational constraints extracted from each dark energy scenario.

Let us first focus on the general dark energy parametrization given in
equation (\ref{model-general}). We have constrained this dark energy
scenario using different cosmological datasets such as CMB+BAO, CMB+BAO+Pantheon and CMB+BAO+Pantheon+CC, the results of which
are summarized in Table \ref{tab:model-gen}. From Table \ref{tab:model-gen},
a general conclusion that one might quickly observe is that, the inclusion of Pantheon to CMB+BAO significantly improves the error bars on all the parameters, and not only that, some of the parameters are significantly improved concerning their mean valus. 
In fact, the best constraints on the model parameters are achieved
for the combination CMB+BAO+Pantheon. The inclusion of CC to CMB+BAO+Pantheon although does not add much significant insight on the cosmological constraints, however, the effects on $w_a$ are pronounced when CC data are added to CMB+BAO+Pantheon. 
Let us now focus on the constraints on individual model parameters.  
As one can see from Table \ref{tab:model-gen} that 
the mean value of the dark energy equation of state
at present, i.e., $w_{0}$ is always in the quintessential regime: $ w_{0}=-0.963_{-0.082}^{+0.060}$ at 68\% CL for CMB+BAO+Pantheon and $w_{0}=-0.933_{-0.070}^{+0.071}$ at 68\% CL for CMB+BAO+Pantheon+CC. Although from statistical point of view, one can argue that the constraints on $w_0$ mildly suggest for a crossing of the phantom divide line $w_{0}=-1$, however, $w_0 = -1$ is the most consistent scenario.  Concerning the remaining key parameter, $w_{a}$, we find that it may assume nonnull values, however, $w_{a}=0$ is allowed within 68\% CL of course. In Fig. \ref%
{fig:contour-gen}, we have shown the one dimensional posterior distributions
for some selected parameters of this model as well as the two dimensional
contour plots considering various combinations of the model parameters. 
From Fig. \ref{fig:contour-gen}, one can see that the parameters shown in this figure are correlated with each other. Specifically, we find a strong correlation
between $w_{0}$, $w_{a}$ and $H_{0}$. Finally, we focus on the estimation of the Hubble constant $H_0$ for all three datasets. One can strikingly see that for CMB+BAO, $H_0$ assumes a very lower value ($H_0 = 64.14_{- 3.80}^{+ 2.51}$ at 68\% CL, CMB+BAO) compared to the $\Lambda$CDM based Planck's estimation \cite{Ade:2015xua} and this naturally increases the tension with the local measurements \cite{Riess:2016jrr}. However, for the remaining datasets, we find that $H_0$ takes higher values with slightly higher error bars compared to the $\Lambda$CDM based Planck's estimation \cite{Ade:2015xua}, thus, it slightly decreases the tension on it.

We now consider the first extension of the general parametrization (\ref%
{model-general}) that leads to the well known CPL parametrization of (\ref%
{extension1}). The cosmic scenario driven by this parametrization has been constrained 
using the same observational datasets applied to the general DE parametrization
and the numerical results are summarized in Table \ref{tab:ext1CPL}. One can 
clearly see that the Hubble constant takes similar values compared to the previous scenario (see Table \ref{tab:model-gen}). In fact, concerning the key parameters, namely,  $w_{0}$ and $w_{0}$, our conclusion  remains same, that means the constraints on $w_0$ and $w_a$ are almost similar to what we have found
with the general parametrization (\ref{model-general}). So, effectively we see
that the first approximation (\ref{extension1}) of the original
parametrization (\ref{model-general}) returns similar fit to the original
parametrization (\ref{model-general}). Finally, 
in Fig. \ref{fig:contour-ext1CPL} we have shown the graphical behaviour of various model parameters 
containing the one dimensional
marginalized posterior distributions as well as the two dimensional contour
plots at 68\% and 95\% CL.

Then we move to the observational constraints of the next parametrization
given in eqn. \ref{extension2}. The results for this parametrization are
shown in Table \ref{tab:ext2} and  Fig. \ref{fig:contour-ext2}. We do not find any notable changes due to the extension of one more term in the DE parametrization. That means, the
parametrization behaves similarly to the previous two parametrizations.

Finally, we focus on the last parametrization of this series, namely
Ext3 shown in eqn. (\ref{extension3}). We have summarized the results
in Table \ref{tab:ext3} and Fig. \ref{fig:contour-ext3}, using the same 
combinations of the cosmological datasets that have been used for the previous parametrizations. It is interesting to note that even if we successively increase the terms in the Taylor expansion of the generalized parametrization (\ref{model-general}), but that does not attribute to any change in the constraints on the key parameters as well as on the derived parameters, for instance the Hubble constant.  For a better understanding  on the key parameters $(w_0, w_a)$ obtained from various combined datasets, in Table \ref{tab:comparison}, we have presented their 68\% CL constraints and in Fig. \ref{fig:w0-wa}, we have shown their two dimensional contour plots. The Table \ref{tab:comparison} and Fig. \ref{fig:w0-wa} clearly emphasize that at the background level, none of the extensions can be distinguished from the general parametrization. 
However, as we will show below that, at the level of perturbations, the inclusion of higher order terms certainly exhibits some changes.

\begin{figure*}%[tbp]
\includegraphics[width=0.3\textwidth]{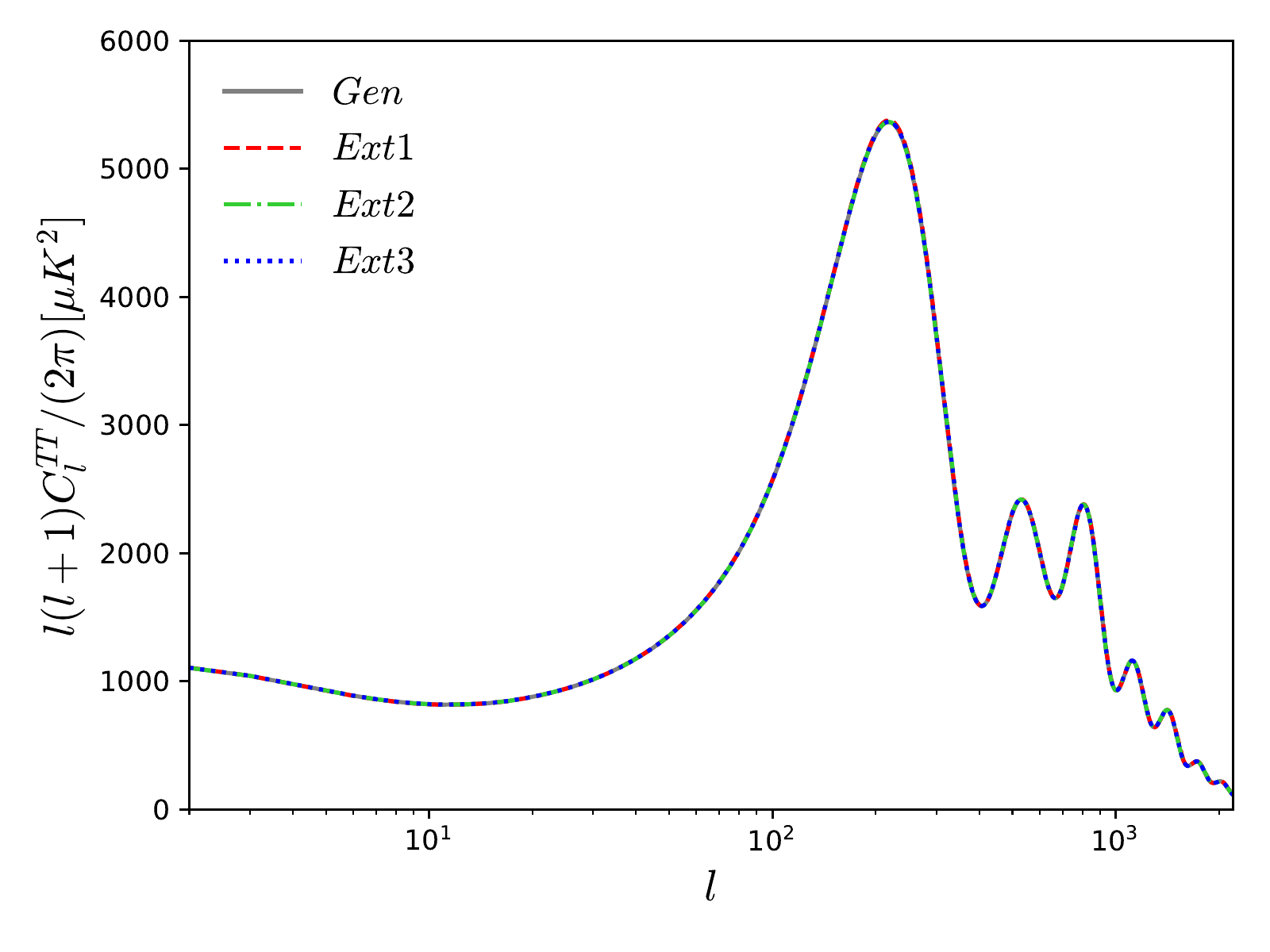} %
\includegraphics[width=0.3\textwidth]{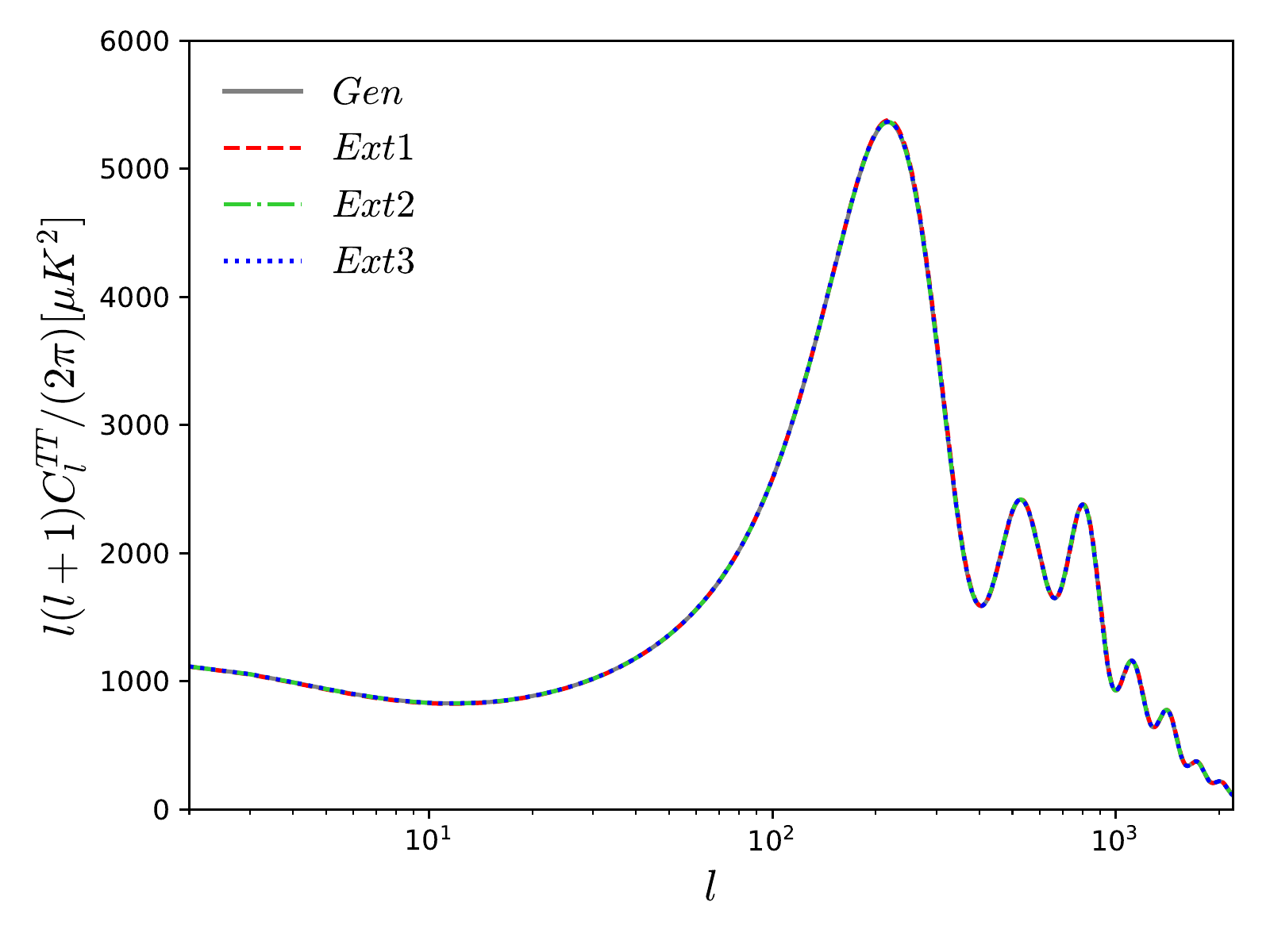} %
\includegraphics[width=0.3\textwidth]{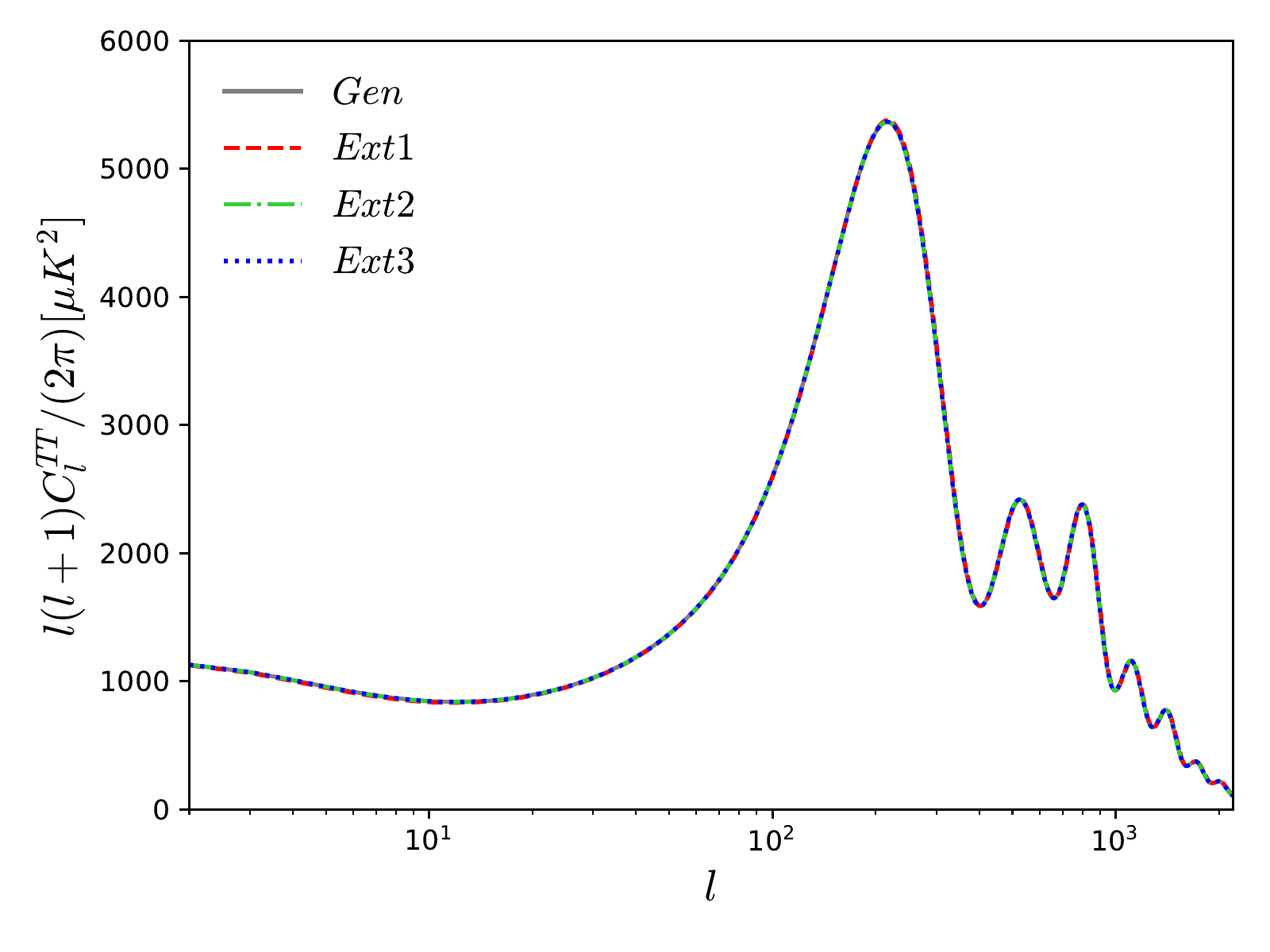}\newline
\includegraphics[width=0.3\textwidth]{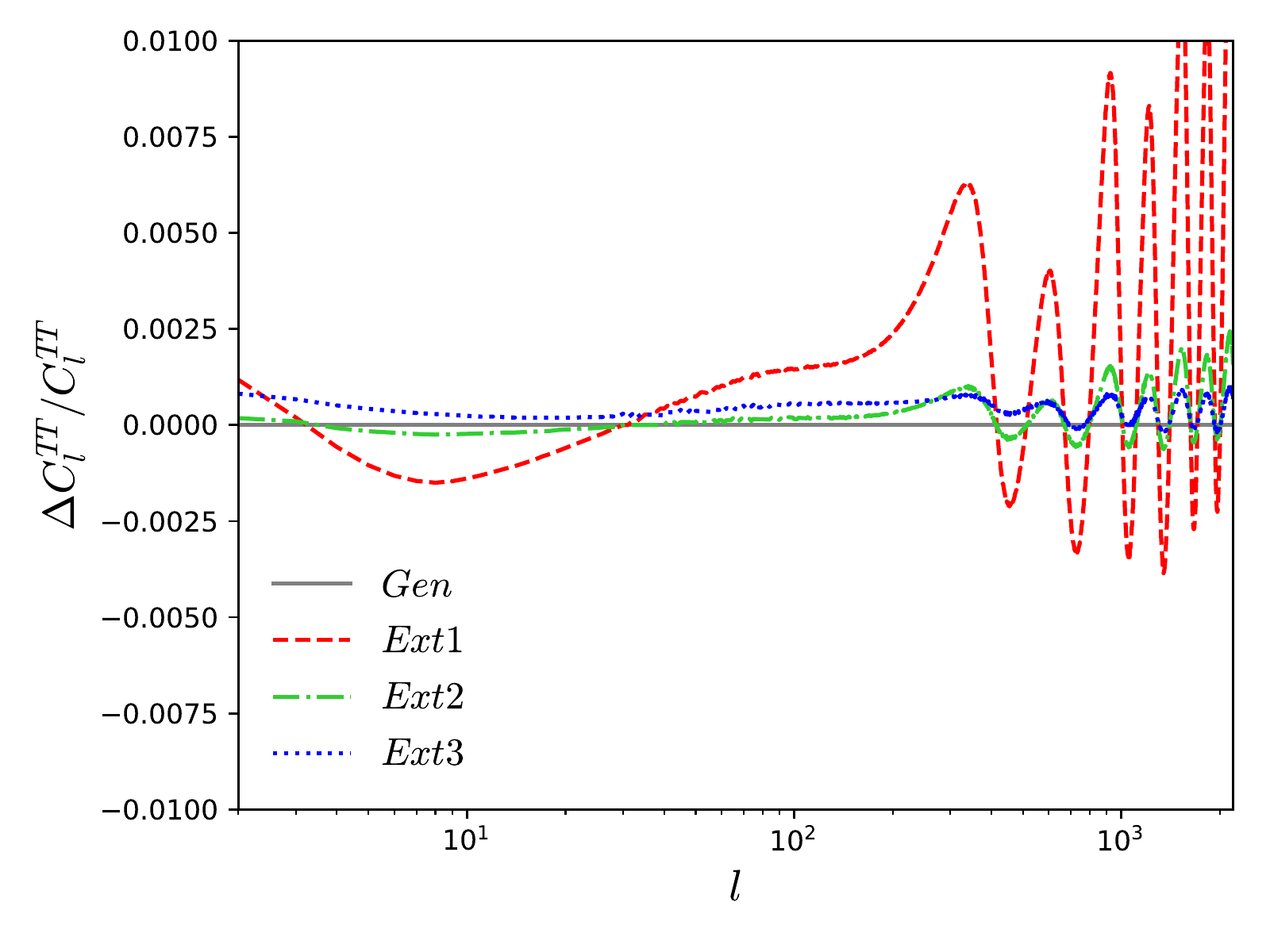} %
\includegraphics[width=0.3\textwidth]{CMBpower_wa01_ratio.pdf} %
\includegraphics[width=0.3\textwidth]{CMBpower_wa01_ratio.pdf}\newline
\includegraphics[width=0.3\textwidth]{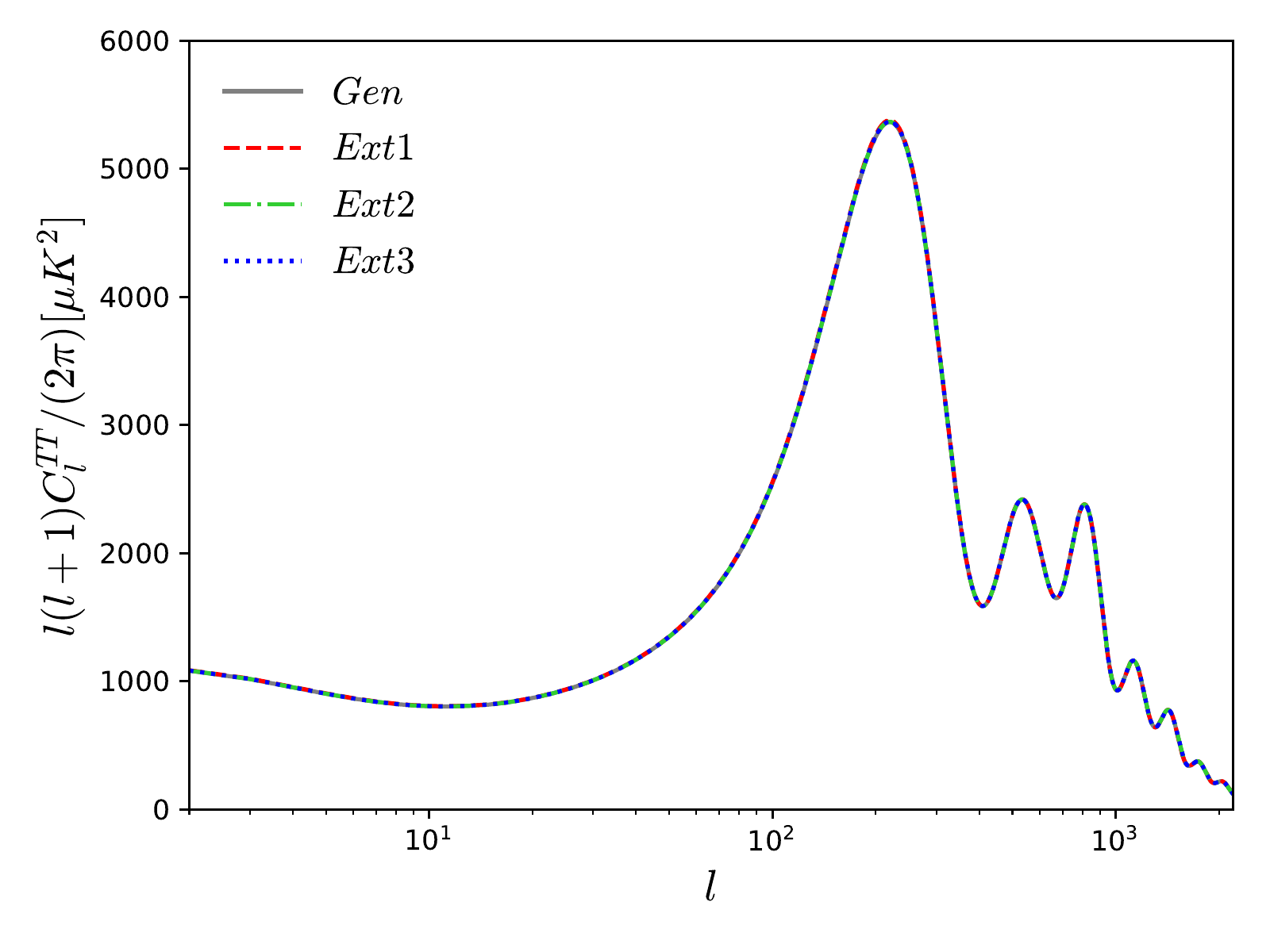} %
\includegraphics[width=0.3\textwidth]{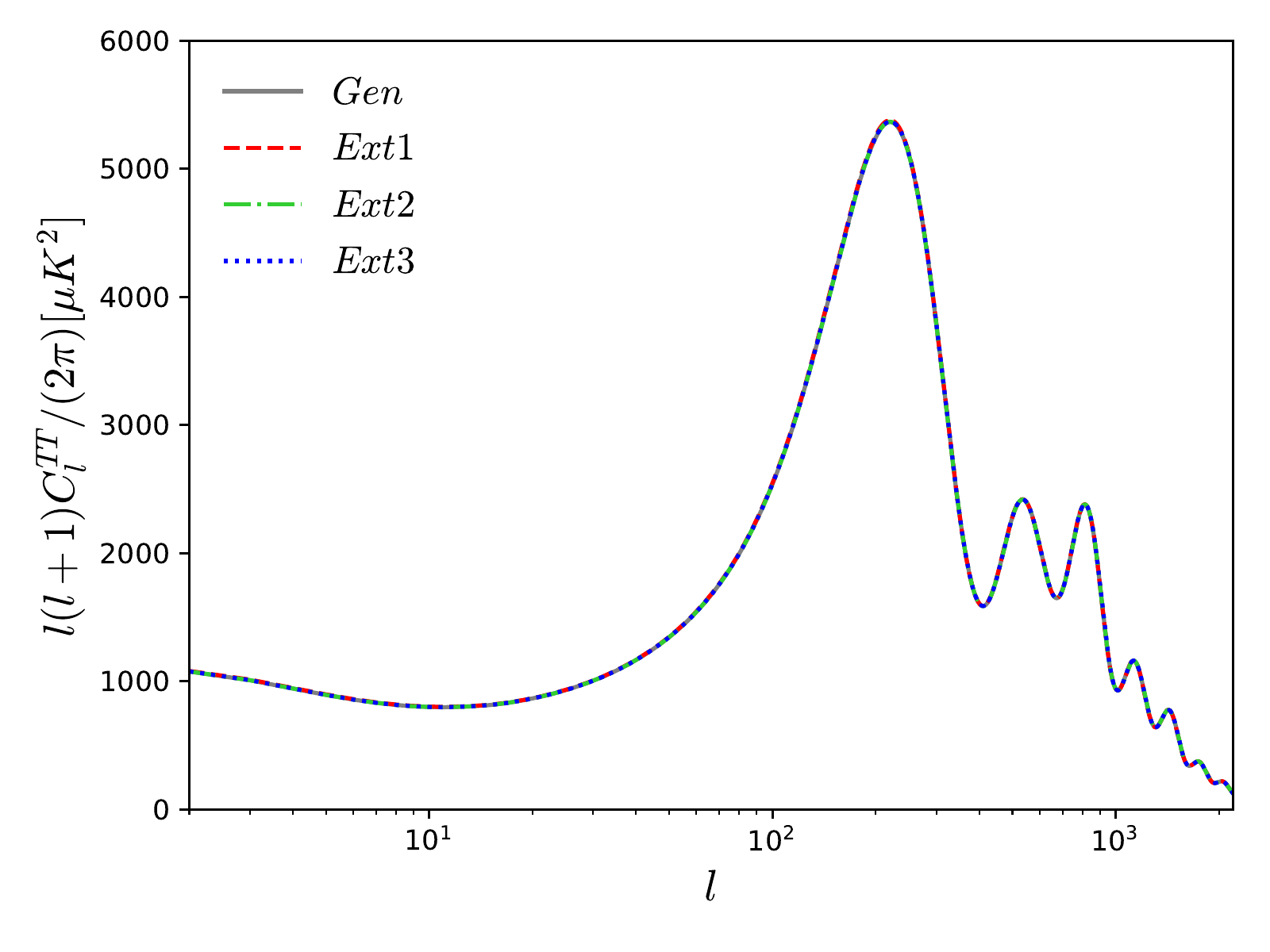} %
\includegraphics[width=0.3\textwidth]{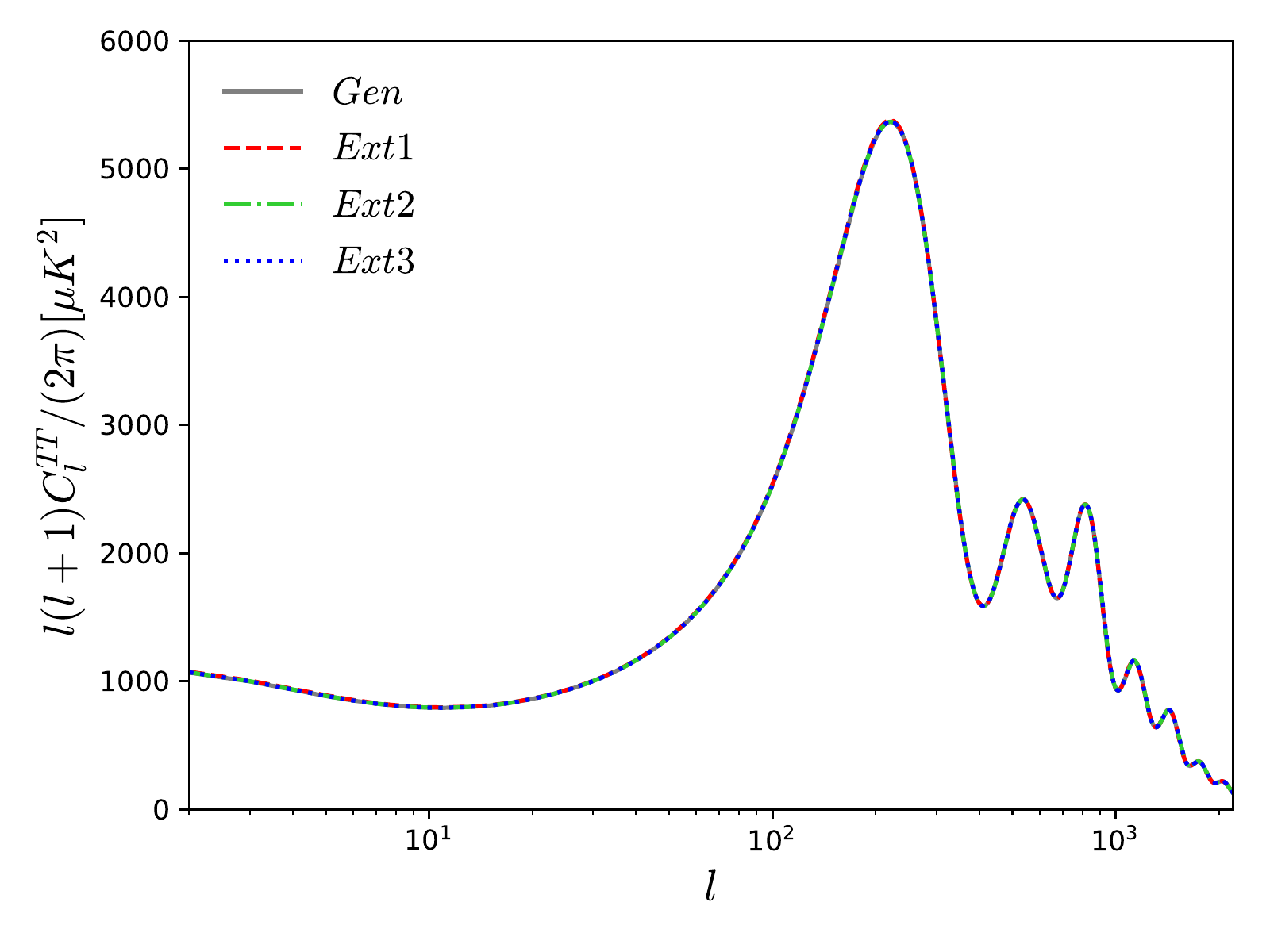}
\includegraphics[width=0.3\textwidth]{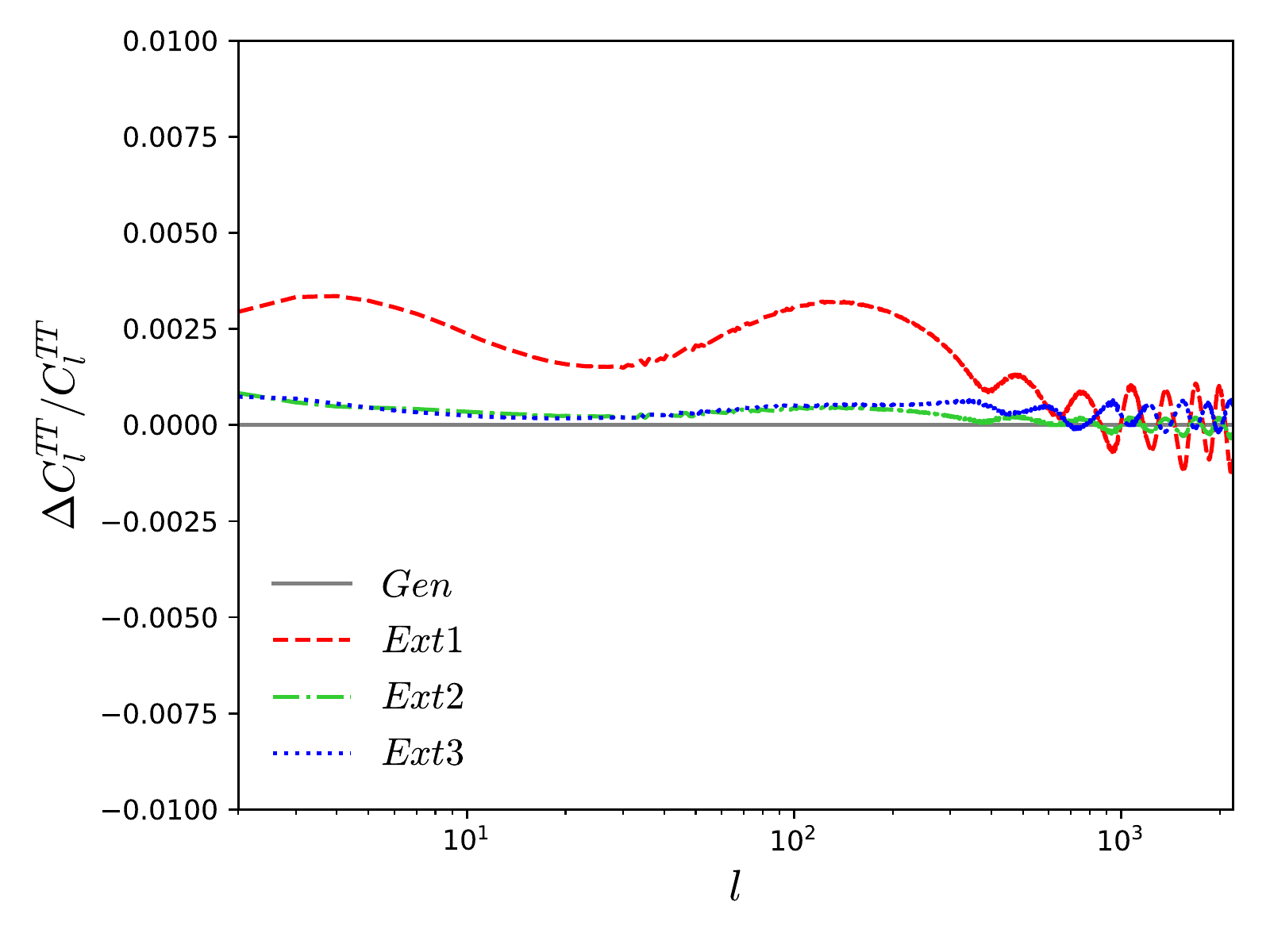} %
\includegraphics[width=0.3\textwidth]{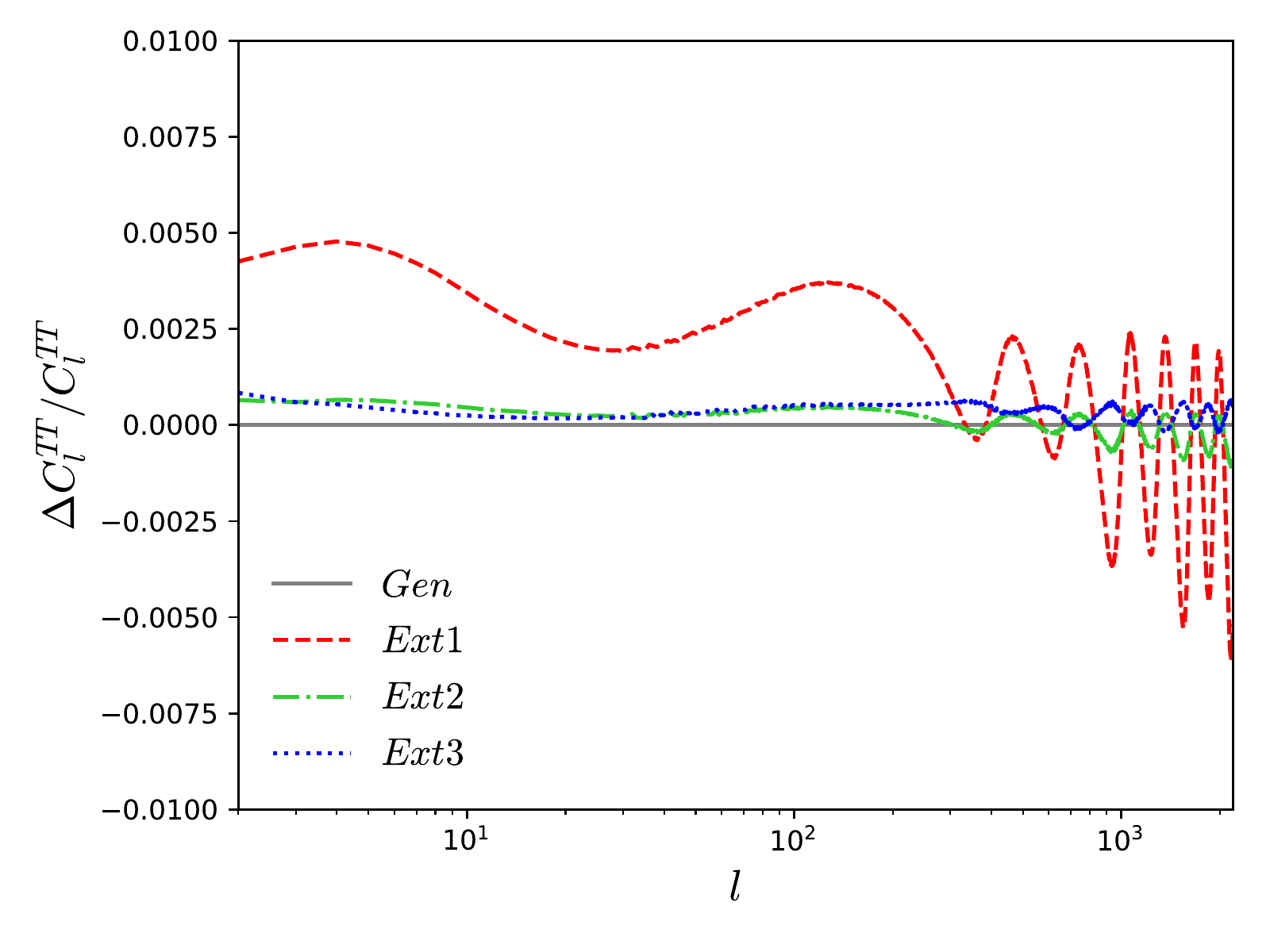} %
\includegraphics[width=0.3\textwidth]{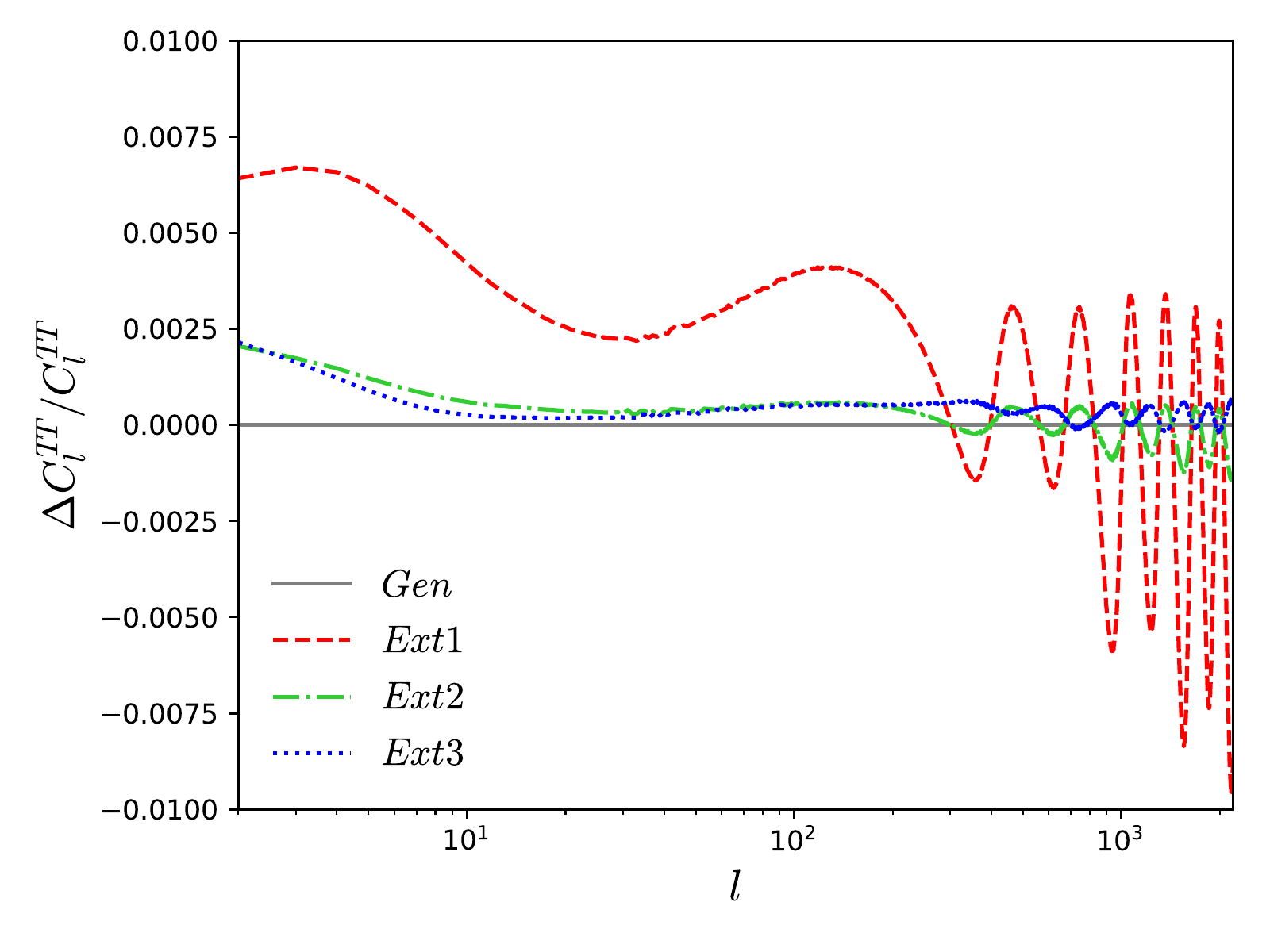}
\caption{We show the cosmic microwave background  spectra and the corresponding residual plots for the present dynamical dark energy parameterizations using various values of the $w_a$ parameter with a fixed $w_0 =-0.95$.  The plots in the first row present the cosmic microwave background spectra and the plots in the second row present the corresponding residual plots. The plots from left to right in the first and second panels of this figure respectively stand for $w_a  = 0.1, 0.2,$ and $0.3$. Similarly, the plots in the third row (showing the cosmic microwave background spectra) and last row (residual plots of the third row) stand for $w_a < 0$ in which the plots from left to right for both the above rows (third and last rows) of this figure respectively stand for $w_a  = -0.1, -0.2,$ and $-0.3$. }
\label{fig:cmbplots}
\end{figure*}

Thus, we investigate how the present dark energy parametrization, namely, the
new dark energy parameterization in eqn. (\ref{model-general}), and its
extensions in equations (\ref{extension1}), (\ref{extension2}) and (\ref%
{extension3}) affect various observables, such as the temperature anisotropy in the cosmic microwave
background spectra as well as the matter power spectra. Such an
investigation is important since this  enables one to understand
how the higher order extensions of the original dark energy parametrization (%
\ref{model-general}) affect the structure formation of the universe.
Thus, in Fig. \ref{fig:cmbplots} we show the
temperature anisotropy in the CMB spectra and the residual plots for different dark energy
parametrizations for various values of $w_{a}$ parameter with a fixed value of $w_{0}=-0.95~$. We have actually fixed $w_{0}=-0.95$ since from the observational analyses of
the models presented in various tables of this article, $w_0$ assumes values close to $-0.95$. For completeness, we have considered both the possibilities namely $w_a > 0$ and $w_a< 0$. The plots in the first row of Fig. \ref{fig:cmbplots} depict the temperature anisotropy in the CMB spectra for $w_a > 0$ and the 
plots in the second row of Fig. \ref{fig:cmbplots} describe the corresponding residual plots.  Let us note that 
the plots from left to right in both the first and second rows of Fig. \ref{fig:cmbplots} respectively stand for $w_a = 0.1,~ 0.2,$ and $0.3$.  In a similar fashion, the plots in the third row of Fig. \ref{fig:cmbplots} stands for the CMB spectra assuming $w_a< 0$ and the plots in the last row of Fig. \ref{fig:cmbplots} represent the corresponding residual plots. And the plots from left to right in both the third and last rows of Fig. \ref{fig:cmbplots} respectively stand for $w_a  = -0.1,~-0.2$ and $-0.3$. From the first row of Fig. \ref{fig:cmbplots} one can clearly notice that the DE parametrizations cannot be distinguished from one another, 
even if we increase the magnitude of $w_a$, however, when we look at the corresponding residual plots shown in the second row of Fig. \ref{fig:cmbplots}, we realize the differences. It is clear that that Ext3 is more close to the original parametriation (\ref{model-general}) compared to Ext1 and Ext2.  The same conclusion can be drawn from the last row of Fig. Fig. \ref{fig:cmbplots}. So, effectively, independently of the sign of $w_a$, the conclusion remains same. 

\begin{figure*}%[tbp]
\includegraphics[width=0.3\textwidth]{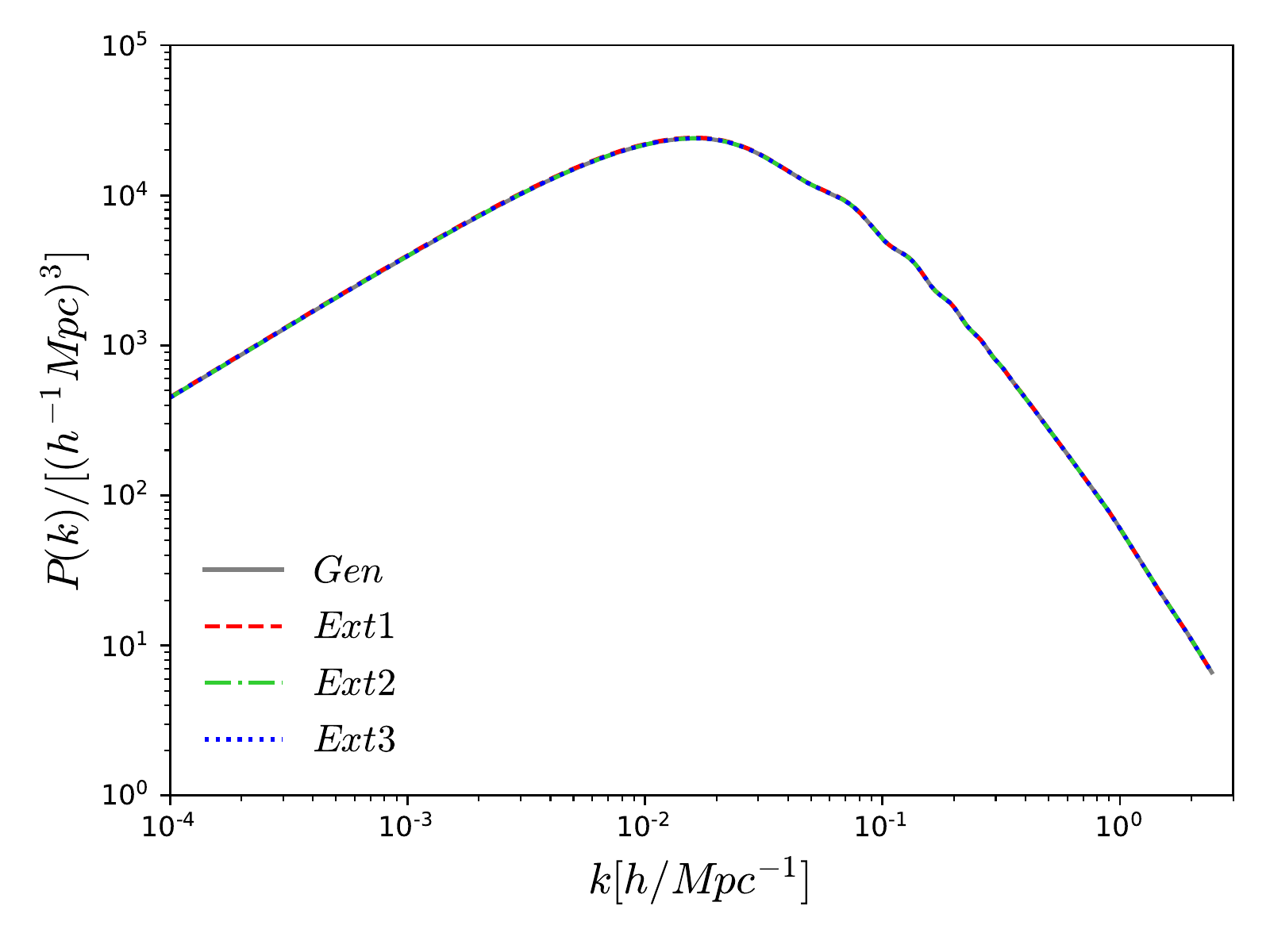} %
\includegraphics[width=0.3\textwidth]{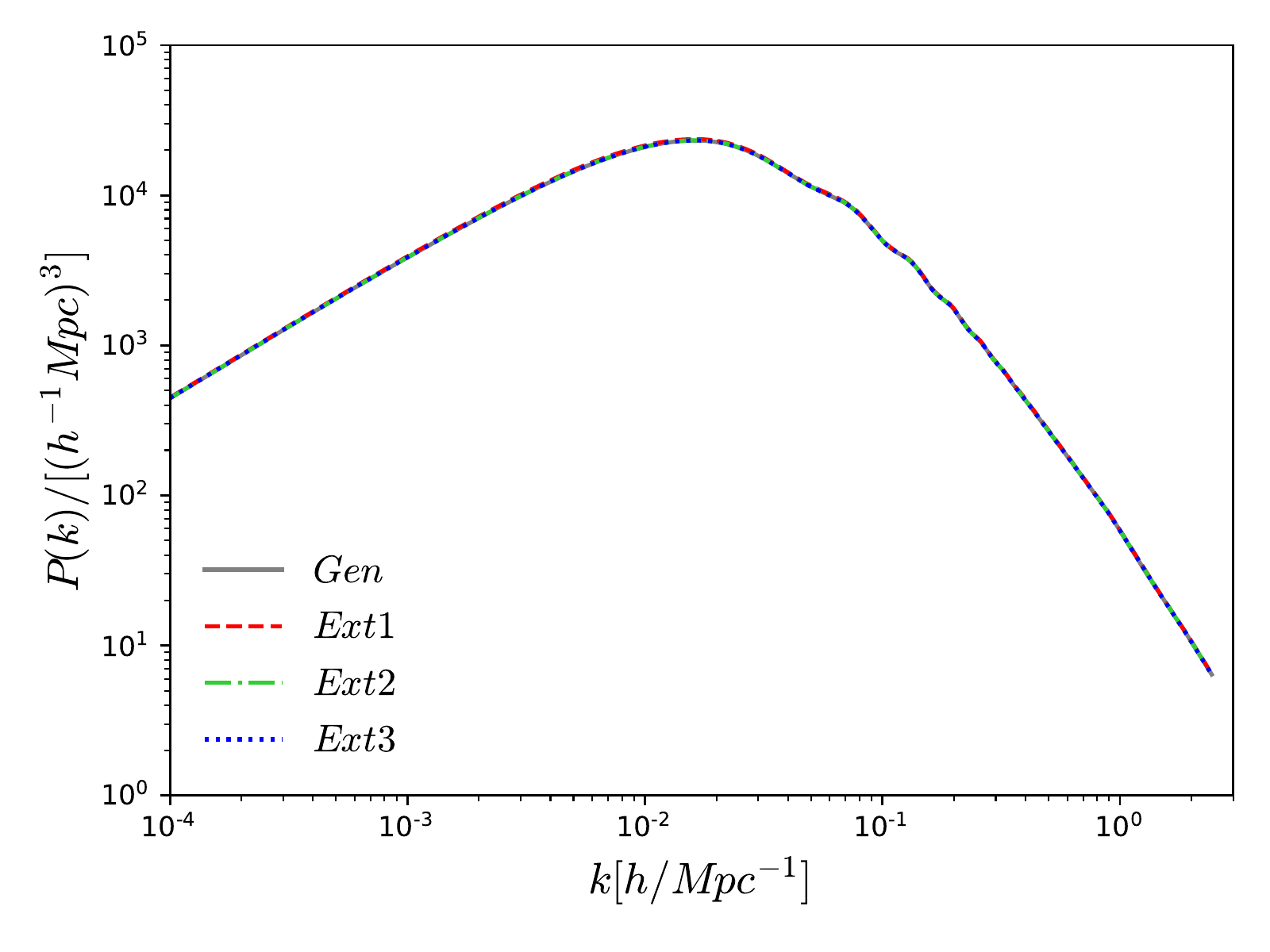} %
\includegraphics[width=0.3\textwidth]{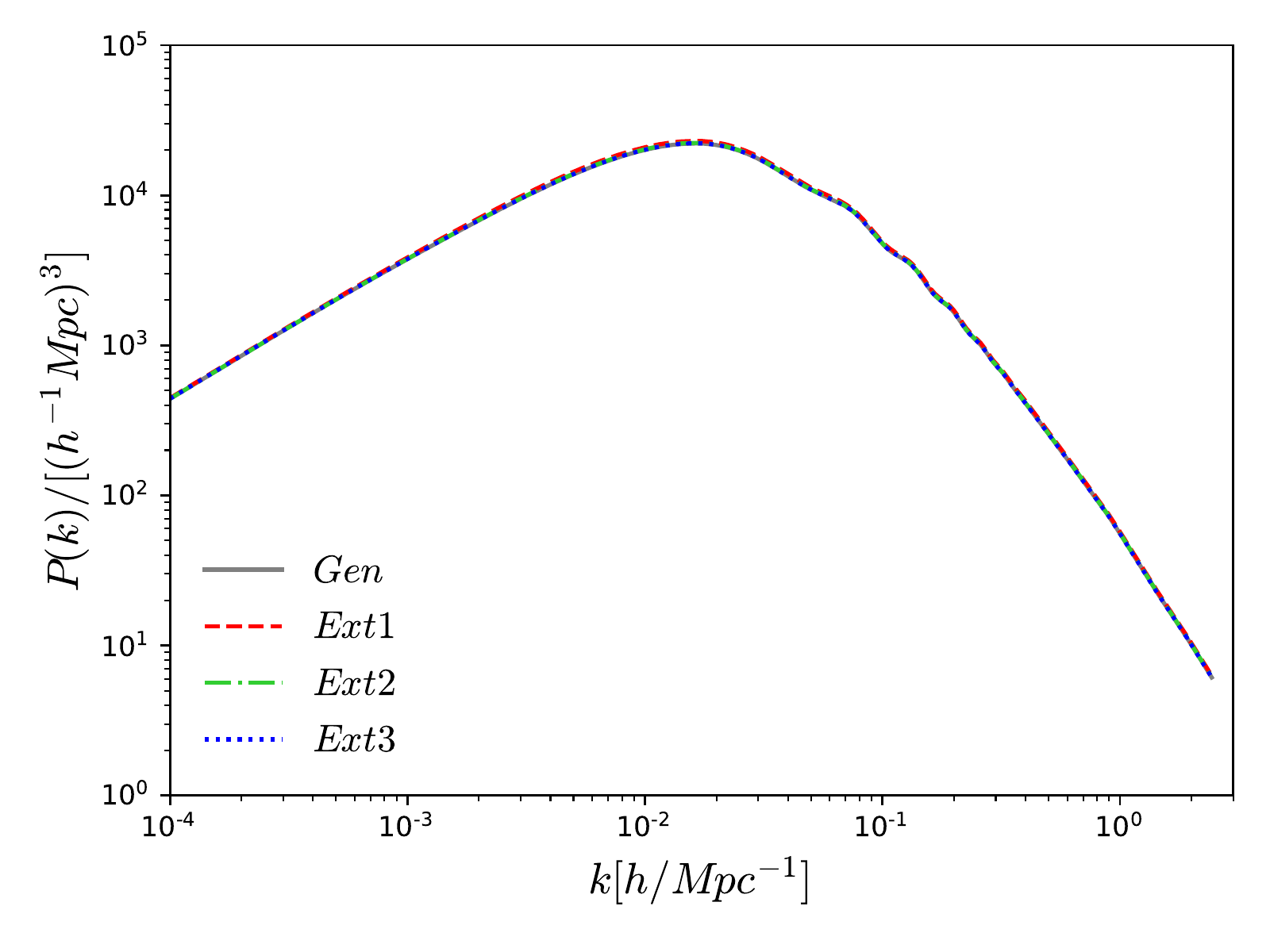}\newline
\includegraphics[width=0.3\textwidth]{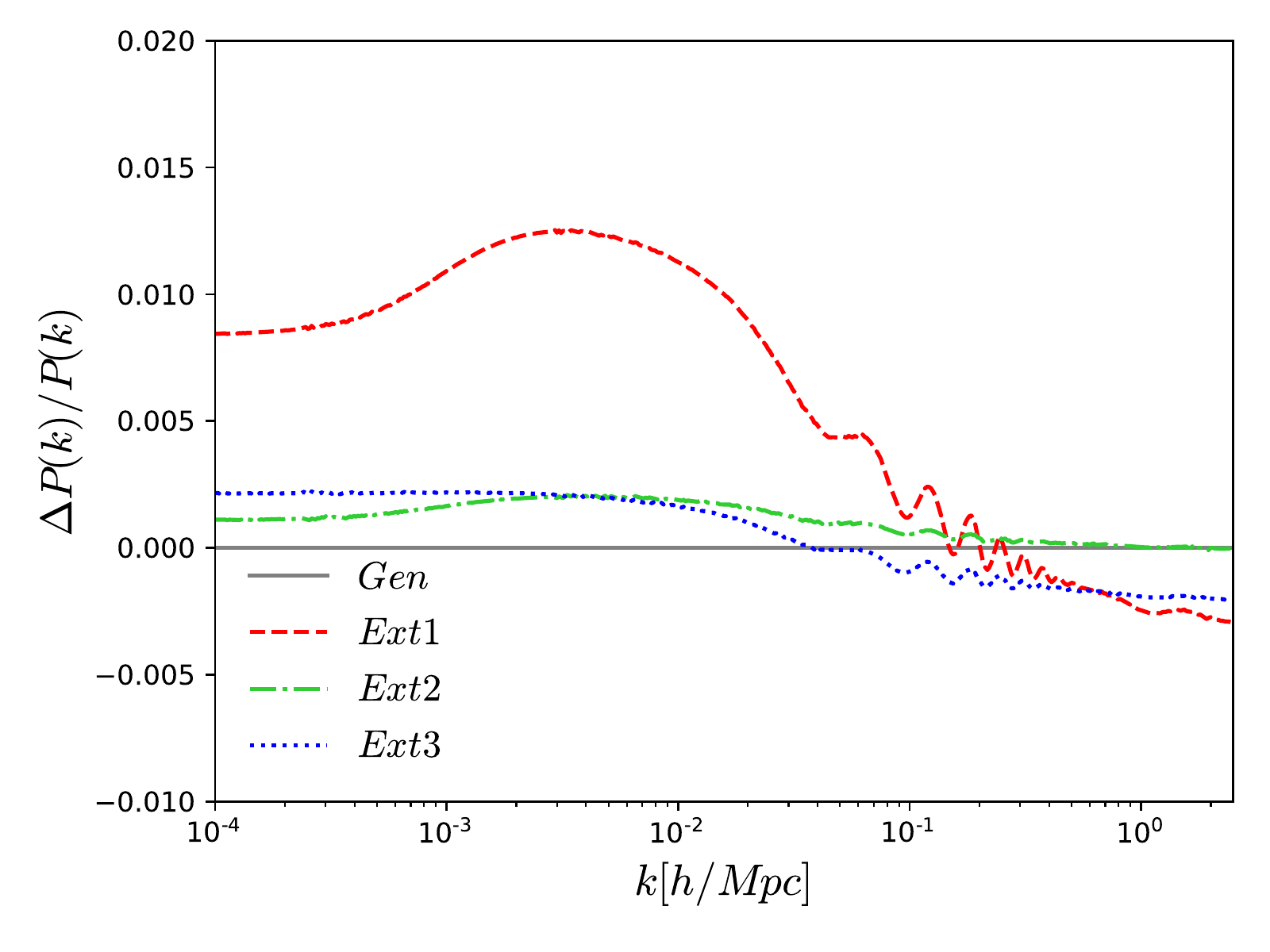} %
\includegraphics[width=0.3\textwidth]{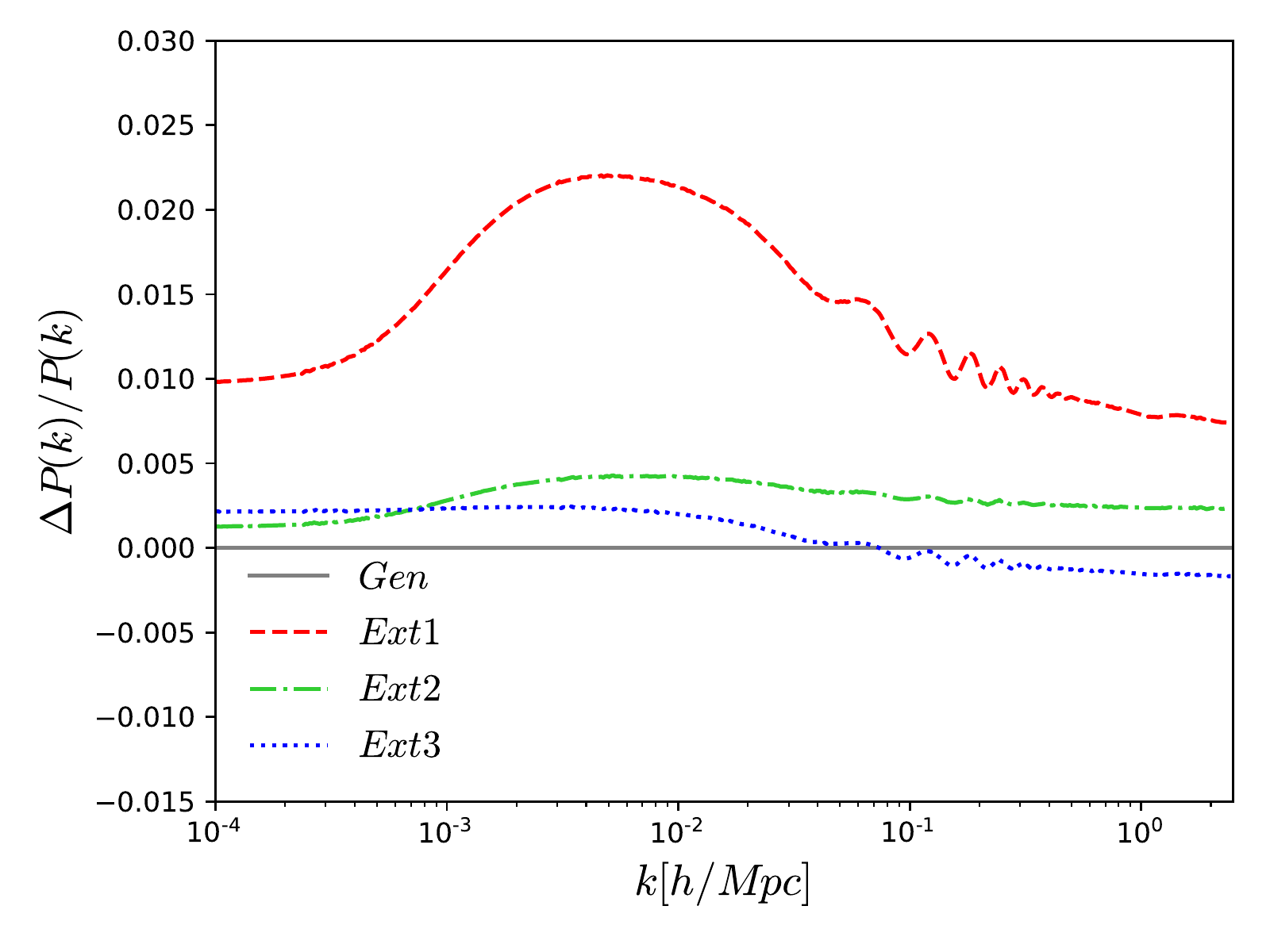} %
\includegraphics[width=0.3\textwidth]{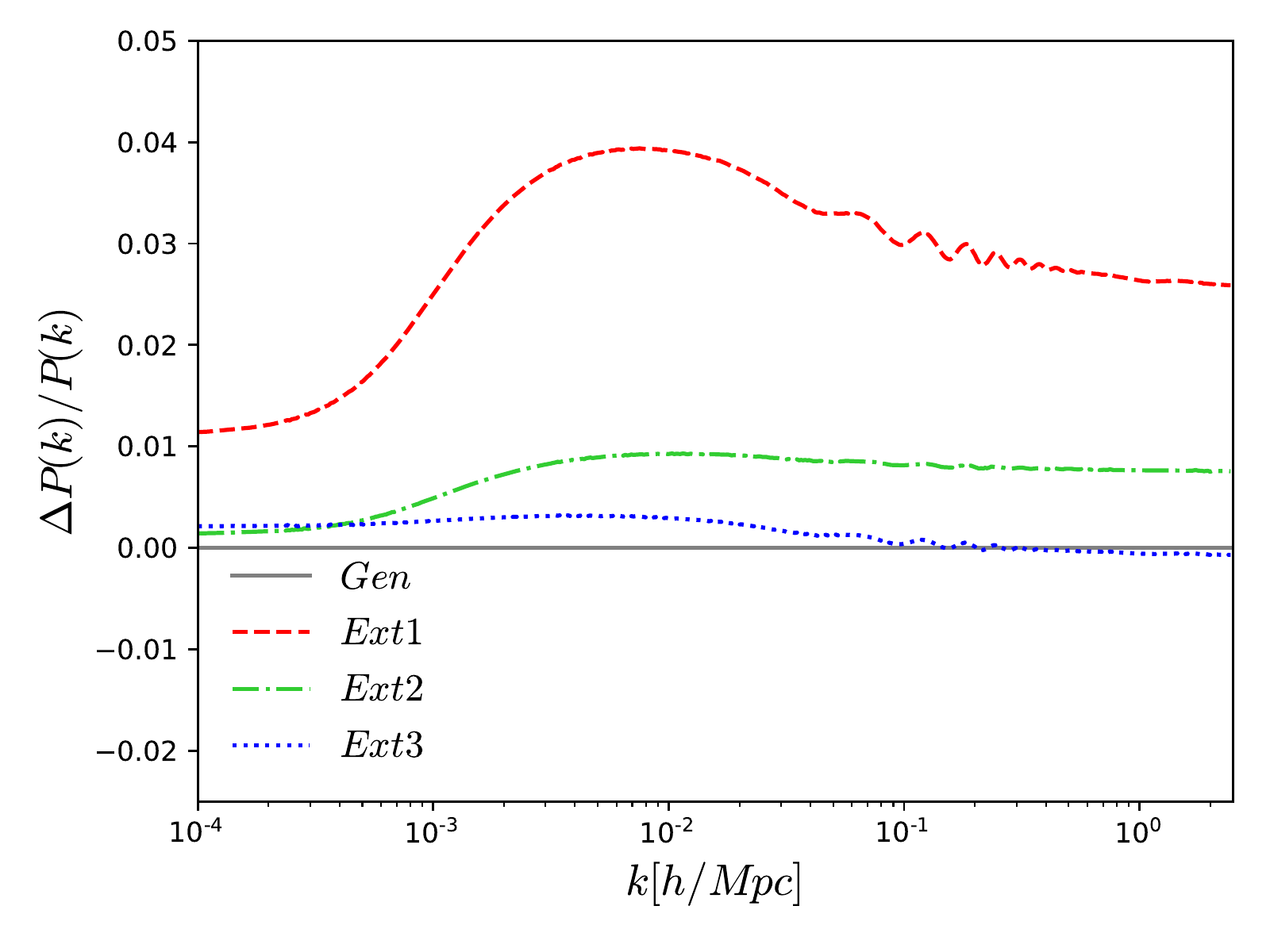}\newline
\includegraphics[width=0.3\textwidth]{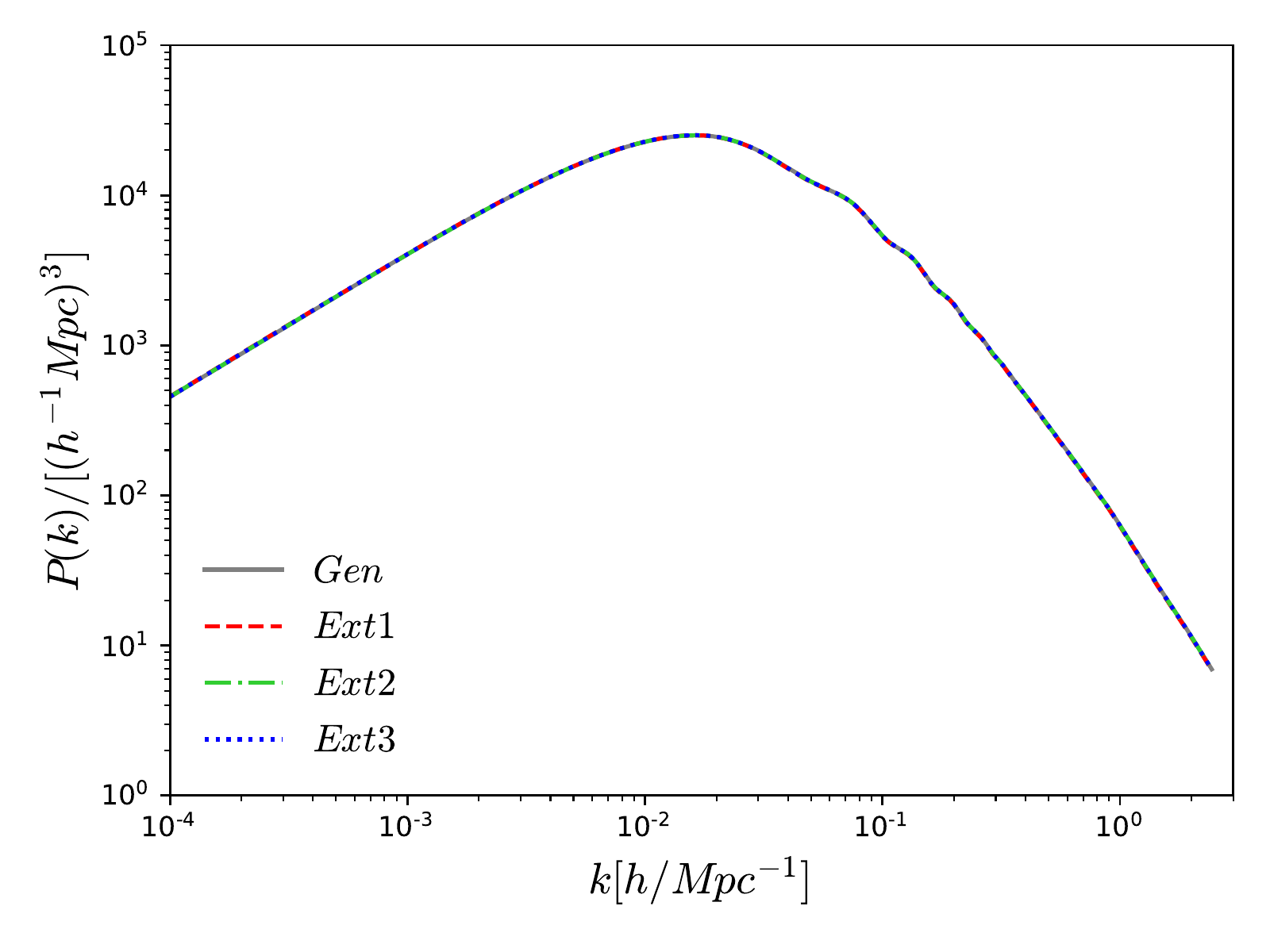} %
\includegraphics[width=0.3\textwidth]{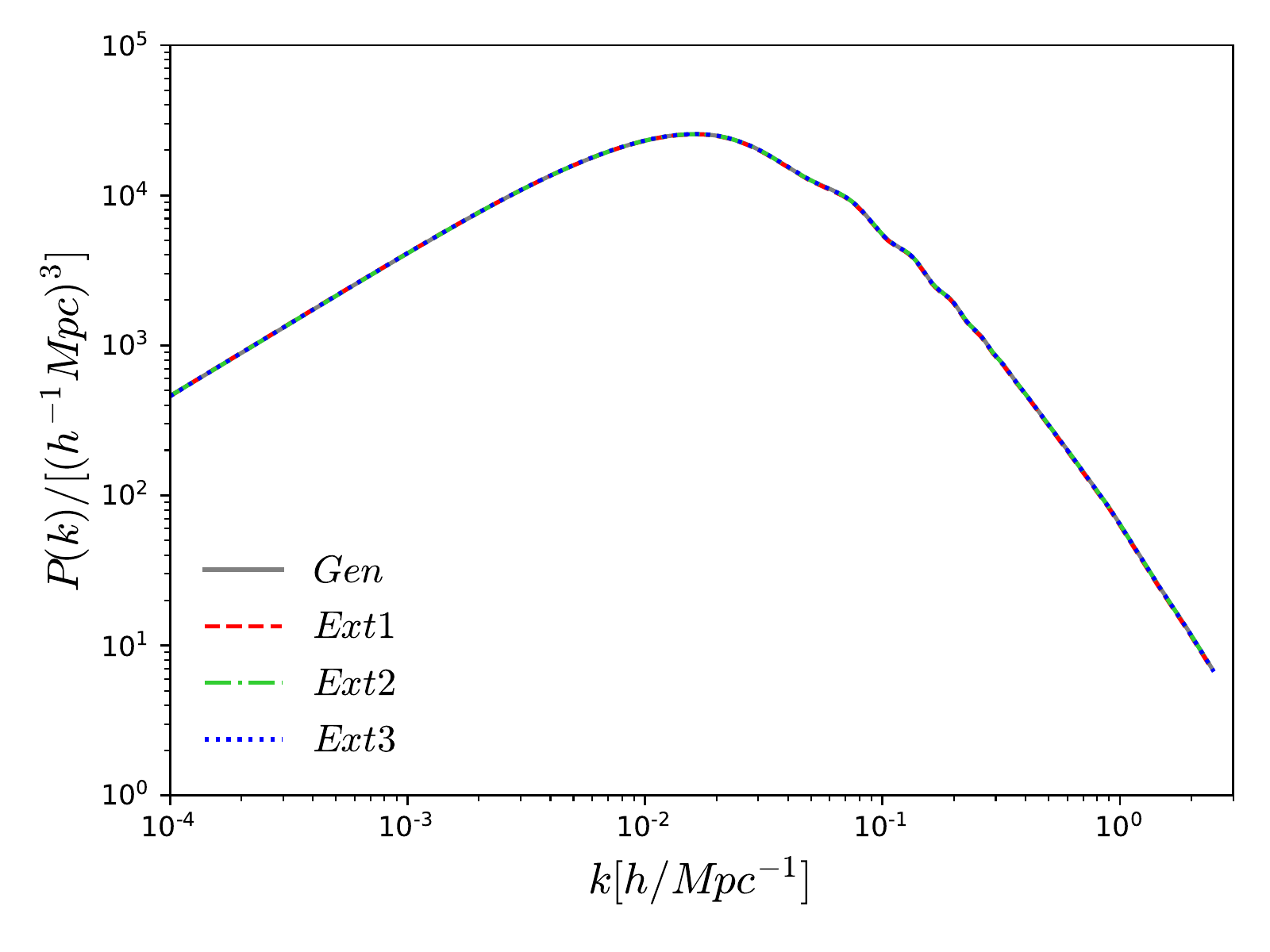} %
\includegraphics[width=0.3\textwidth]{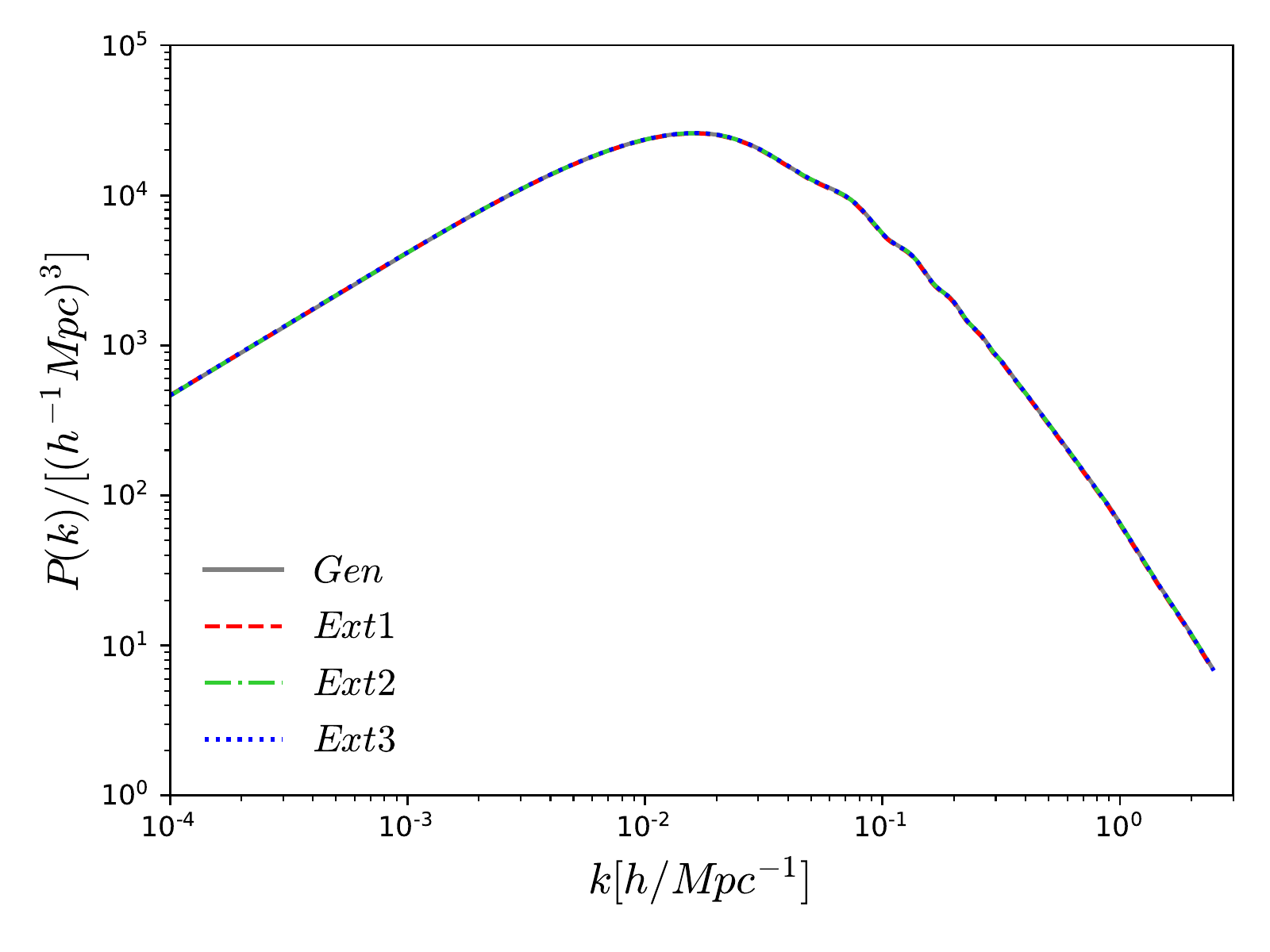}
\includegraphics[width=0.3\textwidth]{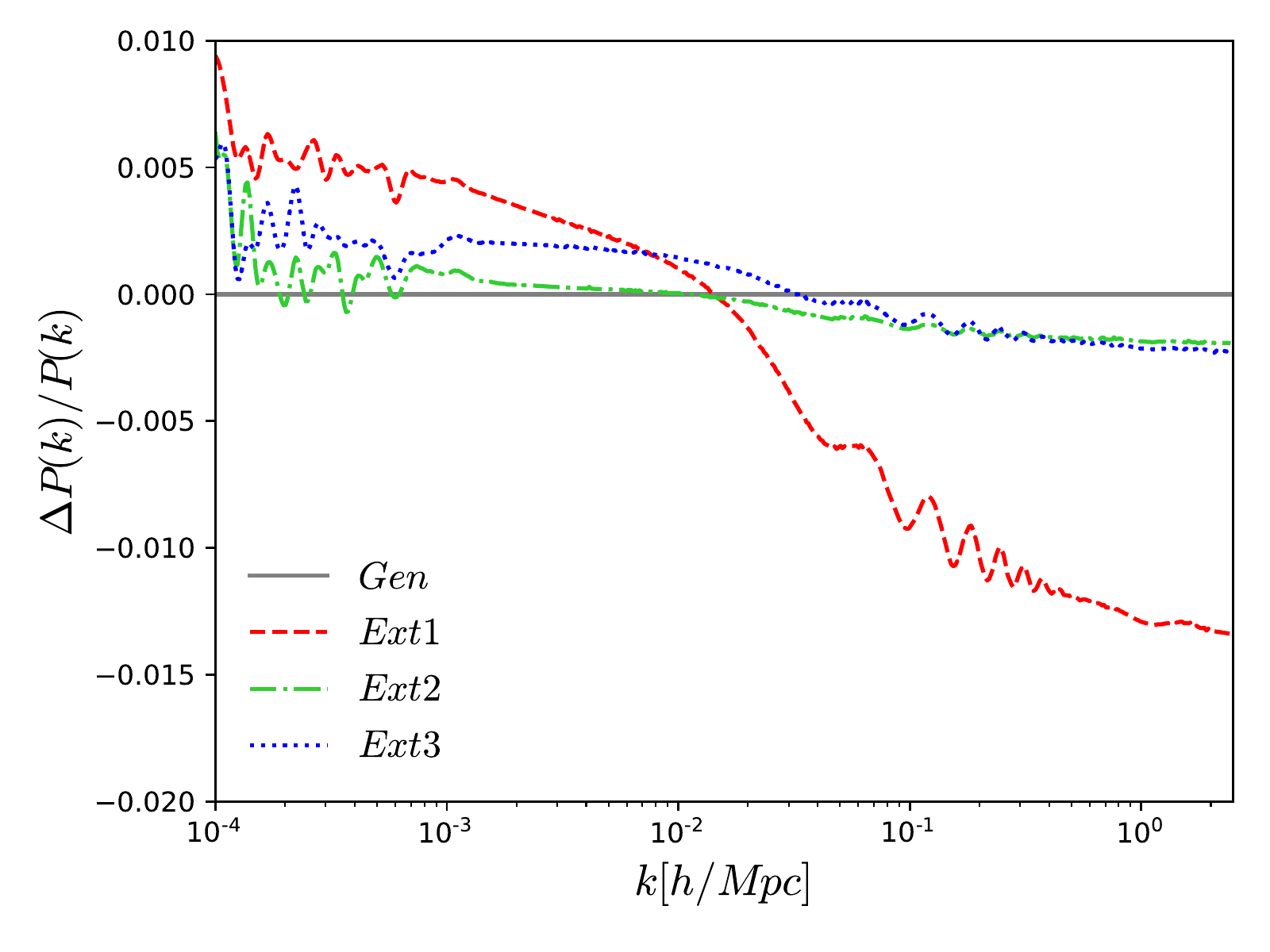} %
\includegraphics[width=0.3\textwidth]{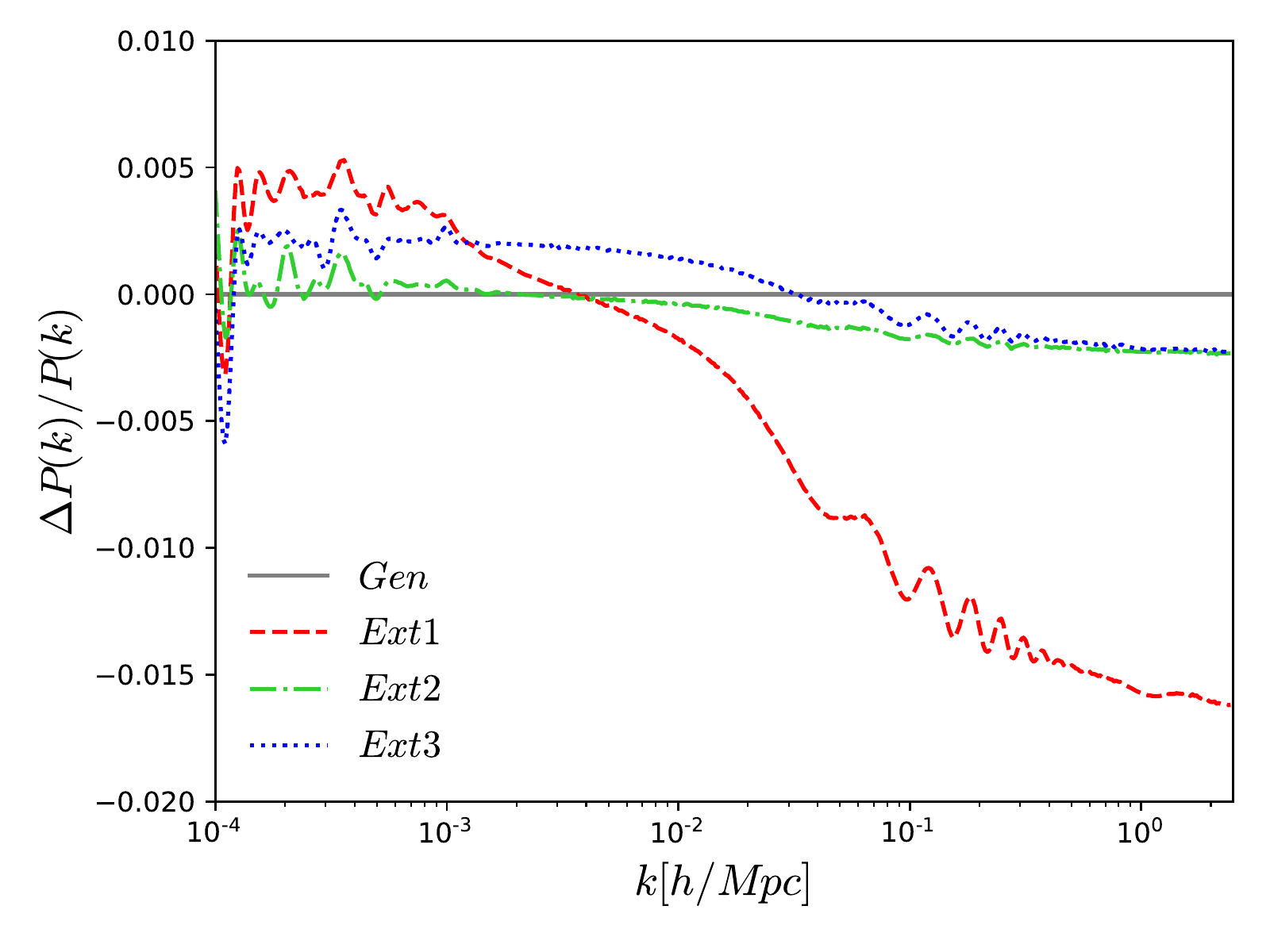} %
\includegraphics[width=0.3\textwidth]{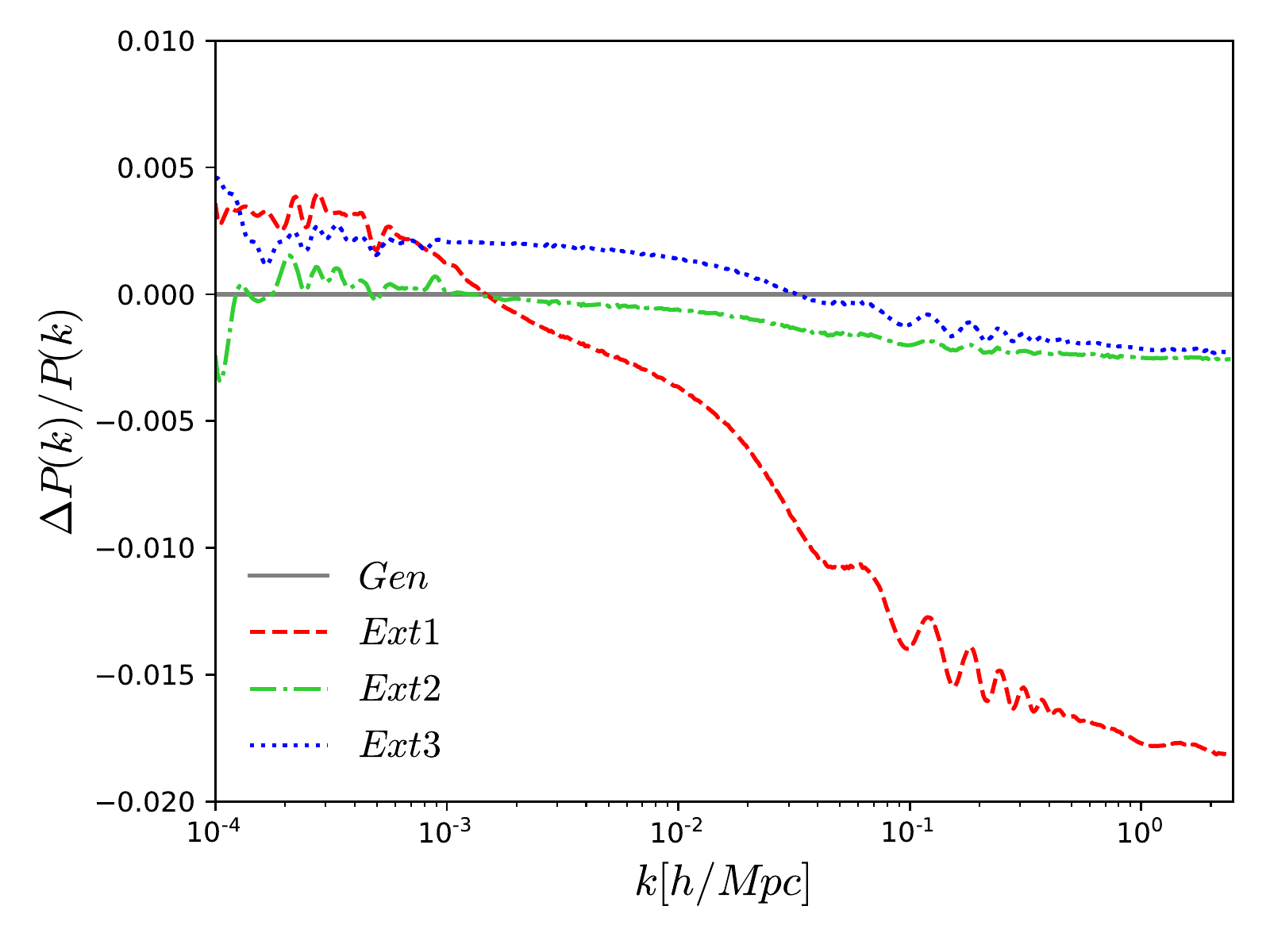}
\caption{We show the matter power  spectra and the corresponding residual plots for the present dynamical dark energy parameterizations using various values of the $w_a$ parameter with a fixed $w_0 =-0.95$.  The plots in the first row present the matter power spectra and the plots in the second row present the corresponding residual plots. The plots from left to right in the first and second panels of this figure respectively stand for $w_a  = 0.1, 0.2,$ and $0.3$. Similarly, the plots in the third row (showing the matter power spectra) and last row (residual plots of the third row) stand for $w_a < 0$ in which the plots from left to right for both the above rows (third and last rows) of this figure respectively stand for $w_a  = -0.1, -0.2,$ and $-0.3$. }
\label{fig:mpowerplots}
\end{figure*}

Following a similar graphical strategy applied to matter power spectra plots as shown in Fig. \ref{fig:mpowerplots}, we arrive at the same conclusion that the models are only distingusihed from one another if we look at the residual plots, that means the plots summarized in the second and last rows of Fig. \ref{fig:mpowerplots}. 
\begin{center}
\begin{table}[!h]
\begin{tabular}{cc}
\hline\hline
$\ln B_{ij}$ & $\ \ $ Evidence for model ${M}_i$ $\ \ $ \\ \hline
$0 \leq \ln B_{ij} < 1$ & Weak \\ 
$1 \leq \ln B_{ij} < 3$ & Definite/Positive \\ 
$3 \leq \ln B_{ij} < 5$ & Strong \\ 
$\ln B_{ij} \geq 5$ & Very strong \\ \hline\hline
\end{tabular}%
\caption{Revised Jeffreys scale quantifying the observational viability of
any cosmological model $M_i$ compared to some reference model $M_j$ 
\protect\cite{Kass:1995loi}. }
\label{tab:jeffreys}
\end{table}
\end{center}
\begin{center}
\squeezetable
\begin{table*}[tbp]
\begin{center}
\begin{tabular}{ccccccccc}
\hline\hline
Dataset & Model & $\ln B_{ij}$ & Evidence for $\Lambda$CDM  \\ \hline
CMB+BAO & Gen & $-1.4$ & Definite \\ 
CMB+BAO+Pantheon & Gen & $-3.4$ & Strong \\ 
CMB+BAO+Pantheon+CC & Gen &  $-3.6$ & Strong \\ \hline\hline

CMB+BAO & Ext1 & $-1.1$ & Definite \\ 
CMB+BAO+Pantheon & Ext1 & $-3.3$ & Strong \\ 
CMB+BAO+Pantheon+CC & Ext1 &  $-3.4$ & Strong \\ \hline\hline

CMB+BAO & Ext2 &  $-1.7$ & Definite \\ 
CMB+BAO+Pantheon & Ext2 & $-3.2$ & Strong \\ 
CMB+BAO+Pantheon+CC & Ext2 & $-3.3$ & Strong \\ \hline\hline

CMB+BAO & Ext3 & $-1.6$ & Definite \\ 
CMB+BAO+Pantheon & Ext3 & $-3.9$ & Strong \\ 
CMB+BAO+Pantheon+CC & Ext3 & $-3.5$ & Strong \\ 
\hline\hline\hline
\end{tabular}%
\end{center}
\caption{The values of $\ln B_{ij}$ computed for the present dark energy
parametrizations ($M_i$) where the reference scenario is the $\Lambda$CDM model ($M_j$). We note that the negative value of $\ln B_{ij}$ means that $\Lambda$CDM is preferred. }
\label{tab:bayesian}
\end{table*}
\end{center}

\section{Bayesian evidence}
\label{sec-bayesian}

A general and natural question that we will be looking for in this section is that, 
how the models are efficient 
compared to the standard $\Lambda$CDM cosmology. 
Thus, we need a statistical comparison between all four dynamical DE parametrizations where the base model will be fixed as $\Lambda$CDM. This statsitical comparison comes through the Bayesian evidence.  
Here we apply publicly available code \texttt{MCEvidence} 
\cite{Heavens:2017hkr,Heavens:2017afc}\footnote{See
\href{https://github.com/yabebalFantaye/MCEvidence}{github.com/yabebalFantaye/MCEvidence}.} to compute the evidences of the models. The use of 
\texttt{MCEvidence} is very easy since 
the code only needs the MCMC chains used to extract the free parameters of the DE parametrizations.

While dealing with Bayesian analysis we need the posterior probability of the model 
parameters (denoted by $\theta$), given a specific observational data ($x$) with any prior information for a model ($M$). Following Bayes theorem, one can write,
\begin{eqnarray}\label{BE}
p(\theta|x, M) = \frac{p(x|\theta, M)\,\pi(\theta|M)}{p(x|M)},
\end{eqnarray}
where $p(x|\theta, M)$ is the likelihood as a function of $\theta$ and  
$\pi(\theta|M)$ refers to the prior information. 
Here, the quantity $p(x|M)$ appearing in the denominator of (\ref{BE}) is the Bayesian evidence that we actually need for the model comparison. 
Now, for two  cosmological models $M_i$, $M_j$ where $M_j$ is acting as the reference 
model \footnote{The reference model should be the most well motivated cosmological model that must be highly sound to the observational data; and without any doubt, $\Lambda$CDM is the best choice for such a model comparison. },  
the posterior probability is, 

\begin{eqnarray}
\frac{p(M_i|x)}{p(M_j|x)} = \frac{\pi(M_i)}{\pi(M_j)}\,\frac{p(x| M_i)}{p(x|M_j)} = 
\frac{\pi(M_i)}{
\pi(M_j)}\, B_{ij},
\end{eqnarray}
in which $B_{ij} = \frac{p(x| M_i)}{p(x|M_j)}$, is the Bayes factor of the 
model $M_i$ relative to $M_j$. And based on the values of 
$B_{ij}$ (alternatively, $\ln B_{ij}$) 
we quantify the 
observational support of the underlying model $M_i$ relative to $M_j$. 
The quantification is done through the 
widely  accepted Jeffreys scales \cite{Kass:1995loi} (see Table \ref{tab:jeffreys}). 
We also note that the negative values of $\ln B_{ij}$ indicate that the reference model ($M_j$) is preferred over the underlying model ($M_i$).  
 
In Table \ref{tab:bayesian} we have shown the values of $\ln B_{ij}$ computed 
for all DE parametrizations considering all the datasets. We find that the values of 
$\ln B_{ij}$ are all negative indicating that $\Lambda$CDM is always preferred and this is true for all the observational datasets.

\section{Concluding remarks}

\label{sec-conclu}

The dark energy, a hypothetical fluid in Einstein gravity is the main
concern of this work. This dark energy, as examined by many investigators
since the year 1998, could be anything obeying only one condition that the
pressure of the fluid should be negative. Thereafter, a cluster of dark
energy models have been introduced and confronted with the observational
data, see \cite{Copeland:2006wr} to get an overview of the models.

Among them an interesting construction of the dark energy models comes
through the equation of state of dark energy, $w_{x}=p_{x}/\rho _{x}$ which
in principle is the function of the underlying cosmological time parameter,
usually the function of the redshift. Technically, there is no such
restriction to pick up any specific functional form for $w_{x}$, however,
the viability of the model is only tested through the observational data and
its effects on the large scale structure of the universe indeed. According
to the investigations performed in the last couple of years, the
Chevallier-Polarski-Linder parametrization is a feasible and well
functioning dark energy parametrization with the observational data. The
present work is motivated in the same direction whilst we have investigated
something different as follows.

We have introduced a new dark energy parametrization (\ref{model-general})
having a novel feature. The model recovers the well known CPL
parametrization in its first order Taylor series expansion around $z=0$.
Thus, the model actually presents a generalized version of the CPL
parametrization. Since the model is a nonlinear generalized version of the
CPL model, thus, a natural inquiry one may ask for is, how its higher order
corrections are important for the expansion history of the universe, and
moreover, how the higher order corrections could affect the evolution of the
universe at the level of background and perturbations. In order to
investigate these issues, we have considered the generalized model (\ref%
{model-general}) together with its first, second and third order Taylor
approximations around the present cosmic epoch $z=0$, given in equations (%
\ref{extension1}), (\ref{extension2}) and (\ref{extension3}). Since the
original model (\ref{model-general}) contains only two free parameters $w_{0}
$ (current value of the dark energy equation of state) and $w_{a}$
(parameter quantifying the dynamical nature of the DE), thus its extensions
contain the same free parameters. We then constrain all the models using a
class of astronomical data, such as CMB, BAO, Pantheon from SNIa and the
Hubble parameter measurements (summarized in section \ref{sec-data}).

The observational constraints are summarized in Table \ref{tab:model-gen}
(for eqn. (\ref{model-general})), Table \ref{tab:ext1CPL} (for eqn. (\ref%
{extension1})), Table \ref{tab:ext2} (for eqn. (\ref{extension2})), Table %
\ref{tab:ext3} (for eqn. (\ref{extension3})) and the graphical variations of
the model parameters are also shown in Fig. \ref{fig:contour-gen}, Fig. \ref%
{fig:contour-ext1CPL}, Fig. \ref{fig:contour-ext2} and Fig. \ref%
{fig:contour-ext3}, respectively for the general, Ext1, Ext2, and Ext3 
parametrizations. From
the analyses, it is clear that the cosmological parameters
assume similar constraints and according to the employed 
observational data applied to the present models, the  
dark energy equation of state at present, $w_0$, is consistent to $w_0 = -1$ scenario. 
In addition, we find that,  for CMB+BAO data, $H_0$ for all parametrizations, assumes very lower values, if we disregard its error bars, however, for CMB+BAO+Pantheon and CMB+BAO+Pantheon+CC, $H_0$ increases with slightly higher error bars compared to the $\Lambda$CDM based Planck's estimation \cite{Ade:2015xua}, and as a result the tension on $H_0$ is slightly reduced. However, at the level of background, the models cannot be distinguished from one another while from the investigations at perturbations stage, one can distinguish between the models, see the residual plots in Fig. \ref{fig:cmbplots} and Fig. \ref{fig:mpowerplots}.

We also performed the Bayesian evidence analysis using the \texttt{MCEvidence} 
and compared the models with respect to the standard $\Lambda $CDM
reference scenario. Our analysis reveals that $\Lambda $CDM is
favored over all the dynamical DE parametrizations. This is an expected result because
the parameters space of the leading cosmic scenarios driven by the present dynamical DE parametrizations are of eight
dimensional while  $\Lambda $CDM has only six parameters.
 
Last but not least, we would like to comment that the model (\ref%
{model-general}), so far we are aware of the literature, is a new one in the
field of dark energy which naturally recovers CPL parametrization in its 
first order approximation and sounds good with the Bayesian evidence.  
Therefore, a number of investigations can be performed in various contexts of current interests. A quite straightforward and appealing investigation would be to measure the mass bounds of neutrinos in such a generalized framework. Moreover, it will be further interesting to consider a number of upcoming cosmological surveys, such as, Simons Observatory Collaboration (SOC) \cite{Ade:2018sbj}, 
Cosmic Microwave Background Stage-4 (CMB-S4) \cite{Abitbol:2017nao}, EUCLID Collaboration \cite{Scaramella:2015rra,Laureijs:2011gra}, 
Dark Energy Spectroscopic Instrument  (DESI) \cite{Aghamousa:2016zmz}, Large Synoptic Survey Telescope (LSST) \cite{Newman:2019doi,Hlozek:2019vjs,Mandelbaum:2019zej}, in order to forecast the present dark energy parametrizations. The inclusion of gravitational waves data from various sources, such as, Laser Interferometer Space Antenna (LISA) \cite{Audley:2017drz}, Deci-hertz Interferometer Gravitational wave Observatory (DECIGO) \cite{Kawamura:2011zz,Sato:2017dkf}, TianQin \cite{Luo:2015ght}, is also an appealing direction of research in this direction.

\section*{Acknowledgments}
The authors thank the referee for some useful and important comments that helped to improve the quality of the manuscript. 
SP acknowledges the financial support through the Faculty Research and
Professional Development Fund (FRPDF) Scheme of Presidency University,
Kolkata, India. WY was supported by the financial support from the National
Natural Science Foundation of China under Grants No. 11705079 and No.
11647153. AP acknowledges financial support of FONDECYT grant no. 3160121.

%------------------------------------------------------------------

\end{document}